\documentclass[aps,twocolumn,superscriptaddress,a4paper,floatfix]{revtex4}

\usepackage{amsmath}
\usepackage{graphicx}
\usepackage{epsfig}
\usepackage{color}
\usepackage{enumerate}
\usepackage{bm}
\usepackage{amssymb}
\usepackage{relsize} 
\usepackage{lipsum} 
\usepackage{cancel}
\usepackage[normalem]{ulem}
\usepackage{soul} 
\usepackage{float}

\usepackage{rotating}
\usepackage{adjustbox}
\usepackage{xurl}
\usepackage{scalerel,stackengine}

\makeatletter
\let\NR@nopatch@sectioning\relax
\makeatother

\usepackage{silence}
\usepackage{hyperref}
\usepackage{bookmark}
\usepackage{mathrsfs}

\hypersetup{
	bookmarksopen=false,
    unicode=false,          
    pdftoolbar=true,        
    pdfmenubar=true,        
    pdffitwindow=true,     
    pdfstartview={FitH},    
    pdfpagemode=UseNone,		
    pdftitle={Modified Kalman Filtering Derived from Non-Maxwellian Distribution Functions in Open Systems},    
	pdfauthor={Olivier Izacard},     
    pdfnewwindow=true,      
    colorlinks=true,       
    linkcolor=blue,          
    citecolor=blue,        
    filecolor=magenta,      
    urlcolor=blue           
}

\newcommand{\bgeqa}{\begin{eqnarray}}
\newcommand{\edeqa}{\end{eqnarray}}
\newcommand{\vpartb}[1]{{\bf{{#1}}}{}}	

\makeatletter
\let\oldthebibliography\thebibliography
\let\endoldthebibliography\endthebibliography

\renewenvironment{thebibliography}[1]
  {%
   \begingroup
   \let\section\@gobbletwo
   \oldthebibliography{#1}%
  }
  {%
   \endoldthebibliography
   \endgroup
  }
\makeatother

\stackMath
\newcommand\reallywidehat[1]{%
\savestack{\tmpbox}{\stretchto{%
  \scaleto{%
    \scalerel*[\widthof{\ensuremath{#1}}]{\kern-.6pt\bigwedge\kern-.6pt}%
    {\rule[-\textheight/2]{1ex}{\textheight}}
  }{\textheight}%
}{1.25ex}}
\stackon[1pt]{#1}{\tmpbox}%
}

\begin{document}

\title{Modified Kalman Filtering Derived\\ from Non-Maxwellian Distribution Functions in Open Systems}
\author{Olivier Izacard}
\affiliation{San Diego, California, USA}

\begin{abstract}
Kalman filtering (KF) recursively infers plasma quantities, represented by a state, from noisy diagnostics while propagating uncertainty in the inferred state separately from diagnostic noise.
For linear dynamical and measurement models with Gaussian probability density functions (PDFs), the state mean and covariance provide the KF description.
We modify this KF for open plasmas with particle and energy sources by extending the physical state from Maxwellian variables to retained NMDF coordinates, whose evolution follows a projection of the nonlinear Landau--Fokker--Planck equation.
The same NMDF state is propagated through a fixed diagnostic response to obtain the corresponding measurement PDF.
Because a non-Gaussian likelihood can drive the posterior outside the Gaussian family, the mean-covariance representation is extended to a finite moment closure whose coordinates are related to the retained moments through their Jacobian.
The recursive prediction-correction structure is preserved, recovering conventional KF results for linear models with Gaussian PDFs.
As a proof of concept, we use seven published non-Gaussian Alcator C-Mod Langmuir-probe current PDFs.
For each target PDF, NMDF response parameters are calibrated from the other probe PDFs and frozen; this excluded target tests independent prediction.
This test shows that different plasma regions require different NMDF structures, instead of the standard assumption of Maxwellian velocity distributions across all regimes and locations.
Because the published measurement PDFs do not retain time ordering, this proof of concept does not yet test recursive uncertainty dynamics.
Implementing the projected Landau--Fokker--Planck prediction with time-resolved diagnostics is the next step toward full validation of the modified KF.
\end{abstract}

\date{\today}
\maketitle

\section{Introduction}
\label{sec:introduction}

Kalman filtering provides a finite-dimensional recursive representation of conditional uncertainty by evolving the mean and covariance of a state estimate~\cite{Kalman_1960}.
For linear dynamics with Gaussian process and observation statistics~\cite{KalmanBucy_1961}, this representation is not an approximation: the complete posterior remains Gaussian~\cite{Jazwinski_1970,AndersonMoore_1979}, and the prediction and measurement steps reduce exactly to matrix evolution and a quadratic optimization.
The same structure explains both the efficiency and the limitation of the method.
Gaussian nonlinear extensions such as the unscented Kalman filter~\cite{JulierUhlmann_2004} and cubature Kalman filter~\cite{ArasaratnamHaykin_2009} improve the propagation of nonlinear moments while retaining a Gaussian state representation.
Accordingly, statistical structure that lies outside the mean-covariance manifold is discarded, absorbed into the process covariance, or reintroduced indirectly through an empirical observation model.
\\
Open systems provide a different statistical setting.
Collisions relax the velocity distribution toward a local Maxwellian state~\cite{Braginskii_1965}, while external heating, particle injection, spatial gradients, intermittent losses, and finite-time relaxation continually drive it away from that state.
Strong gradients can additionally produce nonlocal kinetic transport~\cite{LucianiMoraVirmont_1983} and transport coefficients that depart from local Maxwellian closures~\cite{EpperleinHaines_1986}.
Finite-moment closures retain only selected information from the velocity distribution~\cite{Grad_1949,ChapmanCowling_1970}, whereas source-driven non-Maxwellian coordinates can retain part of the missing kinetic transport explicitly~\cite{Izacard_2017,Izacard_2026_INMDF}.
A closure restricted to evolving Maxwellian parameters can reproduce prescribed changes of density, flow, or temperature without representing the kinetic redistribution through which those changes are produced~\cite{Izacard_2017}.
A diagnostic inversion restricted to that Maxwellian model can consequently return effective Maxwellian-like quantities even when the diagnostic remains sensitive to a non-Maxwellian population~\cite{Izacard_2016}.
The discarded structure may then be absorbed into residuals, empirical transport terms, or inferred uncertainty.
A Gaussian posterior introduces a second, independent reduction by retaining only the posterior mean and covariance.
Non-Gaussian structure can therefore enter both through the physical kinetic state and measurement likelihood and through the posterior used to infer that state.
\\
Existing non-Gaussian filters address parts of this problem through robust or heavy-tailed likelihoods~\cite{Huber_1964,MasreliezMartin_1977,Roth_2013}, skewed distributions~\cite{Azzalini_1985}, Gaussian-mixture representations~\cite{AlspachSorenson_1972}, particle ensembles~\cite{Gordon_1993,Kitagawa_1996,Doucet_2000,Arulampalam_2002}, or projection onto a selected parametric family~\cite{BrigoHanzonLeGland_1998}.
These methods establish that non-Gaussian filtering is possible.
The question addressed here is more specific: which non-Gaussian coordinates are generated by the open-system kinetic dynamics and remain observable through a fixed diagnostic?
Accordingly, the diagnostic likelihood and state-prediction dynamics are derived from the kinetic model before a finite posterior closure is selected, while unresolved state-space diffusion is specified independently of the held-out diagnostic statistics.
The first explicit realization uses an INMDF channel because its amplitude, support center, and kinetic widths already possess source--collision dynamics.
A Kappa posterior is retained separately as an alternative representation of broad power-law tails~\cite{Vasyliunas_1968,SummersThorne_1991,LivadiotisMcComas_2009,PierrardLazar_2010,LazarFichtnerYoon_2016} and is not included in the INMDF coordinate vector.
\\
The construction is organized around five core objects; \(\mathbf X_s\) and \(\mathbf U_s\) are coordinate vectors attached to these objects rather than additional probability laws.
The kinetic manifold specifies the physical distribution \(f_s(\mathbf X_s,\mathbf v)\), whose normalized form is \(K_s=f_s/n_s\).
The diagnostic map \(\mathcal H_D\) converts \(K_s\) into either a measurement PDF \(p_{D,s}(y\mid\mathbf X_s)\) or a scalar diagnostic signal whose PDF follows from temporal or ensemble variation.
The projected kinetic equation evolves the physical coordinates \(\mathbf X_s\), while \(\rho(\mathbf X,t\mid\mathcal D_t)\) is their posterior after conditioning on the measurements.
When \(\rho\) is represented by a finite posterior closure, its coordinates are denoted by \(\mathbf U_s\).
Thus \(f_s\), \(K_s\), \(\mathcal H_D\), \(p_{D,s}\), and \(\rho\) retain one meaning throughout the derivation.
Table~\ref{tab:construction_roadmap} summarizes the maps connecting these objects and the sequence used below.
\begin{table*}[t]
\centering
\small
\renewcommand{\arraystretch}{1.18}
\caption{
Roadmap of the physical-to-inferential construction.
The kinetic distribution, measurement PDF, and posterior remain distinct throughout.
}
\label{tab:construction_roadmap}
\begin{tabular}{|
p{0.110\textwidth}|
p{0.125\textwidth}|
p{0.185\textwidth}|
p{0.285\textwidth}|
p{0.205\textwidth}|
}
\hline
\textbf{Stage}
&
\textbf{Object}
&
\textbf{Map or tool}
&
\textbf{Assumption introduced}
&
\textbf{Output}
\\
\hline
\begin{tabular}{l}Physical\\ kinetic state\end{tabular}
&
$f_s(\mathbf X_s,\mathbf v)$
&
Open kinetic equation and retained velocity-space manifold
&
Choice of MDF, INMDF, Kappa, two-population, or another positive kinetic family
&
Physical coordinates $\mathbf X_s(t)$
\\
\begin{tabular}{l}Normalized\\ kinetic state\end{tabular}
&
$K_s=f_s/n_s$
&
Normalization in velocity space
&
No measurement model is introduced
&
Probability entering the diagnostic map
\\
\begin{tabular}{l}Diagnostic\\ map\end{tabular}
&
$\mathcal H_D$
&
Collection, transmission, or spectral-response map
&
Geometry and calibration are fixed independently of the kinetic manifold
&
$p_{D,s}(y\mid\mathbf X_s)$ or a scalar signal $y_{D,s}(t)$
\\
Measurement statistics
&
$p_{D,s}(y)$
&
Direct diagnostic law or time/ensemble pushforward of $y_{D,s}(t)$
&
No independently fitted residual family is required
&
Likelihood, moments, cumulants, and correlations
\\
\begin{tabular}{l}State\\ prediction\end{tabular}
&
$\rho^{\mathrm{pred}}(\mathbf X,t)$
&
Projected kinetic dynamics and Fokker--Planck evolution
&
Unresolved diffusion is fixed independently of held-out measurements
&
Predicted posterior
\\
Bayesian correction
&
$\rho^{\mathrm{exact}}\rightarrow\rho_s(\mathbf U_s)$
&
Bayes' rule; admissible moment matching or, when required, constrained KL projection
&
The posterior closure must remain normalized, positive, and identifiable
&
Corrected posterior coordinates $\mathbf U_s$
\\
\begin{tabular}{l}Gaussian\\ specialization\end{tabular}
&
$(\widehat{\mathbf X},\mathsf P)$
&
Gaussian invariant closure
&
Affine state dynamics, state-independent diffusion, linear observation map, and Gaussian observation diffusion
&
Ordinary Kalman--Bucy and discrete Kalman equations
\\
\hline
\end{tabular}
\end{table*}
Section~\ref{sec:kinetic_to_measurement} defines the kinetic manifolds and diagnostic map.
Sections~\ref{sec:continuous_filtering} and \ref{sec:structured_closure} derive the continuous state dynamics and finite posterior closure, and Section~\ref{sec:continuous_discrete} converts them into an executable prediction--correction cycle.
Section~\ref{sec:nmdf_uncertainty} then distinguishes process noise, measurement noise, predicted-measurement uncertainty, and posterior state uncertainty, and derives how the full non-Gaussian diagnostic likelihood modifies the inferred uncertainty beyond the Gaussian Kalman covariance formula.
Section~\ref{sec:entropy} separates kinetic entropy from posterior uncertainty and measurement information.
Section~\ref{sec:inmdf_measurement} specializes the diagnostic map to the Langmuir probe, Sections~\ref{sec:validation} and \ref{sec:kube_results} define and apply the validation tests, and Section~\ref{sec:conclusion} discusses the resulting physical interpretation.

\section{From open-system non-Maxwellian kinetics to diagnostic statistics}
\label{sec:kinetic_to_measurement}

\subsection{Open kinetic system and retained velocity-space manifolds}
We consider a local plasma distribution
\begin{equation}
f_p
\doteq
f_p
\left(
\mathbf X_p(\mathbf x,t),
\mathbf v
\right)
\label{eq:local_distribution_parameterization}
\end{equation}
interacting through nonlinear Landau--Fokker--Planck collisions with itself and with a prescribed source population $f_{\rm src}(\mathbf x,\mathbf v,t)$.
The vector $\mathbf X_p(\mathbf x,t)$ contains the finite local coordinates used to parameterize the retained velocity distribution.
All spatial and temporal dependence of these retained parameters is carried by $\mathbf X_p(\mathbf x,t)$ unless an additional explicit dependence is required.
The local kinetic equation is
\begin{equation}
\partial_t f_p
+
\mathbf v\cdot\nabla_{\mathbf x}f_p
+
\dot{\mathbf v}\cdot\nabla_{\mathbf v}f_p
=
C_{pp}[f_p,f_p]
+
C_{p{\rm src}}[f_p,f_{\rm src}].
\label{eq:open_lfp}
\end{equation}
For two distributions $f_a$ and $f_b$, the bilinear Landau collision operator~\cite{Landau_1937} can be written in Rosenbluth form~\cite{Rosenbluth_1957}; its use for collisional magnetized-plasma transport is reviewed systematically in Ref.~\cite{HelanderSigmar_2002}.
It is
\begin{align}
C_{ab}[f_a,f_b](\mathbf v)
\doteq{}&
\gamma_{ab}
\frac{\partial}{\partial v_i}
\int_{-\infty}^{\infty}
\int_{-\infty}^{\infty}
\int_{-\infty}^{\infty}
U_{ij}(\mathbf v-\mathbf v')
\nonumber\\
&\times
\left[
f_b(\mathbf v')
\frac{\partial f_a(\mathbf v)}{\partial v_j}
-
f_a(\mathbf v)
\frac{\partial f_b(\mathbf v')}{\partial v'_j}
\right]
d^3\vpartb{v},
\label{eq:landau_operator}
\end{align}
where $d^3\vpartb{v} \doteq dv_\parallel dv_{\perp1} dv_{\perp2}$.
Here $\gamma_{ab}>0$ denotes the standard species-dependent Coulomb collision prefactor.
Its conventional dimensional form does not alter the projection geometry developed below.
Repeated velocity indices are summed and
\begin{equation}
U_{ij}(\mathbf w)
\doteq
\frac{
\delta_{ij}|\mathbf w|^2-w_iw_j
}{
|\mathbf w|^3
},
\quad
\mathbf w
\doteq
\mathbf v-\mathbf v'.
\label{eq:landau_tensor}
\end{equation}
With the open kinetic generator specified, the next step is to choose the finite velocity-space manifolds onto which that dynamics will be projected.
The Maxwellian and first-INMDF manifolds are parameterized respectively by
\begin{align}
\mathbf X_M(\mathbf x,t)
\doteq{}&
\left(
n,
u_\parallel,
\mathbf u_\perp,
T
\right),
\nonumber\\
\mathbf X_I(\mathbf x,t)
\doteq{}&
\left(
n,
u_\parallel,
\mathbf u_\perp,
T,
\Gamma,
c,
W_\parallel,
W_\perp
\right).
\label{eq:kinetic_coordinate_vectors}
\end{align}
Thus $\mathbf X_I$ contains the complete Maxwellian backbone $\mathbf X_M$ together with the localized non-Maxwellian coordinates $(\Gamma,c,W_\parallel,W_\perp)$.
\\
The corresponding Maxwellian is
\begin{equation}
f_M
\left(
\mathbf X_M,
\mathbf v
\right)
\doteq
n
\left(
\frac{m}{2\pi T}
\right)^{3/2}
\exp\left[
-\frac{
m|\mathbf v-\mathbf u|^2
}{
2T
}
\right],
\label{eq:maxwellian_distribution}
\end{equation}
where
\begin{equation}
\mathbf u
\doteq
u_\parallel\widehat{\mathbf b}
+
\mathbf u_\perp.
\end{equation}
The first localized INMDF branch introduced for analytic non-Maxwellian kinetic corrections~\cite{Izacard_2016} and subsequently used to construct non-Maxwellian fluid closures~\cite{Izacard_2017} is
\begin{equation}
f_I(\mathbf X_I,\mathbf v)
\doteq
f_M(\mathbf X_M,\mathbf v)
+
\Gamma\eta G_I(\mathbf v),
\label{eq:first_inmdf}
\end{equation}
where
\begin{equation}
\eta
\doteq
v_\parallel-c,
\label{eq:eta_definition}
\end{equation}
and
\begin{equation}
G_I(\mathbf v)
\doteq
\frac{1}{\Delta_I}
\exp\left[
-\frac{m\eta^2}{2W_\parallel}
-\frac{
m|\mathbf v_\perp-\mathbf u_\perp|^2
}{
2W_\perp
}
\right],
\label{eq:GI_definition}
\end{equation}
with
\begin{equation}
\Delta_I
\doteq
\frac{
(2\pi)^{3/2}
W_\perp
W_\parallel^{3/2}
}{
m^{5/2}
}.
\label{eq:DeltaI_definition}
\end{equation}
The notation no longer repeats $n(t)$, $\Gamma(t)$, $c(t)$, $W_\parallel(t)$, or $W_\perp(t)$ inside every expression because their complete local dependence is already contained in $\mathbf X_M(\mathbf x,t)$ and $\mathbf X_I(\mathbf x,t)$.
\\
A Kappa kinetic distribution defines a different velocity-space manifold and is not included in $\mathbf X_I$.
The explicit Kappa kinetic family used in the experimental benchmark is given in Appendix~\ref{app:benchmark_families}, while Section~\ref{sec:structured_closure} introduces a separate Kappa posterior closure in state space.

\subsection{Diagnostic map and measurement statistics}

Having specified the kinetic manifold, we next map it to a measurement.
The diagnostic does not observe \(f_s\) directly; diagnostic \(D\) acts through a fixed map \(\mathcal H_D\) containing its collection, transmission, spectral, geometric, and instrumental response.
When \(\mathcal H_D\) is an integral transform, its weighting function is the response kernel and remains part of the same diagnostic map rather than a second stochastic model.
\\
Before applying this map, define the normalized kinetic probability
\begin{equation}
K_s(\mathbf v\mid\mathbf X_s)
\doteq
\frac{
f_s(\mathbf X_s,\mathbf v)
}{
n_s
}.
\label{eq:conditional_instrument_response}
\end{equation}
By construction,
\begin{equation}
\int_{-\infty}^{\infty}
K_s(\mathbf v\mid\mathbf X_s)
\,d^3\vpartb{v}
=
1.
\label{eq:response_normalization}
\end{equation}
The kinetic family determines \(K_s\); the diagnostic determines \(\mathcal H_D\).
For a diagnostic that directly resolves a measurement coordinate \(y\),
\begin{equation}
p_{D,s}(y\mid\mathbf X_s)
\doteq
\mathcal H_D
\left[
K_s
\right](y).
\label{eq:event_resolved_pushforward}
\end{equation}
We refer to \(p_{D,s}\) as the measurement PDF, or equivalently as the likelihood when it enters Bayes' rule.
For a scalar diagnostic, \(\mathcal H_D\) instead returns an instantaneous signal whose PDF arises from temporal or ensemble variation of \(\mathbf X_s\).
The Langmuir probe in Section~\ref{sec:inmdf_measurement} provides the explicit scalar example.
For a Doppler-resolved laser-induced-fluorescence measurement~\cite{SternJohnson_1975} or a charge-exchange spectroscopic measurement~\cite{Isler_1994}, the resolved measurement coordinate is the line-of-sight velocity, \(y = v_{\mathrm{los}} = \widehat{\mathbf l}\cdot\mathbf v\).
The corresponding diagnostic map is
\begin{equation}
\mathcal H_{\mathrm{Dop}}
\left[
K
\right](y)
\doteq
\int_{-\infty}^{\infty}
G_{\sigma_{\mathrm{Dop}}}
\left(
y-\widehat{\mathbf l}\cdot\mathbf v
\right)
K(\mathbf v)
\,d^3\vpartb{v},
\label{eq:doppler_forward_functional}
\end{equation}
so that the measurement PDF generated by kinetic manifold \(s\) is
\begin{equation}
p_{\mathrm{Dop},s}(y\mid\mathbf X_s)
\doteq
\mathcal H_{\mathrm{Dop}}
\left[
K_s
\right](y).
\label{eq:doppler_response_example}
\end{equation}
Here $\widehat{\mathbf l}$ is the viewing direction and
\begin{equation}
G_\sigma(a)
\doteq
\frac{1}{\sqrt{2\pi\sigma^2}}
\exp\left(
-\frac{a^2}{2\sigma^2}
\right)
\label{eq:gaussian_instrument_function}
\end{equation}
represents the calibrated spectral resolution.
The same separation applies to other diagnostics without introducing new notation.
For Thomson scattering, \(y=\Delta\omega\) is a resolved spectral coordinate and \(\mathcal H_{\mathrm{TS}}\) contains the known scattering and spectral response~\cite{EvansKatzenstein_1969,Sheffield_2010}; for a retarding-field analyzer, \(V_R\) is a controlled threshold and \(\mathcal H_{\mathrm{RFEA}}\) returns the transmitted fraction~\cite{Simpson_1961,Hutchinson_2002}.
Thus \(K_s\) carries the kinetic-state dependence, whereas \(\mathcal H_D\) carries the diagnostic dependence.
With this distinction fixed, the localized INMDF correction can be propagated through the same diagnostic map.
For the first-INMDF branch,
\begin{equation}
K_I(\mathbf v\mid\mathbf X_I)
=
K_M(\mathbf v\mid\mathbf X_M)
+
\frac{\Gamma}{n}
\eta G_I(\mathbf v).
\label{eq:inmdf_probability_decomposition}
\end{equation}
For a linear diagnostic transform,
\begin{equation}
p_{D,I}(y\mid\mathbf X_I)
=
p_{D,M}(y\mid\mathbf X_M)
+
\frac{\Gamma}{n}
\Delta p_D
(y\mid c,W_\parallel,W_\perp),
\label{eq:inmdf_pushforward_decomposition}
\end{equation}
where
\begin{equation}
\Delta p_D
(y\mid c,W_\parallel,W_\perp)
\doteq
\mathcal H_D
\left[
\eta G_I
\right](y).
\label{eq:diagnostic_inmdf_correction}
\end{equation}
The quantity $\Delta p_D$ is a signed correction, not a separate probability density.
Its integral over $y$ vanishes because the first-INMDF correction preserves particle number.
\\
Thus \(\mathcal H_D\) always denotes the fixed diagnostic map, whereas \(K_s\) carries the kinetic-state dependence.
Once the measurement PDF \(p_{D,s}(y\mid\mathbf X_s)\) is known, its moments are
\begin{equation}
M_{r,D}(\mathbf X_s)
\doteq
\int_{-\infty}^{\infty}
y^r
p_{D,s}(y\mid\mathbf X_s)
\,dy,
\quad
r\in\mathbb N_0.
\label{eq:measurement_raw_moments}
\end{equation}
These moments introduce no new fitting parameters; they are derived quantities of the same kinetic coordinates and diagnostic response.
The simplest analytic example is an ideal measurement of \(v_\parallel\), for which the forward functional reduces to
\begin{align}
p_{v_\parallel,s}(y\mid\mathbf X_s)
\doteq{}&
\mathcal H_{v_\parallel}
\left[
K_s
\right](y)
\nonumber\\
={}&
\int_{-\infty}^{\infty}
\delta(y-v_\parallel)
K_s(\mathbf v\mid\mathbf X_s)
\,d^3\vpartb{v}.
\label{eq:ideal_parallel_velocity_kernel}
\end{align}
It gives for the Maxwellian
\begin{align}
M_{1,M}
={}&
u_\parallel,
\nonumber\\
M_{2,M}
={}&
u_\parallel^2+\frac{T}{m},
\nonumber\\
M_{3,M}
={}&
u_\parallel^3
+
3u_\parallel\frac{T}{m},
\label{eq:maxwellian_velocity_measurement_moments}
\end{align}
whereas the first-INMDF gives
\begin{align}
M_{1,I}
={}&
u_\parallel+\frac{\Gamma}{n},
\nonumber\\
M_{2,I}
={}&
u_\parallel^2
+
\frac{T}{m}
+
2\frac{\Gamma}{n}c,
\nonumber\\
M_{3,I}
={}&
u_\parallel^3
+
3u_\parallel\frac{T}{m}
+
3\frac{\Gamma}{n}
\left(
c^2+\frac{W_\parallel}{m}
\right).
\label{eq:inmdf_velocity_measurement_moments}
\end{align}
The diagnostic therefore converts the same finite kinetic coordinates $(\Gamma,c,W_\parallel)$ into explicit changes of measurable moments.
To separate changes of location from changes of fluctuation shape, the corresponding raw moments can be rewritten as cumulants.
The first three are
\begin{align}
\kappa_1
\doteq{}&
M_1,
\nonumber\\
\kappa_2
\doteq{}&
M_2-M_1^2,
\nonumber\\
\kappa_3
\doteq{}&
M_3-3M_2M_1+2M_1^3.
\label{eq:conditional_cumulants}
\end{align}
These cumulants remain predictions of the selected physical distribution propagated through the diagnostic response; they are not independent residual-shape parameters.

\section{Continuous filtering theory and the Gaussian limit}
\label{sec:continuous_filtering}

Section~\ref{sec:kinetic_to_measurement} defined how a retained kinetic state is observed.
We now derive how its coordinates evolve from the kinetic equation rather than assigning their dynamics independently.
For every retained coordinate $X_A$ of $\mathbf X_I$, define the kinetic tangent function
\begin{equation}
\Phi_A(\mathbf v;\mathbf X_I)
\doteq
\frac{\partial
f_I(\mathbf X_I,\mathbf v)}
{\partial X_A}.
\label{eq:kinetic_tangent_functions}
\end{equation}
The tangent function is the velocity-space deformation produced by changing one physical coordinate while the others are held fixed.
For an infinitesimal coordinate displacement,
\begin{equation}
\delta f_I(\mathbf v)
=
\Phi_A(\mathbf v;\mathbf X_I)
\delta X_A
+
O(|\delta\mathbf X|^2).
\label{eq:kinetic_tangent_linearization}
\end{equation}
Thus the tangent functions provide the local velocity-space directions associated with density, flow, temperature, amplitude, support position, and support width.
For the first-INMDF manifold of Eqs.~\eqref{eq:first_inmdf}--\eqref{eq:DeltaI_definition}, representative tangent functions are
\begin{align}
\Phi_n
={}&
\frac{f_M}{n},
\nonumber\\
\Phi_{u_\parallel}
={}&
\frac{m(v_\parallel-u_\parallel)}{T}
f_M,
\nonumber\\
\Phi_T
={}&
\left[
\frac{m|\mathbf v-\mathbf u|^2}{2T^2}
-
\frac{3}{2T}
\right]
f_M,
\nonumber\\
\Phi_\Gamma
={}&
\eta G_I,
\nonumber\\
\Phi_c
={}&
\Gamma
\left(
\frac{m\eta^2}{W_\parallel}
-
1
\right)
G_I,
\label{eq:kinetic_tangent_examples_a}
\end{align}
and
\begin{align}
\Phi_{u_{\perp j}}
={}&
\frac{
m(v_{\perp j}-u_{\perp j})
}{
T
}
f_M
+
\Gamma\eta
\frac{
m(v_{\perp j}-u_{\perp j})
}{
W_\perp
}
G_I,
\nonumber\\
\Phi_{W_\parallel}
={}&
\Gamma\eta
\left[
\frac{m\eta^2}{2W_\parallel^2}
-
\frac{3}{2W_\parallel}
\right]
G_I,
\nonumber\\
\Phi_{W_\perp}
={}&
\Gamma\eta
\left[
\frac{
m|\mathbf v_\perp-\mathbf u_\perp|^2
}{
2W_\perp^2
}
-
\frac{1}{W_\perp}
\right]
G_I,
\label{eq:kinetic_tangent_examples_b}
\end{align}
where $j\in\{1,2\}$ labels the two perpendicular components.
These examples make the geometry explicit.
$\Phi_n$ changes the density of the Maxwellian backbone, $\Phi_{u_\parallel}$ translates its parallel center, and $\Phi_T$ changes its thermal width.
The non-Maxwellian directions have different roles: $\Phi_\Gamma$ changes the amplitude of the localized odd channel, $\Phi_c$ translates its support center, and $\Phi_{W_\parallel}$ and $\Phi_{W_\perp}$ change its parallel and perpendicular widths.
The dual tangent functions perform the inverse local operation.
They extract the coordinate displacement associated with a velocity-space deformation.
Define $\Psi_A(\mathbf v;\mathbf X_I)$ by
\begin{equation}
\int_{-\infty}^{\infty}
\Psi_A(\mathbf v;\mathbf X_I)
\Phi_B(\mathbf v;\mathbf X_I)
\,d^3\vpartb{v}
\doteq
\delta_{AB}.
\label{eq:kinetic_dual_functions}
\end{equation}
Combining Eqs.~\eqref{eq:kinetic_tangent_linearization} and~\eqref{eq:kinetic_dual_functions} gives
\begin{equation}
\delta X_A
=
\int_{-\infty}^{\infty}
\Psi_A(\mathbf v;\mathbf X_I)
\delta f_I(\mathbf v)
\,d^3\vpartb{v}
+
O(|\delta\mathbf X|^2).
\label{eq:kinetic_dual_coordinate_extraction}
\end{equation}
The tangent therefore maps a coordinate change into a distribution change, whereas its dual maps a distribution change back onto that coordinate.
The Maxwellian submanifold provides a simple explicit example.
For $\Gamma=0$ and $i\in\{\parallel,\perp1,\perp2\}$, the tangent functions
\begin{align}
\Phi_n^{(M)}
={}&
\frac{f_M}{n},
\nonumber\\
\Phi_{u_i}^{(M)}
={}&
\frac{m(v_i-u_i)}{T}f_M,
\nonumber\\
\Phi_T^{(M)}
={}&
\left[
\frac{m|\mathbf v-\mathbf u|^2}{2T^2}
-
\frac{3}{2T}
\right]f_M
\end{align}
have the explicit duals
\begin{align}
\Psi_n^{(M)}
={}&
1,
\nonumber\\
\Psi_{u_i}^{(M)}
={}&
\frac{v_i-u_i}{n},
\nonumber\\
\Psi_T^{(M)}
={}&
\frac{m}{3n}
\left(
|\mathbf v-\mathbf u|^2
-
\frac{3T}{m}
\right).
\label{eq:maxwellian_kinetic_dual_examples}
\end{align}
These functions satisfy
\begin{equation}
\int_{-\infty}^{\infty}
\Psi_A^{(M)}
\Phi_B^{(M)}
\,d^3\vpartb{v}
=
\delta_{AB}.
\end{equation}
On the full first-INMDF manifold, $\Psi_n=1$ remains an exact density dual because
\begin{align}
\int_{-\infty}^{\infty}
f_I\,d^3\vpartb{v}
=
n
\end{align}
and every non-density tangent integrates to zero.
The other duals generally mix several velocity moments because one measured moment can depend on several kinetic coordinates.
For example,
\begin{equation}
\int_{-\infty}^{\infty}
v_\parallel
f_I(\mathbf X_I,\mathbf v)
\,d^3\vpartb{v}
=
nu_\parallel+\Gamma.
\label{eq:parallel_momentum_coordinate_mixing}
\end{equation}
A parallel-momentum moment alone therefore cannot distinguish a change of $u_\parallel$ from a change of $\Gamma$.
The dual tangent family removes this coordinate mixing.
More generally, let $N_X$ denote the number of retained physical coordinates, equal to the dimension of $\mathbf X_I$ for the first-INMDF branch.
Choose $N_X$ velocity-space test functions $\chi_r(\mathbf v)$, with $r\in\{1,\ldots,N_X\}$, whose moment responses to the retained tangent directions are linearly independent.
The selected set \(\{\chi_r\}\) is part of the definition of the finite kinetic projection and is held fixed once that projection is chosen; changing this set defines a different projection of dynamics that lie outside the retained tangent manifold.
The corresponding kinetic moment Jacobian is
\begin{equation}
J^{(\mathrm{kin})}_{rA}
\doteq
\int_{-\infty}^{\infty}
\chi_r(\mathbf v)
\Phi_A(\mathbf v;\mathbf X_I)
\,d^3\vpartb{v}.
\label{eq:kinetic_moment_jacobian}
\end{equation}
Each $\chi_r$ acts as a velocity-space moment weight, while $J^{(\mathrm{kin})}_{rA}$ measures the response of that moment to an infinitesimal change of coordinate $X_A$.
Whenever this matrix is non-singular, these moment responses distinguish all retained local coordinate directions and an explicit dual family is
\begin{equation}
\Psi_A(\mathbf v;\mathbf X_I)
=
\left[
J^{(\mathrm{kin})-1}
\right]_{Ar}
\chi_r(\mathbf v),
\label{eq:kinetic_dual_jacobian_construction}
\end{equation}
because
\begin{align}
\int_{-\infty}^{\infty}
\Psi_A
\Phi_B
\,d^3\vpartb{v}
&=
\left[
J^{(\mathrm{kin})-1}
\right]_{Ar}
J^{(\mathrm{kin})}_{rB}
\nonumber\\
&=
\delta_{AB}.
\end{align}
This provides an explicit construction of the kinetic duals rather than assuming them abstractly.
It is also the velocity-space counterpart of the finite-moment dual construction used later for the state-space posterior in Section~\ref{sec:structured_closure}.
With the tangent and dual tangent families now specified, the deterministic force acting on $X_A$ is obtained by projection of Eq.~\eqref{eq:open_lfp}
\begin{align}
F_A(\mathbf X_I,t)
\doteq{}&
\int_{-\infty}^{\infty}
\Psi_A
\Big[
-\mathbf v\cdot\nabla_{\mathbf x}f_I
-\dot{\mathbf v}\cdot\nabla_{\mathbf v}f_I
\nonumber\\
&\hspace{2em}
+
C_{pp}[f_I,f_I]
+
C_{p{\rm src}}[f_I,f_{\rm src}]
\Big]
\,d^3\vpartb{v}.
\label{eq:projected_state_force}
\end{align}
Equation~\eqref{eq:projected_state_force} supplies the deterministic drift of each retained coordinate through transport, acceleration, self-collision, source-collision, and relaxation~\cite{Izacard_2026_INMDF}.
Dynamics not resolved by this finite projection are represented by
\begin{equation}
dX_A
=
F_A(\mathbf X_I,t)\,dt
+
B_{A\alpha}(\mathbf X_I,t)\,dW_\alpha,
\label{eq:projected_state_sde}
\end{equation}
where $W_\alpha(t)$ are independent standard Wiener processes, $\alpha$ labels the unresolved stochastic channels, and $B_{A\alpha}$ maps channel $\alpha$ into retained coordinate $X_A$.
The stochastic differential is understood in the It\^o sense~\cite{Ito_1944}, with
\begin{equation}
dW_\alpha dW_\beta
=
\delta_{\alpha\beta}\,dt,
\label{eq:wiener_covariance}
\end{equation}
where $\delta_{\alpha\beta}$ is the Kronecker delta.
The corresponding state-space diffusion tensor is
\begin{equation}
D_{AB}(\mathbf X_I,t)
\doteq
\sum_\alpha
B_{A\alpha}(\mathbf X_I,t)
B_{B\alpha}(\mathbf X_I,t).
\label{eq:state_diffusion_tensor}
\end{equation}
A pure random walk is recovered only when all deterministic forces $F_A$ vanish.
The drift $F_A$ is fixed by the kinetic projection of Eq.~\eqref{eq:projected_state_force}, whereas $B_{A\alpha}$ must be derived from unresolved kinetic dynamics or fixed independently from fluctuation information that is not part of the held-out diagnostic test.
In particular, $D_{AB}$ is not re-fitted to reproduce the held-out non-Gaussian current probability density.
\\
Equation~\eqref{eq:projected_state_sde} specifies stochastic trajectories of the retained physical coordinates.
State estimation requires the probability density of those coordinates conditioned on the measurements already acquired.
We therefore define
\begin{equation}
\rho(\mathbf X,t\mid\mathcal D_t)
\end{equation}
as the conditional probability density of the retained physical coordinates given all diagnostic measurements collected up to time $t$ (i.e.\ the posterior probability density in standard Bayesian theory), where $\mathcal D_t$ denotes the complete measurement history.
The equations below are written in an unconstrained local state-space chart.
When a retained physical coordinate has a restricted domain, such as \(n>0\), \(T>0\), or a positive kinetic width, the same probability dynamics are understood on the corresponding admissible state domain with vanishing normal probability flux at a finite physical boundary, or equivalently in a smooth unconstrained coordinate representation.
Between measurement updates, this conditional density is predicted by the Fokker--Planck equation corresponding to Eq.~\eqref{eq:projected_state_sde}~\cite{Risken_1989}
\begin{equation}
\partial_t\rho
=
-
\frac{\partial}{\partial X_A}
(F_A\rho)
+
\frac{1}{2}
\frac{\partial^2}{\partial X_A\partial X_B}
(D_{AB}\rho).
\label{eq:state_fokker_planck}
\end{equation}
The prediction equation alone does not incorporate a new measurement.
The observation map is supplied by the same diagnostic map introduced in Section~\ref{sec:kinetic_to_measurement}.
For channel \(\mu\),
\begin{equation}
H_\mu(\mathbf X,t)
\doteq
\left[
\mathcal H_D
\left[
K_s(\mathbf v\mid\mathbf X)
\right]
\right]_\mu.
\label{eq:state_observation_map}
\end{equation}
The index \(\mu\) denotes either a resolved measurement bin or a controlled diagnostic setting.
For example,
\begin{align}
H_\mu^{\mathrm{Dop}}(\mathbf X,t)
&=
\mathcal H_{\mathrm{Dop}}
\left[
K_s
\right](y_\mu),
\nonumber\\
H_\mu^{\mathrm{LP}}(\mathbf X,t)
&=
\mathcal H_{\mathrm{LP}}^{(0)}
\left[
K_s
\right](U_\mu).
\label{eq:observation_map_examples}
\end{align}
Thus \(H_\mu\) is a component of the fixed diagnostic map evaluated on the kinetic state; it is not an independently chosen statistical function.
The general physical diagnostic law need not be Gaussian.
To recover the continuous Kalman--Bucy specialization~\cite{KalmanBucy_1961}, we now impose the additional Gaussian-diffusion observation model
\begin{equation}
d\mathcal Z_\mu
=
H_\mu(\mathbf X,t)\,dt
+
R_{\mu\nu}^{1/2}(t)\,dV_\nu,
\label{eq:diffusion_observation}
\end{equation}
where $R_{\mu\nu}$ is the covariance per unit time of this Gaussian-diffusion specialization and
\begin{equation}
dV_\mu dV_\nu
=
\delta_{\mu\nu}\,dt.
\label{eq:observation_wiener_covariance}
\end{equation}
Its conditional mean is
\begin{equation}
\widehat H_\mu(t)
\doteq
\int_{-\infty}^{\infty}
\cdots
\int_{-\infty}^{\infty}
H_\mu(\mathbf X,t)
\rho(\mathbf X,t\mid\mathcal D_t)
\,dX_1\cdots dX_{N_X}.
\label{eq:conditional_observation_mean}
\end{equation}
Under this Gaussian-diffusion observation specialization, the conditional state density satisfies the continuous nonlinear filtering equation associated with the Kushner~\cite{Kushner_1964} and Stratonovich~\cite{Stratonovich_1960} formulations
\begin{align}
d\rho
={}&
\left[
-
\frac{\partial}{\partial X_A}
(F_A\rho)
+
\frac{1}{2}
\frac{\partial^2}{\partial X_A\partial X_B}
(D_{AB}\rho)
\right]dt
\nonumber\\
&+
\rho
\left(
H_\mu-\widehat H_\mu
\right)
(R^{-1})_{\mu\nu}
\left(
d\mathcal Z_\nu-\widehat H_\nu\,dt
\right).
\label{eq:kushner_stratonovich}
\end{align}
The first line transports and diffuses the conditional state density according to the projected kinetic dynamics.
The second line is the measurement correction associated specifically with the Wiener observation model of Eq.~\eqref{eq:diffusion_observation}.
It is not imposed as the general fluctuation law of a plasma diagnostic.
To identify the conditions under which this full conditional-density equation closes on the ordinary Kalman variables, we now extract its first two state moments.
The conditional mean and covariance of the retained physical coordinates are\begin{align}
\widehat X_A(t)
\doteq{}&
\int_{-\infty}^{\infty}
\cdots
\int_{-\infty}^{\infty}
X_A
\rho(\mathbf X,t\mid\mathcal D_t)
\,dX_1\cdots dX_{N_X},
\\
P_{AB}(t)
\doteq{}&
\int_{-\infty}^{\infty}
\cdots
\int_{-\infty}^{\infty}
\left(
X_A-\widehat X_A
\right)
\left(
X_B-\widehat X_B
\right) \nonumber\\
&\quad\quad\quad\quad\quad \times
\rho(\mathbf X,t\mid\mathcal D_t)
\,dX_1\cdots dX_{N_X}.
\label{eq:posterior_mean_covariance}
\end{align}
The Gaussian conditional state density associated with these moments is
\begin{equation}
\rho_G(\mathbf X,t)
\doteq
\frac{
\exp\left[
-\frac{1}{2}
(\mathbf X-\widehat{\mathbf X})^T
\mathsf P^{-1}
(\mathbf X-\widehat{\mathbf X})
\right]
}{
(2\pi)^{N_X/2}
\sqrt{\det\mathsf P}
}.
\label{eq:gaussian_posterior}
\end{equation}
For affine state dynamics,
\begin{equation}
F_A(\mathbf X,t)
=
A_{AB}(t)X_B+b_A(t),
\label{eq:affine_state_force}
\end{equation}
a linear diagnostic response,
\begin{equation}
H_\mu(\mathbf X,t)
=
H_{\mu A}(t)X_A,
\label{eq:linear_observation}
\end{equation}
and state-independent tensors
\begin{equation}
Q_{AB}(t)
\doteq
D_{AB}(t),
\label{eq:kalman_process_covariance}
\end{equation}
and $R_{\mu\nu}(t)$, the Gaussian family is invariant.
The exact conditional moments then satisfy
\begin{align}
d\widehat X_A
={}&
\left(
A_{AB}\widehat X_B+b_A
\right)dt
+
K_{A\mu}
\left(
d\mathcal Z_\mu
-
H_{\mu B}\widehat X_B\,dt
\right),
\label{eq:kalman_bucy_mean}
\\
\dot P_{AB}
={}&
A_{AC}P_{CB}
+
P_{AC}A_{BC}
+
Q_{AB}
-
K_{A\mu}R_{\mu\nu}K_{B\nu},
\label{eq:kalman_bucy_covariance}
\end{align}
where the continuous Kalman gain is
\begin{equation}
K_{A\mu}
\doteq
P_{AB}H_{\nu B}
(R^{-1})_{\nu\mu}.
\label{eq:continuous_kalman_gain}
\end{equation}
Thus $Q_{AB}$ measures the covariance growth produced by unresolved state dynamics, $R_{\mu\nu}$ is the covariance of the Gaussian-diffusion diagnostic noise, and $K_{A\mu}$ determines how a residual in diagnostic channel $\mu$ modifies state coordinate $X_A$.
The Wiener observation model used in Eqs.~\eqref{eq:diffusion_observation}--\eqref{eq:continuous_kalman_gain} is retained only to recover the standard continuous Kalman--Bucy limit.
The physical diagnostic probability law itself is obtained from the normalized kinetic probability $K_s=f_s/n$ through the fixed diagnostic functional $\mathcal H_D$ introduced in Section~\ref{sec:kinetic_to_measurement}.

\section{Structured non-Gaussian closure from open-system probability currents}
\label{sec:structured_closure}

Section~\ref{sec:continuous_filtering} evolves the complete posterior \(\rho(\mathbf X,t)\), so the exact filtering problem remains infinite-dimensional.
To obtain a finite estimator, we represent this posterior by a closure family \(\rho_s(\mathbf X,t)\) with coordinates
\begin{equation}
\mathbf U_s
\doteq
\left(
U_{s,1},
\ldots,
U_{s,N_s}
\right),
\label{eq:closure_coordinates}
\end{equation}
where the label $s$ identifies the selected posterior closure.
This closure label need not imply that the posterior family has the same functional form as the kinetic family carrying the corresponding physical state; for example, a non-Maxwellian kinetic state can still be represented by a Gaussian posterior when that posterior closure is adequate.
The distinction is essential: \(\mathbf X\) contains the physical kinetic coordinates, whereas \(\mathbf U_s\) parameterizes only the finite representation of their posterior.
\\
For the Gaussian family,
\begin{equation}
\mathbf U_G
\doteq
\left(
\widehat{\mathbf X},
\mathsf P
\right),
\label{eq:gaussian_parameters}
\end{equation}
and $\rho_G$ is given by Eq.~\eqref{eq:gaussian_posterior}.
Its tangent functions are
\begin{equation}
\Phi_A^{(G)}(\mathbf X;\mathbf U_G)
\doteq
\frac{\partial\rho_G(\mathbf X,t)}
{\partial U_{G,A}}.
\label{eq:gaussian_tangent_functions}
\end{equation}
Let $\Psi_A^{(G)}$ be the corresponding dual functions satisfying
\begin{align}
&
\int_{-\infty}^{\infty}
\cdots
\int_{-\infty}^{\infty}
\Psi_A^{(G)}(\mathbf X)
\Phi_B^{(G)}(\mathbf X)
\,dX_1\cdots dX_{N_X}
\doteq
\delta_{AB}.
\label{eq:gaussian_dual_functions}
\end{align}
Following the finite-dimensional projection-filter construction~\cite{BrigoHanzonLeGland_1998}, the Gaussian tangent projector is
\begin{align}
\mathcal P_G g
\doteq{}&
\sum_A
\Phi_A^{(G)}
\int_{-\infty}^{\infty}
\cdots
\int_{-\infty}^{\infty}
\Psi_A^{(G)}(\mathbf X)
g(\mathbf X)
\,dX_1\cdots dX_{N_X}.
\label{eq:gaussian_projector}
\end{align}
The part of the prediction dynamics that cannot be represented by the Gaussian coordinates is
\begin{align}
\mathcal R_G(\mathbf X,t)
\doteq{}&
\left(
1-\mathcal P_G
\right)
\Bigg[
-
\frac{\partial}{\partial X_A}
\left(
F_A\rho_G
\right)
\nonumber\\
&\hspace{7em}
+
\frac{1}{2}
\frac{\partial^2}{\partial X_A\partial X_B}
\left(
D_{AB}\rho_G
\right)
\Bigg].
\label{eq:gaussian_generator_residual}
\end{align}
The Gaussian family is dynamically invariant only when $\mathcal R_G=0$.
When $\mathcal R_G\neq0$, the residual identifies probability evolution that cannot be represented by the posterior mean and covariance, but it does not by itself specify which additional coordinates should be retained.
We therefore first formulate the finite non-Gaussian projection for an arbitrary multidimensional posterior manifold in vector and tensor form, including the existence and local invertibility of its retained coordinates and the role of the state-space drift and diffusion tensors in their projected dynamics.
The physical state $\mathbf X$ contains the coordinates of the selected kinetic manifold.
For example, Eq.~\eqref{eq:kinetic_coordinate_vectors} gives the Maxwellian state $\mathbf X_M=(n,u_\parallel,\mathbf u_\perp,T)$ and the first-INMDF state $\mathbf X_I=(n,u_\parallel,\mathbf u_\perp,T,\Gamma,c,W_\parallel,W_\perp)$.
For a Tsallis kinetic manifold in which the tail coordinate $q$ is retained dynamically, the corresponding state can be written schematically as $\mathbf X_q=(n,u_\parallel,\mathbf u_\perp,T,q)$; the associated width $w_q$ used in Appendix~\ref{app:benchmark_families} is determined by $q$ and is not an independent coordinate.
Thus $\mathbf X$ always identifies a possible physical kinetic state, whereas $\mathbf U_s$ parameterizes the posterior probability assigned to those states.
The scalar first-INMDF realization below provides a concrete example: its physical state variable is $Y=\Gamma$, while the five coordinates $(Y_G,P_G,\Gamma_X,c_X,W_X)$ parameterize the posterior of that one physical variable rather than five additional kinetic variables.
For a posterior family $\rho_s(\mathbf X;\mathbf U_s)$ with $N_s$ retained coordinates, choose $N_s$ independent state-space moment functions $\chi_r(\mathbf X)$ and define
\begin{align}
M_r^{(s)}(\mathbf U_s)
\doteq{}&
\int_{-\infty}^{\infty}
\cdots
\int_{-\infty}^{\infty}
\chi_r(\mathbf X)
\rho_s(\mathbf X;\mathbf U_s)
\,dX_1\cdots dX_{N_X},
\nonumber\\
\mathbf M_s
\doteq{}&
\left(
M_1^{(s)},
\ldots,
M_{N_s}^{(s)}
\right)^T,
\qquad
r\in\{1,\ldots,N_s\}.
\label{eq:general_posterior_moment_vector}
\end{align}
The moment vector $\mathbf M_s$ and coordinate vector $\mathbf U_s$ therefore have the same finite dimension but different meanings: $\mathbf U_s$ parameterizes the retained posterior manifold, whereas $\mathbf M_s$ contains the state-space moments used to identify a point on that manifold.
For the scalar first-INMDF example below, $\chi_r(Y)=Y^r$ for $r\in\{1,\ldots,5\}$, so the first five raw moments identify the five posterior coordinates $(Y_G,P_G,\Gamma_X,c_X,W_X)$.
Writing
$\Phi_A^{(s)}\doteq\partial\rho_s/\partial U_{s,A}$,
the local conversion from posterior-coordinate changes to moment changes is determined by the moment Jacobian
\begin{equation}
\mathsf J_{rA}^{(s)}
\doteq
\frac{\partial M_r^{(s)}}{\partial U_{s,A}}
=
\int_{-\infty}^{\infty}
\cdots
\int_{-\infty}^{\infty}
\chi_r(\mathbf X)
\Phi_A^{(s)}(\mathbf X;\mathbf U_s)
\,dX_1\cdots dX_{N_X}.
\label{eq:general_moment_jacobian}
\end{equation}
The Jacobian therefore has a direct geometric meaning: $\mathsf J^{(s)}$ maps motion along the retained posterior coordinates into changes of their retained moments, while $(\mathsf J^{(s)})^{-1}$ performs the inverse local conversion from moment space back to posterior-coordinate space.
Let $\mathcal A_s$ denote the admissible posterior-coordinate domain associated with the selected closure family, with its branch-specific normalization and positivity conditions specified below.
For a prescribed retained moment vector $\mathbf M^\star$, an admissible solution can exist only when
\begin{equation}
\mathbf M^\star
\in
\mathbf M_s(\mathcal A_s),
\label{eq:general_moment_existence}
\end{equation}
where $\mathbf M_s(\mathcal A_s)$ denotes the image of the admissible posterior manifold in retained-moment space.
At an interior admissible solution $\mathbf U_s^\star$, local invertibility additionally requires
\begin{equation}
\det
\mathsf J^{(s)}(\mathbf U_s^\star)
\neq
0.
\label{eq:general_moment_local_invertibility}
\end{equation}
The inverse-function theorem then gives a locally unique relation between $\mathbf M_s$ and $\mathbf U_s$ around that solution.
In a general multidimensional closure, $(\mathsf J^{(s)})^{-1}$ need not be available in closed analytic form and may instead be evaluated numerically as a matrix operator, provided the retained moment map remains sufficiently far from its singular and ill-conditioned surfaces.
The scalar first-INMDF branch below is a special case for which the determinant, singular surfaces, and factorized inverse can all be obtained analytically.
The same inverse Jacobian supplies the dual tangent directions and hence the finite projector.
Specifically,
\begin{align}
&\Psi_A^{(s)}(\mathbf X;\mathbf U_s)
\doteq{}
(\mathsf J^{(s)-1})_{Ar}
\chi_r(\mathbf X),
\nonumber\\
\mathcal P_s g
\doteq{}&
\Phi_A^{(s)}
\int_{-\infty}^{\infty}
\cdots
\int_{-\infty}^{\infty}
\Psi_A^{(s)}(\mathbf X;\mathbf U_s)
g(\mathbf X)
\,dX_1\cdots dX_{N_X}.
\label{eq:general_moment_projector}
\end{align}
By construction, the dual and tangent directions satisfy $\int\Psi_A^{(s)}\Phi_B^{(s)}\,d^{N_X}\mathbf X=\delta_{AB}$.
Thus the same moment Jacobian controls both reconstruction of the posterior coordinates and projection of the probability dynamics onto those coordinates.
\\
The physical evolution entering this projection remains the state-space Fokker--Planck dynamics of Eq.~\eqref{eq:state_fokker_planck}, with deterministic drift $F_A$ and diffusion tensor $D_{AB}$.
These quantities act on the physical state $\mathbf X$, not on an independently chosen statistical residual.
For an MDF state, the components of $F_A$ include the projected evolution of density, flow, and temperature.
For the first-INMDF state they additionally include $F_\Gamma$, $F_c$, $F_{W_\parallel}$, and $F_{W_\perp}$, while a dynamically retained Tsallis coordinate would introduce the corresponding component $F_q$.
The projected drift contains transport, acceleration, source, collision, and relaxation contributions inherited from Eq.~\eqref{eq:projected_state_force}.
Its collisional and relaxation components provide friction-like directed evolution toward preferred kinetic states, whereas source and transport terms can maintain or drive departures from those states.
The diffusion tensor $D_{AB}$ instead represents unresolved stochastic state evolution.
Its diagonal components broaden the conditional density along individual physical coordinates, while its off-diagonal components generate correlated spreading between different state directions; for example, $D_{\Gamma T}$ couples unresolved INMDF-amplitude and temperature fluctuations, whereas $D_{qT}$ would provide the analogous coupling for a retained Tsallis-tail coordinate.
The competition between $F_A$ and $D_{AB}$ therefore determines whether the predicted state probability is transported, locally concentrated, broadened, or correlated across the retained kinetic coordinates before a new diagnostic measurement is applied.
Under the decay or vanishing-normal-flux boundary condition specified above, projection of this physical drift--diffusion dynamics onto retained moment $M_r^{(s)}$ gives
\begin{align}
\mathcal A_r^{(s)}
\doteq{}&
\int_{-\infty}^{\infty}
\cdots
\int_{-\infty}^{\infty}
\Bigg[
\frac{\partial\chi_r}{\partial X_A}
F_A
+
\frac{1}{2}
\frac{\partial^2\chi_r}{\partial X_A\partial X_B}
D_{AB}
\Bigg]
\nonumber\\
&\hspace{10em}\times
\rho_s(\mathbf X,t)
\,dX_1\cdots dX_{N_X}.
\label{eq:general_moment_prediction}
\end{align}
The first contribution propagates each retained moment through the resolved deterministic kinetic dynamics, including its friction-like collisional and relaxation components, while the second transfers unresolved tensorial state-space diffusion into the same moment hierarchy.
Thus $\mathcal A_r^{(s)}$ is not an independently prescribed posterior force; it is the moment-space image of the physical $F_A$ and $D_{AB}$ already defined by the state prediction.
The measurement enters through a different physical map.
Within the Gaussian-diffusion observation specialization of Eq.~\eqref{eq:diffusion_observation}, define $\widehat H_{\mu,s}(t)\doteq\int H_\mu(\mathbf X,t)\rho_s(\mathbf X,t)\,d^{N_X}\mathbf X$ and $dI_{\mu,s}\doteq d\mathcal Z_\mu-\widehat H_{\mu,s}dt$.
Here $R_{\mu\nu}$ is the diagnostic-noise covariance already introduced in Section~\ref{sec:continuous_filtering}: it appears in the observation increment through $R^{1/2}_{\mu\nu}dV_\nu$, and in the discrete Gaussian limit the same distinction becomes $S_{\mu\nu}=H_{\mu A}P_{AB}^{\mathrm{pred}}H_{\nu B}+R_{\mu\nu}$.
Thus $R_{\mu\nu}$ characterizes diagnostic noise at fixed state, whereas $P_{AB}$ characterizes uncertainty of the physical state and $S_{\mu\nu}$ characterizes the resulting predicted-measurement uncertainty.
The coupling between retained moment $M_r^{(s)}$ and diagnostic channel $\mu$ is
\begin{align}
\mathcal C_{r\mu}^{(s)}
\doteq
&
\int_{-\infty}^{\infty}
\cdots
\int_{-\infty}^{\infty}
\chi_r(\mathbf X)
\left[
H_\mu(\mathbf X,t)-\widehat H_{\mu,s}(t)
\right]
\nonumber\\
&\quad\quad\quad\quad\times
\rho_s(\mathbf X,t)
\,dX_1\cdots dX_{N_X}.
\label{eq:general_moment_observation_coupling}
\end{align}
Thus $\mathcal C_{r\mu}^{(s)}$ measures how diagnostic channel $\mu$ constrains retained posterior moment $M_r^{(s)}$.
For a Langmuir probe, for example, strong sensitivity of the current likelihood to the INMDF amplitude produces a strong coupling to moments that distinguish $\Gamma$, whereas weak sensitivity to a retained Tsallis coordinate $q$ would correspondingly produce a weak measurement correction along that state direction.
The multidimensional moment increment is therefore
\begin{equation}
dM_r^{(s)}
=
\mathcal A_r^{(s)}\,dt
+
\mathcal C_{r\mu}^{(s)}
(R^{-1})_{\mu\nu}
dI_{\nu,s}.
\label{eq:general_moment_filter}
\end{equation}
The two terms have different origins: $\mathcal A_r^{(s)}$ is generated by physical state drift and diffusion, while $\mathcal C_{r\mu}^{(s)}(R^{-1})_{\mu\nu}dI_{\nu,s}$ is generated by the diagnostic innovation.
Because the filter evolves $\mathbf U_s$ rather than $\mathbf M_s$, the inverse Jacobian converts both moment-space contributions back to posterior-coordinate space.
The corresponding coordinate response to the diagnostic innovation is
\begin{equation}
\mathcal K_{s,A\nu}
\doteq
(\mathsf J^{(s)-1})_{Ar}
\mathcal C_{r\mu}^{(s)}
(R^{-1})_{\mu\nu}.
\label{eq:general_coordinate_response}
\end{equation}
Here $\mathcal K_{s,A\nu}$ is the response of posterior coordinate $U_{s,A}$ to diagnostic channel $\nu$; it is distinct from the normalized kinetic probability $K_s=f_s/n_s$.
Because the retained moments can depend nonlinearly on $\mathbf U_s$, define inline the moment-map Hessian $\mathsf H_{rAB}^{(s)}\doteq\partial^2M_r^{(s)}/(\partial U_{s,A}\partial U_{s,B})$, which is distinct from the diagnostic response $H_\mu$.
Its only role here is the It\^{o} correction generated by the stochastic coordinate response.
The resulting multidimensional coordinate dynamics are
\begin{align}
dU_{s,A}
={}&
(\mathsf J^{(s)-1})_{Ar}
\Bigg[
\mathcal A_r^{(s)}
-
\frac{1}{2}
\mathsf H_{rBC}^{(s)}
\mathcal K_{s,B\mu}
R_{\mu\nu}
\mathcal K_{s,C\nu}
\Bigg]dt
\nonumber\\
&+
\mathcal K_{s,A\nu}
\,dI_{\nu,s}.
\label{eq:general_coordinate_dynamics}
\end{align}
The deterministic contribution first propagates the physical drift and diffusion through the retained moments and then converts those moment changes into motion on the posterior manifold through $(\mathsf J^{(s)})^{-1}$.
The stochastic contribution instead passes the diagnostic innovation through $R^{-1}$, the moment--diagnostic coupling $\mathcal C^{(s)}$, and the same inverse Jacobian.
State prediction and measurement correction therefore arise from different physical mechanisms but share one posterior geometry.
For example, when the physical coordinate is $\Gamma$, this construction evolves the posterior attached to the INMDF amplitude; when the retained physical coordinate is a Tsallis tail parameter $q$, the same tensorial construction evolves the posterior attached to that tail coordinate.
\\
The first explicit realization is now constructed in one scalar physical coordinate, for which the complete moment map and its singular structure can be obtained analytically.
For the first Langmuir-probe implementation, that physical coordinate is the first-INMDF amplitude
\begin{equation}
Y
\doteq
\Gamma,
\quad
F(Y,t)
\doteq
F_{\Gamma}(\mathbf X_I,t),
\quad
D(Y,t)
\doteq
D_{\Gamma\Gamma}(\mathbf X_I,t),
\label{eq:scalar_state_identification}
\end{equation}
where $F_{\Gamma}$ is the $\Gamma$ component of the projected kinetic force in Eq.~\eqref{eq:projected_state_force} and $D_{\Gamma\Gamma}$ is the corresponding state-space diffusion coefficient.
The remaining physical coordinates $(n,\mathbf u,T,c,W_\parallel,W_\perp)$ are conditioned on their current estimates in this first scalar realization.
The posterior-shape coordinates introduced next describe uncertainty in the physical variable \(Y=\Gamma\), not additional velocity-space parameters.
In particular, \(\Gamma_X\) is distinct from the physical INMDF amplitude \(\Gamma\).
We begin with a Gaussian core,
\begin{equation}
\rho_{G,I}(Y,t)
\doteq
\frac{1}{
\sqrt{2\pi P_G(t)}
}
\exp\left[
-\frac{
(Y-Y_G(t))^2
}{
2P_G(t)
}
\right],
\label{eq:scalar_gaussian_core}
\end{equation}
where $Y_G$ and $P_G$ are respectively the center and variance of the Gaussian core. They are not assumed to be the mean and variance of the complete non-Gaussian posterior.
The first state-space INMDF posterior is
\begin{equation}
\rho_I(Y,t)
\doteq
\rho_{G,I}(Y,t)
+
\Gamma_X(t)\eta_Y(t)G_Y(Y,t),
\label{eq:state_inmdf_ansatz}
\end{equation}
where
\begin{equation}
\eta_Y(t)
\doteq
Y-c_X(t),
\label{eq:state_inmdf_shift}
\end{equation}
and
\begin{equation}
G_Y(Y,t)
\doteq
\frac{1}{
\sqrt{2\pi}\,W_X^{3/2}(t)
}
\exp\left[
-\frac{
\eta_Y^2(t)
}{
2W_X(t)
}
\right].
\label{eq:state_inmdf_support}
\end{equation}
The support normalization gives
\begin{align}
\int_{-\infty}^{\infty}
\eta_YG_Y
\,dY
&=
0,
\nonumber\\
\int_{-\infty}^{\infty}
\eta_Y^2G_Y
\,dY
&=
1.
\label{eq:state_inmdf_normalization}
\end{align}
Consequently, $\Gamma_X$ is the first-moment coordinate of the localized correction, $c_X$ is its support center, and $W_X$ is its support width.
The total posterior mean is
\begin{equation}
\widehat X_I(t)
\doteq
\int_{-\infty}^{\infty}
Y\rho_I(Y,t)
\,dY
=
Y_G(t)+\Gamma_X(t),
\label{eq:state_inmdf_total_mean}
\end{equation}
and the total posterior variance is
\begin{align}
P_I(t)
\doteq{}&
\int_{-\infty}^{\infty}
\left[
Y-\widehat X_I(t)
\right]^2
\rho_I(Y,t)
\,dY
\nonumber\\
={}&
P_G(t)
+
2\Gamma_X(t)
\left[
c_X(t)-Y_G(t)
\right]
-
\Gamma_X^2(t).
\label{eq:state_inmdf_total_variance}
\end{align}
The Gaussian-core coordinates $(Y_G,P_G)$ and the total posterior moments $(\widehat X_I,P_I)$ therefore coincide only in the Gaussian limit $\Gamma_X=0$.
The retained state-space INMDF coordinates are
\begin{equation}
\mathbf U_I
\doteq
\left(
Y_G,
P_G,
\Gamma_X,
c_X,
W_X
\right).
\label{eq:inmdf_parameters}
\end{equation}
The tangent functions generated by these coordinates are
\begin{align}
\Phi_{Y_G}^{(I)}
\doteq{}&
\frac{Y-Y_G}{P_G}
\rho_{G,I},
\nonumber\\
\Phi_{P_G}^{(I)}
\doteq{}&
\left[
\frac{(Y-Y_G)^2}{2P_G^2}
-
\frac{1}{2P_G}
\right]
\rho_{G,I},
\nonumber\\
\Phi_{\Gamma_X}^{(I)}
\doteq{}&
\eta_YG_Y,
\nonumber\\
\Phi_{c_X}^{(I)}
\doteq{}&
\Gamma_X
\left(
-1+
\frac{\eta_Y^2}{W_X}
\right)
G_Y,
\nonumber\\
\Phi_{W_X}^{(I)}
\doteq{}&
\Gamma_X
\left(
\frac{\eta_Y^3}{2W_X^2}
-
\frac{3\eta_Y}{2W_X}
\right)
G_Y.
\label{eq:inmdf_tangent_functions}
\end{align}
The scalar first-INMDF posterior now realizes the general moment construction of Eqs.~\eqref{eq:general_posterior_moment_vector}--\eqref{eq:general_coordinate_dynamics} explicitly.
Because $\mathbf U_I$ contains five coordinates, choose the five moment functions $\chi_r(Y)=Y^r$ with $r\in\{1,\ldots,5\}$ and define
\begin{equation}
M_r^{(I)}(t)
\doteq
\int_{-\infty}^{\infty}
Y^r\rho_I(Y,t)
\,dY.
\label{eq:state_inmdf_raw_moment_definition}
\end{equation}
Direct Gaussian integration gives
\begin{align}
M_1^{(I)}
={}&
Y_G+\Gamma_X,
\nonumber\\
M_2^{(I)}
={}&
Y_G^2+P_G+2\Gamma_Xc_X,
\nonumber\\
M_3^{(I)}
={}&
Y_G^3+3Y_GP_G
+3\Gamma_X(c_X^2+W_X),
\nonumber\\
M_4^{(I)}
={}&
Y_G^4+6Y_G^2P_G+3P_G^2
\nonumber\\
&+
4\Gamma_X(c_X^3+3c_XW_X),
\nonumber\\
M_5^{(I)}
={}&
Y_G^5+10Y_G^3P_G+15Y_GP_G^2
\nonumber\\
&+
5\Gamma_X
(c_X^4+6c_X^2W_X+3W_X^2).
\label{eq:state_inmdf_raw_moments}
\end{align}
The general moment Jacobian and Hessian then reduce to
\begin{align}
\mathsf J_{rA}^{(I)}
\doteq{}&
\frac{\partial M_r^{(I)}}{\partial U_{I,A}}
=
\int_{-\infty}^{\infty}
Y^r\Phi_A^{(I)}(Y)
\,dY,
\nonumber\\
\mathsf H_{rAB}^{(I)}
\doteq{}&
\frac{\partial^2M_r^{(I)}}
{\partial U_{I,A}\partial U_{I,B}}.
\label{eq:state_inmdf_moment_jacobian}
\end{align}
To determine where this moment-to-coordinate map loses invertibility, it is useful to express the Jacobian in terms of the relative center and width differences
\begin{equation}
\Delta_c
\doteq
c_X-Y_G,
\quad
\Delta_P
\doteq
P_G-W_X,
\label{eq:moment_map_differences}
\end{equation}
and the determinant factor
\begin{equation}
\mathcal D_J
\doteq
9\Delta_P^3
-
9\Delta_P^2\Delta_c^2
-
3\Delta_P\Delta_c^4
-
\Delta_c^6.
\label{eq:moment_map_determinant_factor}
\end{equation}
Direct differentiation of Eq.~\eqref{eq:state_inmdf_raw_moments} gives the explicit Jacobian in Appendix~\ref{app:projection} and
\begin{equation}
\det\mathsf J^{(I)}
=
-60\Gamma_X^2\mathcal D_J.
\label{eq:state_inmdf_identifiability}
\end{equation}
The five-moment map is therefore locally invertible only when
\begin{equation}
\Gamma_X
\neq
0,
\quad
\mathcal D_J
\neq
0.
\label{eq:state_inmdf_local_invertibility}
\end{equation}
Within the positivity domain $\Delta_P>0$, write $r\doteq\Delta_c^2/\Delta_P$.
The additional singular surface is the unique positive root
\begin{equation}
r^3+3r^2+9r-9
=
0,
\quad
r
\simeq
0.7592298.
\label{eq:moment_map_additional_singularity}
\end{equation}
Thus $c_X=Y_G$ is not itself singular when $\Gamma_X\neq0$ and $P_G\neq W_X$.
The conditions in Eq.~\eqref{eq:state_inmdf_local_invertibility} are the explicit scalar realization of the general local-invertibility requirement in Eq.~\eqref{eq:general_moment_local_invertibility}.
The exact factorized inverse given in Appendix~\ref{app:projection} is an additional analytic property of this five-coordinate branch and is not required by the general multidimensional construction.
Whenever Eq.~\eqref{eq:state_inmdf_local_invertibility} holds, the general dual construction in Eq.~\eqref{eq:general_moment_projector} becomes
\begin{equation}
\Psi_A^{(I)}(Y;\mathbf U_I)
\doteq
(\mathsf J^{(I)-1})_{Ar}Y^r,
\label{eq:inmdf_dual_functions}
\end{equation}
because
\begin{equation}
\int_{-\infty}^{\infty}
\Psi_A^{(I)}(Y)
\Phi_B^{(I)}(Y)
\,dY
=
\delta_{AB}.
\label{eq:inmdf_dual_biorthogonality}
\end{equation}
The associated moment-preserving tangent projector is
\begin{equation}
\mathcal P_I g
\doteq
\Phi_A^{(I)}
(\mathsf J^{(I)-1})_{Ar}
\int_{-\infty}^{\infty}
Y^r g(Y)
\,dY.
\label{eq:inmdf_moment_projector}
\end{equation}
Two distinct requirements must now be separated: whether the first-INMDF posterior represents the probability evolution better than the Gaussian closure, and whether its coordinates can be reconstructed stably.
The velocity-space choice \(f_I\) does not by itself require an INMDF posterior.
Let \(\mathcal F[\rho]\) denote the Fokker--Planck prediction operator on the right-hand side of Eq.~\eqref{eq:state_fokker_planck}.
For posterior closure \(s\), define
\begin{equation}
\varepsilon_s
\doteq
\frac{
\left\|
\mathcal F[\rho]
-
\mathcal P_s\mathcal F[\rho]
\right\|_2
}{
\left\|
\mathcal F[\rho]
\right\|_2
},
\label{eq:closure_residual_metric}
\end{equation}
where \(\|\cdot\|_2\) is the state-space \(L^2\) norm.
The first-INMDF posterior is retained when its additional tangent directions reduce this unresolved evolution, for example when \(\varepsilon_I<\varepsilon_G\).
If not, the first-INMDF posterior is not the appropriate finite closure even when the physical velocity distribution is non-Maxwellian.
Even when the representation is appropriate and the moment map is exactly invertible, the reconstruction can become numerically unstable because the inverse entries associated with $c_X$ and $W_X$ diverge as $\Gamma_X\to0$, while all inverse entries diverge as $\mathcal D_J\to0$.
To measure this numerical sensitivity independently of units, let $s_A>0$ and $m_r>0$ be characteristic scales of $U_{I,A}$ and $M_r^{(I)}$ and define the dimensionless Jacobian
\begin{equation}
\overline{\mathsf J}_{rA}^{(I)}
\doteq
\frac{s_A}{m_r}
\mathsf J_{rA}^{(I)}.
\label{eq:dimensionless_moment_jacobian}
\end{equation}
The numerical identifiability measure is
\begin{equation}
\kappa_2
\left(
\overline{\mathsf J}^{(I)}
\right)
\doteq
\frac{
\sigma_{\max}
\left(
\overline{\mathsf J}^{(I)}
\right)
}{
\sigma_{\min}
\left(
\overline{\mathsf J}^{(I)}
\right)
}.
\label{eq:moment_jacobian_condition_number}
\end{equation}
The five-coordinate inversion is accepted only while this condition number remains below a stated numerical threshold; otherwise the weakly identifiable support coordinates are removed or the hierarchy is changed.
The regular Gaussian limit therefore removes $(c_X,W_X)$ before inversion when $\Gamma_X=0$.
The localized first-INMDF closure is not intended to represent every form of non-Gaussian posterior structure.
As a contrasting broad-tail posterior closure, a Kappa family motivated by nonthermal power-law populations~\cite{Vasyliunas_1968,SummersThorne_1991,PierrardLazar_2010} defines the separate state-space family
\begin{equation}
\rho_\kappa(Y,t)
\doteq
\frac{1}{Z_\kappa(t)}
\left[
1+
\frac{
(Y-\widehat X_\kappa(t))^2
}{
\kappa(t)W_\kappa(t)
}
\right]^{-[\kappa(t)+1]},
\label{eq:kappa_closure}
\end{equation}
where
\begin{equation}
Z_\kappa(t)
\doteq
\int_{-\infty}^{\infty}
\left[
1+
\frac{
(Y-\widehat X_\kappa(t))^2
}{
\kappa(t)W_\kappa(t)
}
\right]^{-[\kappa(t)+1]}
\,dY.
\label{eq:kappa_normalization}
\end{equation}
Its coordinates are
\begin{equation}
\mathbf U_\kappa
\doteq
\left(
\widehat X_\kappa,
W_\kappa,
\kappa
\right).
\label{eq:kappa_parameters}
\end{equation}
For the normalized state-space closure with finite mean and variance used here, the admissible branch requires \(W_\kappa>0\) and \(\kappa>1/2\); retaining higher posterior moments would impose the corresponding stronger moment-existence conditions.
This state-space Kappa posterior is distinct from the velocity-space Kappa kinetic manifold used in Section~\ref{sec:kube_results} and defined explicitly in Appendix~\ref{app:benchmark_families}.
The Kappa posterior is retained as a separate broad symmetric-tail alternative and does not share the localized coordinates $(\Gamma_X,c_X,W_X)$ of Eq.~\eqref{eq:state_inmdf_ansatz}.
The executable non-Gaussian filter is derived below for the first-INMDF closure, to which we now return.
Within the Gaussian-diffusion observation specialization of Eq.~\eqref{eq:diffusion_observation}, its scalar conditional state density has diagnostic mean
\begin{equation}
\widehat H_{\mu,I}(t)
\doteq
\int_{-\infty}^{\infty}
H_\mu(Y,t)
\rho_I(Y,t)
\,dY,
\label{eq:inmdf_conditional_observation_mean}
\end{equation}
and innovation increment
\begin{equation}
dI_{\mu,I}
\doteq
d\mathcal Z_\mu
-
\widehat H_{\mu,I}\,dt.
\label{eq:inmdf_innovation}
\end{equation}
Its quadratic variation is
\begin{equation}
dI_{\mu,I}dI_{\nu,I}
=
R_{\mu\nu}\,dt.
\label{eq:inmdf_innovation_covariance}
\end{equation}
With $\chi_r(Y)=Y^r$, the general multidimensional moment equation~\eqref{eq:general_moment_filter} reduces for $r\in\{1,\ldots,5\}$ to
\begin{equation}
dM_r^{(I)}
=
\mathcal A_r^{(I)}\,dt
+
\mathcal C_{r\mu}^{(I)}
(R^{-1})_{\mu\nu}
dI_{\nu,I},
\label{eq:inmdf_moment_filter}
\end{equation}
where Eq.~\eqref{eq:general_moment_prediction} becomes
\begin{align}
\mathcal A_r^{(I)}
\doteq{}&
r
\int_{-\infty}^{\infty}
Y^{r-1}F(Y,t)\rho_I(Y,t)
\,dY
\nonumber\\
&+
\frac{r(r-1)}{2}
\int_{-\infty}^{\infty}
Y^{r-2}D(Y,t)\rho_I(Y,t)
\,dY,
\label{eq:inmdf_moment_prediction}
\end{align}
and Eq.~\eqref{eq:general_moment_observation_coupling} becomes
\begin{equation}
\mathcal C_{r\mu}^{(I)}
\doteq
\int_{-\infty}^{\infty}
Y^r
\left[
H_\mu(Y,t)-\widehat H_{\mu,I}(t)
\right]
\rho_I(Y,t)
\,dY.
\label{eq:inmdf_moment_observation_covariance}
\end{equation}
Here $F(Y,t)=F_\Gamma(\mathbf X_I,t)$ is the directed evolution of the physical INMDF amplitude and contains the projected transport, source, collisional, and relaxation dynamics, while $D(Y,t)=D_{\Gamma\Gamma}(\mathbf X_I,t)$ is its unresolved scalar state-space diffusion.
The factors $rY^{r-1}$ and $r(r-1)Y^{r-2}$ are precisely the first and second derivatives of the scalar moment function $\chi_r(Y)=Y^r$, so Eq.~\eqref{eq:inmdf_moment_prediction} is the one-dimensional realization of the tensorial drift--diffusion projection in Eq.~\eqref{eq:general_moment_prediction}.
For the physical diagnostic probability law generated from $K_s$ through $\mathcal H_D$, the prediction moments remain generated by $\mathcal A_r^{(I)}$, while the discrete observation contribution is supplied by the exact Bayesian correction and finite-family projection in Section~\ref{sec:continuous_discrete}.
The filter, however, evolves the finite coordinates $\mathbf U_I$ rather than the five raw moments directly.
For this explicit scalar realization, it is convenient to denote the deterministic coordinate drift by $\mathcal A_{I,A}$ and write
\begin{equation}
dU_{I,A}
=
\mathcal A_{I,A}\,dt
+
\mathcal K_{I,A\nu}\,dI_{\nu,I},
\label{eq:inmdf_coordinate_filter}
\end{equation}
where $U_{I,A}$ runs over $(Y_G,P_G,\Gamma_X,c_X,W_X)$.
Applying It\^{o}'s formula to $M_r^{(I)}(\mathbf U_I)$ gives the explicit five-coordinate realization of Eq.~\eqref{eq:general_coordinate_dynamics},
\begin{align}
dM_r^{(I)}
={}&
\mathsf J_{rA}^{(I)}dU_{I,A}
+
\frac{1}{2}
\mathsf H_{rAB}^{(I)}
dU_{I,A}dU_{I,B}.
\label{eq:ito_moment_coordinate_map}
\end{align}
Matching Eqs.~\eqref{eq:inmdf_moment_filter} and~\eqref{eq:ito_moment_coordinate_map} yields the explicit innovation response
\begin{equation}
\mathcal K_{I,A\nu}
\doteq
(\mathsf J^{(I)-1})_{Ar}
\mathcal C_{r\mu}^{(I)}
(R^{-1})_{\mu\nu},
\label{eq:inmdf_coordinate_gain}
\end{equation}
and the prediction drift
\begin{align}
\mathcal A_{I,A}
\doteq{}&
(\mathsf J^{(I)-1})_{Ar}
\Bigg[
\mathcal A_r^{(I)}
-
\frac{1}{2}
\mathsf H_{rBC}^{(I)}
\mathcal K_{I,B\mu}
R_{\mu\nu}
\mathcal K_{I,C\nu}
\Bigg].
\label{eq:inmdf_coordinate_drift}
\end{align}
The complete five-coordinate filter is therefore
\begin{align}
dY_G
={}&
\mathcal A_{I,Y_G}\,dt
+
\mathcal K_{I,Y_G\mu}\,dI_{\mu,I},
\nonumber\\
dP_G
={}&
\mathcal A_{I,P_G}\,dt
+
\mathcal K_{I,P_G\mu}\,dI_{\mu,I},
\nonumber\\
d\Gamma_X
={}&
\mathcal A_{I,\Gamma_X}\,dt
+
\mathcal K_{I,\Gamma_X\mu}\,dI_{\mu,I},
\nonumber\\
dc_X
={}&
\mathcal A_{I,c_X}\,dt
+
\mathcal K_{I,c_X\mu}\,dI_{\mu,I},
\nonumber\\
dW_X
={}&
\mathcal A_{I,W_X}\,dt
+
\mathcal K_{I,W_X\mu}\,dI_{\mu,I}.
\label{eq:inmdf_coordinate_dynamics}
\end{align}
The five equations evolve the Gaussian core and the localized non-Gaussian channel through one moment-projected prediction generator and one common diagnostic innovation.
No empirical gain is assigned independently to a hidden coordinate.
The coordinate equations above define a valid finite posterior only while the represented density remains normalized, positive, and identifiable.
Accordingly, for any selected family $s$, the admissible coordinate domain $\mathcal A_s$ requires
\begin{align}
&
\int_{-\infty}^{\infty}
\cdots
\int_{-\infty}^{\infty}
\rho_s(\mathbf X,t)
\,dX_1\cdots dX_{N_X}
=
1,
\label{eq:closure_admissibility}
\end{align}
with \(\rho_s(\mathbf X,t) \geq 0\) for every admissible \(\mathbf X\), together with linear independence of the retained tangent functions and finiteness of all moments used by the filter.
For the scalar first-INMDF branch, normalization follows from Eq.~\eqref{eq:state_inmdf_normalization}, while positivity requires
\begin{equation}
\rho_{G,I}(Y,t)
+
\Gamma_X\eta_YG_Y(Y,t)
>
0
\quad
\text{for every }Y\in\mathbb R.
\label{eq:first_state_inmdf_positivity}
\end{equation}
Define the one-sided critical amplitudes
\begin{align}
\Gamma_{Y,+}^{\mathrm{crit}}
\doteq{}&
\inf_{\eta_Y<0}
\frac{
\rho_{G,I}(Y,t)
}{
-\eta_YG_Y(Y,t)
},
\nonumber\\
\Gamma_{Y,-}^{\mathrm{crit}}
\doteq{}&
\inf_{\eta_Y>0}
\frac{
\rho_{G,I}(Y,t)
}{
\eta_YG_Y(Y,t)
}.
\label{eq:first_state_inmdf_critical_amplitudes}
\end{align}
A nonzero globally positive first branch requires the localized support to decay faster than the Gaussian core,
\begin{equation}
0<W_X<P_G,
\label{eq:first_state_inmdf_width_domain}
\end{equation}
and its odd amplitude must satisfy
\begin{equation}
-\Gamma_{Y,-}^{\mathrm{crit}}
<
\Gamma_X
<
\Gamma_{Y,+}^{\mathrm{crit}}.
\label{eq:first_state_inmdf_amplitude_domain}
\end{equation}
The admissible first-branch domain is therefore
\begin{align}
\mathcal A_{I,1}
\doteq
\Big\{&
(Y_G,P_G,\Gamma_X,c_X,W_X):
P_G>0,
\ 0<W_X<P_G,
\nonumber\\
&
P_I>0,
\ -\Gamma_{Y,-}^{\mathrm{crit}}<\Gamma_X<\Gamma_{Y,+}^{\mathrm{crit}}
\Big\}.
\label{eq:first_state_inmdf_admissible_domain}
\end{align}
This restriction is the state-space analogue of the positivity-admissible first-INMDF kinetic branch.
If the source-driven prediction or the measurement correction reaches the boundary of $\mathcal A_{I,1}$, the correct continuation is not to extrapolate the first branch into a negative density.
The next INMDF closure level must be activated by retaining additional zero-integral Gaussian--polynomial components,
\begin{align}
\rho_{I,N}(Y,t)
\doteq
\rho_{G,I}(Y,t)
+
\sum_{\ell=1}^{N}
\delta\rho_{I,\ell}(Y,t),
\\
\int_{-\infty}^{\infty}
\delta\rho_{I,\ell}(Y,t)
\,dY
=
0.
\label{eq:state_inmdf_hierarchy}
\end{align}
Loss of first-branch positivity is therefore the criterion for activating the next positive closure level, not a failure of the INMDF filtering framework.
A numerical prediction or measurement correction is admissible only when it remains inside the domain of the retained hierarchy level.

\section{Continuous-discrete modified Kalman filtering}
\label{sec:continuous_discrete}

Sections~\ref{sec:kinetic_to_measurement}--\ref{sec:structured_closure} have defined the kinetic coordinates, their continuous dynamics, the measurement likelihood, and a finite posterior closure.
We now combine these ingredients into one executable prediction--correction cycle.
The state evolves continuously between acquisitions and is corrected only when a new measurement becomes available.
\\
Throughout this section, the superscript $\mathrm{pred}$ denotes a quantity obtained by evolving the physical state and its conditional density to the next measurement time without using that measurement.
The superscript $\mathrm{obs}$ denotes the corresponding quantity after the new measurement has been included.
Starting from the corrected conditional density $\rho_k^{\mathrm{obs}}$ at time $t_k$, the exact predicted density at time $t_{k+1}$ is
\begin{align}
\rho_{k+1}^{\mathrm{pred}}(\mathbf X)
&\doteq{}
\rho_k^{\mathrm{obs}}(\mathbf X)
\nonumber\\
&+
\int_{t_k}^{t_{k+1}}
\left[
-
\frac{\partial}{\partial X_A}
\left(
F_A\rho
\right)
+
\frac{1}{2}
\frac{\partial^2}{\partial X_A\partial X_B}
\left(
D_{AB}\rho
\right)
\right]
\,dt.
\label{eq:continuous_prediction_map}
\end{align}
For a finite posterior family, the corresponding closure coordinates satisfy
\begin{equation}
U_{s,A,k+1}^{\mathrm{pred}}
\doteq
U_{s,A,k}^{\mathrm{obs}}
+
\int_{t_k}^{t_{k+1}}
\dot U_{s,A}(t)
\,dt,
\label{eq:exact_coordinate_prediction}
\end{equation}
where $\dot U_{s,A}$ is obtained from the continuous projected filtering equation.
A Runge--Kutta scheme or another time integrator approximates this integral; it does not define the filtering theory.
\\
After the new diagnostic measurement $y_{k+1}$ is obtained, Bayes' rule gives the exact conditional state density~\cite{Sarkka_2013},
\begin{equation}
\rho_{k+1}^{\mathrm{exact}}(\mathbf X)
\doteq
\frac{
p(y_{k+1}\mid\mathbf X)
\rho_{k+1}^{\mathrm{pred}}(\mathbf X)
}{
\displaystyle
\int_{-\infty}^{\infty}
\cdots
\int_{-\infty}^{\infty}
p(y_{k+1}\mid\mathbf X')
\rho_{k+1}^{\mathrm{pred}}(\mathbf X')
\,dX'_1\cdots dX'_{N_X}
}.
\label{eq:exact_discrete_update}
\end{equation}
For the kinetic construction developed here, the likelihood entering Eq.~\eqref{eq:exact_discrete_update} is not assigned independently.
For diagnostic \(D\) and kinetic family \(s\),
\begin{equation}
p(y_{k+1}\mid\mathbf X_s)
=
p_{D,s}(y_{k+1}\mid\mathbf X_s),
\label{eq:physical_likelihood_discrete_update}
\end{equation}
where \(p_{D,s}\) is generated from the same normalized kinetic probability \(K_s\) and fixed diagnostic map \(\mathcal H_D\) defined in Section~\ref{sec:kinetic_to_measurement}.
For a diagnostic that directly resolves the measurement coordinate,
\begin{equation}
p_{D,s}(y\mid\mathbf X_s)
=
\mathcal H_D
\left[
K_s(\mathbf v\mid\mathbf X_s)
\right](y).
\label{eq:resolved_physical_likelihood_discrete}
\end{equation}
For a scalar diagnostic, \(\mathcal H_D\) instead produces the physical signal
\begin{equation}
y_{D,s}(t)
=
\mathcal H_D
\left[
K_s(\mathbf v\mid\mathbf X_s(t))
\right].
\label{eq:scalar_diagnostic_signal}
\end{equation}
The corresponding measurement PDF \(p_{D,s}\) is obtained from the temporal or ensemble pushforward of this signal.
Successive diagnostic measurements therefore update the retained kinetic coordinates recursively through a likelihood derived from the non-Maxwellian distribution function and the fixed diagnostic response, rather than through an independently assigned residual distribution.
The exact posterior generally lies outside the retained finite family.
Because the continuous closure of Section~\ref{sec:structured_closure} is defined through the retained moment map, the discrete correction first uses the same moment geometry whenever the Bayes-updated retained moments admit an admissible representation.
Define the exact corrected retained moments and their finite-family representation by
\begin{align}
M_{r,s,k+1}^{\mathrm{exact}}
\doteq{}&
\int_{-\infty}^{\infty}
\cdots
\int_{-\infty}^{\infty}
\chi_r(\mathbf X)
\rho_{k+1}^{\mathrm{exact}}(\mathbf X)
\,dX_1\cdots dX_{N_X},
\nonumber\\
M_r^{(s)}
\left(
\mathbf U_{s,k+1}^{\mathrm{obs}}
\right)
={}&
M_{r,s,k+1}^{\mathrm{exact}},
\qquad
r\in\{1,\ldots,N_s\}.
\label{eq:discrete_moment_projection_update}
\end{align}
When the target moment vector belongs to $\mathbf M_s(\mathcal A_s)$ and the moment Jacobian is non-singular on the selected branch, Eq.~\eqref{eq:discrete_moment_projection_update} gives the locally unique corrected coordinates by the same moment map used in the continuous projection.
If no admissible moment-matching solution exists inside the current closure branch, the finite-dimensional representation is instead defined by the constrained relative-entropy projection~\cite{Csiszar_1975}
\begin{equation}
\mathbf U_{s,k+1}^{\mathrm{obs}}
\doteq
\underset{
\mathbf U_s\in\mathcal A_s
}{
\operatorname{arg\,min}
}
\;
D_{\mathrm{KL}}
\left[
\rho_{k+1}^{\mathrm{exact}}
\Vert
\rho_s(\mathbf U_s)
\right].
\label{eq:discrete_projection_update}
\end{equation}
For an interior point of $\mathcal A_s$, define the score functions
\begin{equation}
\mathcal S_{s,A}(\mathbf X;\mathbf U_s)
\doteq
\frac{\partial\ln\rho_s}{\partial U_{s,A}}
=
\frac{
\Phi_A^{(s)}
}{
\rho_s
}.
\label{eq:finite_family_score}
\end{equation}
The exact stationarity equations are
\begin{align}
g_{s,A}(\mathbf U_s)
\doteq{}&
-
\int_{-\infty}^{\infty}
\cdots
\int_{-\infty}^{\infty}
\rho_{k+1}^{\mathrm{exact}}(\mathbf X)
\mathcal S_{s,A}(\mathbf X;\mathbf U_s)
\,dX_1\cdots dX_{N_X}
\nonumber\\
={}&
0.
\label{eq:KL_projection_stationarity}
\end{align}
Their Hessian is
\begin{align}
\mathsf H_{AB}^{\mathrm{KL}}
\doteq{}&
\frac{\partial g_{s,A}}{\partial U_{s,B}}
\nonumber\\
={}&
\int_{-\infty}^{\infty}
\cdots
\int_{-\infty}^{\infty}
\,dX_1\cdots dX_{N_X}
\,\rho_{k+1}^{\mathrm{exact}}(\mathbf X)
\nonumber\\
&\quad\quad\quad\quad \times
\left[
\frac{
\Phi_A^{(s)}\Phi_B^{(s)}
}{
\rho_s^2
}
-
\frac{
\partial_{U_{s,B}}\Phi_A^{(s)}
}{
\rho_s
}
\right]
.
\label{eq:KL_projection_hessian}
\end{align}
At iteration $j$, the interior Newton step is
\begin{align}
\mathsf H_{AB}^{\mathrm{KL}}
\delta U_{s,B}^{(j)}
=
-
g_{s,A}^{(j)},
\\
\mathbf U_s^{(j+1)}
=
\mathbf U_s^{(j)}
+
\alpha_j\delta\mathbf U_s^{(j)},
\label{eq:KL_projection_newton_step}
\end{align}
where $0<\alpha_j\leq1$ is reduced until
$\mathbf U_s^{(j+1)}\in\mathcal A_s$ and the KL divergence decreases.
If no positive step remains inside the first branch, the update activates the next admissible hierarchy level rather than crossing the positivity boundary.
Moment matching follows directly from Eq.~\eqref{eq:KL_projection_stationarity} only when the retained family is an exponential family whose sufficient statistics are the selected moments.
The additive first-INMDF posterior is not an exponential family, so moment projection and relative-entropy projection are not assumed to be identical.
Equation~\eqref{eq:discrete_moment_projection_update} provides the direct finite-step continuation of the continuous moment geometry whenever the Bayes-updated moments remain representable inside the current admissible branch; Eq.~\eqref{eq:discrete_projection_update} supplies the positivity-constrained information projection when they do not.
Equations~\eqref{eq:continuous_prediction_map}--\eqref{eq:KL_projection_newton_step} therefore form one continuous-discrete filter in which the prediction is generated by the physical state dynamics and the correction uses the physically derived likelihood while preserving the retained moment geometry whenever an admissible moment representation exists.
\\
The general prediction--correction cycle is now complete.
To verify that it reduces to the ordinary discrete Kalman predictor under the Gaussian assumptions of Section~\ref{sec:continuous_filtering}, consider the affine continuous state model
\begin{equation}
\dot X_A
=
A_{AB}X_B+b_A.
\label{eq:linear_state_model_discrete_section}
\end{equation}
Let \(\Delta t_k \doteq t_{k+1}-t_k\).
When $\mathsf A_k$, $\mathbf b_k$, and $\mathsf Q_k$ are constant over this interval, the exact finite-time propagator is
\begin{equation}
\mathsf F_k
\doteq
\exp\left(
\mathsf A_k\Delta t_k
\right),
\label{eq:exact_affine_state_transition}
\end{equation}
and
\begin{equation}
\mathbf g_k
\doteq
\int_0^{\Delta t_k}
\exp\left[
\mathsf A_k(\Delta t_k-\tau)
\right]
\mathbf b_k
\,d\tau.
\label{eq:exact_affine_forcing}
\end{equation}
The exact predicted mean is
\begin{equation}
\widehat{\mathbf X}_{k+1}^{\mathrm{pred}}
=
\mathsf F_k
\widehat{\mathbf X}_{k}^{\mathrm{obs}}
+
\mathbf g_k,
\label{eq:exact_affine_mean_prediction}
\end{equation}
and the exact discrete process covariance is the standard matrix-exponential covariance integral~\cite{VanLoan_1978},
\begin{equation}
\mathsf Q_k^{(d)}
\doteq
\int_0^{\Delta t_k}
e^{\mathsf A_k\tau}
\mathsf Q_k
e^{\mathsf A_k^T\tau}
\,d\tau.
\label{eq:exact_discrete_process_covariance}
\end{equation}
Therefore,
\begin{equation}
\mathsf P_{k+1}^{\mathrm{pred}}
=
\mathsf F_k
\mathsf P_k^{\mathrm{obs}}
\mathsf F_k^T
+
\mathsf Q_k^{(d)}.
\label{eq:exact_affine_covariance_prediction}
\end{equation}
The first-order Euler formulas are only the small-step expansions
\begin{align}
\widehat X_{A,k+1}^{\mathrm{pred}}
={}&
\widehat X_{A,k}^{\mathrm{obs}}
+
\Delta t_k
\left(
A_{AB,k}\widehat X_{B,k}^{\mathrm{obs}}
+
b_{A,k}
\right)
+
O(\Delta t_k^2),
\label{eq:mean_prediction_finite_difference}
\end{align}
and
\begin{align}
P_{AB,k+1}^{\mathrm{pred}}
={}&
P_{AB,k}^{\mathrm{obs}}
+
\Delta t_k
\Big(
A_{AC,k}P_{CB,k}^{\mathrm{obs}}
+
P_{AC,k}^{\mathrm{obs}}A_{BC,k}
\nonumber\\
&\hspace{8em}
+
Q_{AB,k}
\Big)
+
O(\Delta t_k^2).
\label{eq:covariance_prediction_finite_difference}
\end{align}
Thus the matrix exponential retains every time order generated by the affine dynamics; it does not introduce nonlinear state dependence.
When the projected closure dynamics is nonlinear, the finite-time prediction instead inherits derivatives of the nonlinear coordinate force.
Writing
\begin{equation}
\dot U_{s,A}
=
\mathcal A_{s,A}(\mathbf U_s,t),
\label{eq:nonlinear_coordinate_prediction_dynamics}
\end{equation}
the local finite-time flow contains the higher nonlinear order
\begin{align}
U_{s,A,k+1}^{\mathrm{pred}}
={}&
U_{s,A,k}^{\mathrm{obs}}
+
\Delta t_k\mathcal A_{s,A,k}
\nonumber\\
&+
\frac{\Delta t_k^2}{2}
\left[
\frac{\partial\mathcal A_{s,A}}{\partial t}
+
\frac{\partial\mathcal A_{s,A}}{\partial U_{s,B}}
\mathcal A_{s,B}
\right]_k
+
O(\Delta t_k^3).
\label{eq:second_order_nonlinear_coordinate_prediction}
\end{align}
A higher-order Runge--Kutta integrator~\cite{Butcher_1963}, implicit method, or structure-preserving integrator approximates this nonlinear flow while enforcing the admissible domain.
For stochastic coordinate equations, the corresponding weak or strong stochastic integrator must reproduce the required It\^o moments~\cite{KloedenPlaten_1992}; replacing Euler by a deterministic high-order formula alone is not sufficient.
\\
Having specialized the prediction step to the Gaussian affine branch, recovery of the ordinary discrete Kalman correction requires the corresponding linear-Gaussian measurement model,
\begin{equation}
y_{\mu,k+1}
=
H_{\mu A,k+1}X_{A,k+1}
+
\nu_{\mu,k+1},
\label{eq:discrete_measurement_model}
\end{equation}
with the Gaussian density
\begin{equation}
p_\nu^{(G)}(\boldsymbol\nu)
\doteq
\frac{
\exp\left[
-\frac{1}{2}
\boldsymbol\nu^T
\mathsf R_{k+1}^{-1}
\boldsymbol\nu
\right]
}{
(2\pi)^{N_y/2}
\sqrt{\det\mathsf R_{k+1}}
}.
\label{eq:discrete_gaussian_error_density}
\end{equation}
Its covariance is
\begin{align}
R_{\mu\nu,k+1}
\doteq{}&
\int_{-\infty}^{\infty}
\cdots
\int_{-\infty}^{\infty}
\nu_\mu\nu_\nu
p_\nu^{(G)}(\boldsymbol\nu)
\,d\nu_1\cdots d\nu_{N_y}.
\label{eq:measurement_covariance_definition}
\end{align}
This Gaussian error density belongs only to the standard Kalman specialization.
The general correction remains Eq.~\eqref{eq:exact_discrete_update} with the physical diagnostic likelihood derived from $K_s$ and $\mathcal H_D$, rather than with an independently assigned residual density.
The measurement residual is
\begin{equation}
r_{\mu,k+1}
\doteq
y_{\mu,k+1}
-
H_{\mu A,k+1}
\widehat X_{A,k+1}^{\mathrm{pred}},
\label{eq:measurement_residual}
\end{equation}
and its predicted covariance is
\begin{equation}
S_{\mu\nu,k+1}
\doteq
H_{\mu A,k+1}
P_{AB,k+1}^{\mathrm{pred}}
H_{\nu B,k+1}
+
R_{\mu\nu,k+1}.
\label{eq:discrete_innovation_covariance}
\end{equation}
The discrete Kalman gain is
\begin{equation}
K_{A\mu,k+1}
\doteq
P_{AB,k+1}^{\mathrm{pred}}
H_{\nu B,k+1}
(S^{-1})_{\nu\mu,k+1}.
\label{eq:discrete_kalman_gain}
\end{equation}
The corrected mean is
\begin{equation}
\widehat X_{A,k+1}^{\mathrm{obs}}
\doteq
\widehat X_{A,k+1}^{\mathrm{pred}}
+
K_{A\mu,k+1}
r_{\mu,k+1}.
\label{eq:corrected_mean}
\end{equation}
The corrected covariance is
\begin{align}
\mathsf P_{k+1}^{\mathrm{obs}}
\doteq{}&
\left(
\mathsf I-\mathsf K_{k+1}\mathsf H_{k+1}
\right)
\mathsf P_{k+1}^{\mathrm{pred}}
\left(
\mathsf I-\mathsf K_{k+1}\mathsf H_{k+1}
\right)^T
\nonumber\\
&+
\mathsf K_{k+1}
\mathsf R_{k+1}
\mathsf K_{k+1}^T.
\label{eq:discrete_corrected_covariance}
\end{align}
Equations~\eqref{eq:measurement_residual}--\eqref{eq:discrete_corrected_covariance} complete the ordinary Gaussian Kalman correction: prediction transports the Gaussian state distribution, while measurement correction contracts or displaces it according to the predicted state and diagnostic uncertainties.
Returning to the non-Gaussian first-INMDF branch, the same exact Bayesian update is followed by the admissible moment-matching correction of Eq.~\eqref{eq:discrete_moment_projection_update} whenever the corrected moment vector can be represented on the retained branch; otherwise the constrained relative-entropy projection of Eq.~\eqref{eq:discrete_projection_update} is used.
Either correction determines the Gaussian-core and localized coordinates
\begin{equation}
\mathbf U_{I,k+1}^{\mathrm{obs}}
\doteq
\left(
Y_{G,k+1}^{\mathrm{obs}},
P_{G,k+1}^{\mathrm{obs}},
\Gamma_{Y,k+1}^{\mathrm{obs}},
c_{Y,k+1}^{\mathrm{obs}},
W_{Y,k+1}^{\mathrm{obs}}
\right).
\label{eq:inmdf_discrete_coordinates}
\end{equation}
The corresponding total posterior mean and variance are not independent coordinates.
They are recovered from
\begin{align}
\widehat X_{I,k+1}^{\mathrm{obs}}
={}&
Y_{G,k+1}^{\mathrm{obs}}
+
\Gamma_{Y,k+1}^{\mathrm{obs}},
\nonumber\\
P_{I,k+1}^{\mathrm{obs}}
={}&
P_{G,k+1}^{\mathrm{obs}}
+
2\Gamma_{Y,k+1}^{\mathrm{obs}}
\left(
c_{Y,k+1}^{\mathrm{obs}}
-
Y_{G,k+1}^{\mathrm{obs}}
\right)
-
\left(
\Gamma_{Y,k+1}^{\mathrm{obs}}
\right)^2.
\label{eq:inmdf_discrete_total_moments}
\end{align}
The measurement may therefore shift the Gaussian core, contract its core covariance, and modify the amplitude, center, or width of the localized non-Gaussian channel.
These changes are obtained from the exact likelihood multiplication followed by the admissible moment map or, when required, its constrained relative-entropy fallback, not from independently assigned empirical gains.
The corrected coordinates must remain inside the first-branch domain $\mathcal A_{I,1}$.
If the measurement correction reaches its positivity boundary, the filter activates the next positive INMDF closure level rather than clipping $\rho_I$ or forcing the first branch beyond its admissible range.
\\
On an admissible moment-matching branch, the continuous and discrete constructions agree in the small-step limit through the common retained moment map.
Expanding the predicted density and Bayesian correction over a short interval $\Delta t$, the terms proportional to $\Delta t$ reproduce the deterministic drift of the continuous moment-projected equation, while the first two conditional moments of the measurement increment reproduce its It\^{o} diffusion or point-process contribution.
This stochastic consistency is defined through conditional moments rather than through a pathwise finite-difference quotient.
When a finite correction instead requires the constrained relative-entropy projection of Eq.~\eqref{eq:discrete_projection_update}, that projection remains a positive finite-family approximation to the exact Bayesian posterior but is not identified with the infinitesimal moment projector outside the exponential-family limit.
At every step, the predicted and corrected coordinate vectors must also remain inside the admissible domain of the retained closure level.

\section{From non-Maxwellian diagnostic statistics to state uncertainty}
\label{sec:nmdf_uncertainty}

The prediction--correction equations above contain two physically distinct stochastic maps.
The first describes how uncertainty propagates through the state dynamics,
whereas the second describes how a physical state produces a diagnostic measurement.
Their Gaussian specialization introduces the familiar process covariance \(Q\), state covariance \(P\), measurement covariance \(R\), and innovation covariance \(S\).
Outside the Gaussian limit, however, the complete conditional densities rather than these covariances alone determine the uncertainty update.
Table~\ref{tab:state_measurement_uncertainty} summarizes this state--measurement symmetry.

\begin{table*}[t]
\centering
\scriptsize
\renewcommand{\arraystretch}{1.18}
\caption{
State and measurement objects in the discrete Gaussian Kalman specialization and their non-Gaussian extensions.
The state uncertainty and measurement likelihood remain distinct objects.
}
\label{tab:state_measurement_uncertainty}
\begin{tabular}{
p{0.185\textwidth}
p{0.365\textwidth}
p{0.365\textwidth}
}
\hline
\textbf{Quantity}
&
\textbf{State}
&
\textbf{Measurement}
\\
\hline
Variable
&
\(\mathbf X_k\)
&
\(\mathbf y_k\)
\\
Meaning
&
True or latent physical state
&
Observed diagnostic data
\\
Plasma example
&
\(\mathbf X=(n,\mathbf u,T,\Gamma,c,W_\parallel,W_\perp,\ldots)\)
&
Langmuir probe:
\(y=I_e^{\rm meas}(U)\);
Thomson scattering:
spectral-channel intensity
\\
Gaussian model
&
\(\mathbf X_k
=
\mathsf F_{k-1}\mathbf X_{k-1}
+
\mathbf g_{k-1}
+
\mathbf w_{k-1}\)
&
\(\mathbf y_k
=
\mathsf H_k\mathbf X_k
+
\boldsymbol\nu_k\)
\\
Conditional PDF
&
\(p(\mathbf X_k\mid\mathbf X_{k-1})\)
&
\(p(\mathbf y_k\mid\mathbf X_k)\)
\\
Interpretation
&
State-evolution probability
&
Measurement-response likelihood
\\
Noise
&
\(\mathbf w_k\)
&
\(\boldsymbol\nu_k\)
\\
Gaussian noise
&
\(\mathbf w_k\sim
\mathcal N(\mathbf 0,\mathsf Q_k^{(d)})\)
&
\(\boldsymbol\nu_k\sim
\mathcal N(\mathbf 0,\mathsf R_k)\)
\\
Noise meaning
&
Unresolved state dynamics
&
Diagnostic or instrumental noise
\\
Prediction
&
\(\widehat{\mathbf X}_k^{\rm pred}
=
\mathsf F_{k-1}
\widehat{\mathbf X}_{k-1}^{\rm obs}
+
\mathbf g_{k-1}\)
&
\(\widehat{\mathbf y}_k^{\rm pred}
=
\mathsf H_k
\widehat{\mathbf X}_k^{\rm pred}\)
\\
Prediction PDF
&
\(p(\mathbf X_k\mid\mathcal D_{k-1})\)
&
\(p(\mathbf y_k\mid\mathcal D_{k-1})\)
\\
Prediction uncertainty
&
\(\mathsf P_k^{\rm pred}\)
&
\(\mathsf S_k\)
\\
Gaussian uncertainty
&
\(\mathsf P_k^{\rm pred}
=
\mathsf F_{k-1}
\mathsf P_{k-1}^{\rm obs}
\mathsf F_{k-1}^{T}
+
\mathsf Q_{k-1}^{(d)}\)
&
\(\mathsf S_k
=
\mathsf H_k
\mathsf P_k^{\rm pred}
\mathsf H_k^T
+
\mathsf R_k\)
\\
Residual
&
\(\mathbf X_k-\widehat{\mathbf X}_k^{\rm pred}\),
normally unobserved
&
\(\mathbf r_k
=
\mathbf y_k^{\rm meas}
-
\widehat{\mathbf y}_k^{\rm pred}\)
\\
Updated PDF
&
\(p(\mathbf X_k\mid\mathcal D_k)\)
&
New \(\mathbf y_k^{\rm meas}\) supplies
\(p(\mathbf y_k^{\rm meas}\mid\mathbf X_k)\)
\\
Updated uncertainty
&
\(\mathsf P_k^{\rm obs}\)
&
In the Gaussian limit, \(\mathsf R_k\) limits the information supplied by the measurement
\\
Non-Gaussian extension
&
\(\rho_s(\mathbf X;\mathbf U_s)\) retains state-space structure beyond \(\mathsf P\)
&
\(p_{D,s}(\mathbf y\mid\mathbf X)\) retains diagnostic structure beyond \(\mathsf R\) and \(\mathsf S\)
\\
\hline
\end{tabular}
\end{table*}

The state prediction and measurement prediction can be written without a Gaussian assumption.
The predicted state density is
\begin{equation}
\rho_k^{\rm pred}(\mathbf X_k)
=
\int
p(\mathbf X_k\mid\mathbf X_{k-1})
\rho_{k-1}^{\rm obs}(\mathbf X_{k-1})
\,d\mathbf X_{k-1}.
\label{eq:general_state_prediction_pdf}
\end{equation}
To separate state-transition noise from propagated state uncertainty, define the conditional state-transition mean
\begin{equation}
\boldsymbol\mu_X(\mathbf X_{k-1})
\doteq
\int
\mathbf X_k\,
p(\mathbf X_k\mid\mathbf X_{k-1})
\,d\mathbf X_k,
\label{eq:general_state_transition_mean}
\end{equation}
and conditional state-transition covariance
\begin{align}
\boldsymbol\Sigma_X(\mathbf X_{k-1})
\doteq{}&
\int
\left[
\mathbf X_k
-
\boldsymbol\mu_X(\mathbf X_{k-1})
\right]
\left[
\mathbf X_k
-
\boldsymbol\mu_X(\mathbf X_{k-1})
\right]^T
\nonumber\\
&\times
p(\mathbf X_k\mid\mathbf X_{k-1})
\,d\mathbf X_k.
\label{eq:general_state_transition_covariance}
\end{align}
The predicted state mean is
\begin{equation}
\widehat{\mathbf X}_k^{\rm pred}
\doteq
\int
\boldsymbol\mu_X(\mathbf X_{k-1})
\rho_{k-1}^{\rm obs}(\mathbf X_{k-1})
\,d\mathbf X_{k-1}.
\label{eq:general_state_prediction_mean_uncertainty}
\end{equation}
The law of total covariance then gives the exact predicted state uncertainty
\begin{align}
\mathsf P_k^{\rm pred}
={}&
\int
\boldsymbol\Sigma_X(\mathbf X_{k-1})
\rho_{k-1}^{\rm obs}(\mathbf X_{k-1})
\,d\mathbf X_{k-1}
\nonumber\\
&+
\int
\left[
\boldsymbol\mu_X(\mathbf X_{k-1})
-
\widehat{\mathbf X}_k^{\rm pred}
\right]
\left[
\boldsymbol\mu_X(\mathbf X_{k-1})
-
\widehat{\mathbf X}_k^{\rm pred}
\right]^T
\nonumber\\
&\qquad\times
\rho_{k-1}^{\rm obs}(\mathbf X_{k-1})
\,d\mathbf X_{k-1}.
\label{eq:total_state_covariance}
\end{align}
The first term is the conditional uncertainty introduced by the state transition.
The second term is the previous state uncertainty propagated through the conditional state-transition mean.
For the affine Gaussian state model,
\begin{equation}
\boldsymbol\mu_X(\mathbf X_{k-1})
=
\mathsf F_{k-1}\mathbf X_{k-1}
+
\mathbf g_{k-1},
\qquad
\boldsymbol\Sigma_X(\mathbf X_{k-1})
=
\mathsf Q_{k-1}^{(d)},
\label{eq:general_state_gaussian_specialization}
\end{equation}
so Eq.~\eqref{eq:total_state_covariance} reduces exactly to
\begin{equation}
\mathsf P_k^{\rm pred}
=
\mathsf F_{k-1}
\mathsf P_{k-1}^{\rm obs}
\mathsf F_{k-1}^T
+
\mathsf Q_{k-1}^{(d)}.
\label{eq:general_state_covariance_gaussian_limit}
\end{equation}
Thus \(\mathsf Q_{k-1}^{(d)}\) is a conditional state-transition noise covariance, whereas \(\mathsf P_k^{\rm pred}\) is the resulting uncertainty of the predicted physical state.
The corresponding predicted measurement density is
\begin{equation}
p_{D,s}^{\rm pred}
(\mathbf y_k\mid\mathcal D_{k-1})
=
\int
p_{D,s}
(\mathbf y_k\mid\mathbf X_k)
\rho_k^{\rm pred}(\mathbf X_k)
\,d\mathbf X_k.
\label{eq:predicted_measurement_density}
\end{equation}
For diagnostic \(D\), define the conditional measurement mean
\begin{equation}
\boldsymbol\mu_{D,s}(\mathbf X_k)
\doteq
\int
\mathbf y\,
p_{D,s}
(\mathbf y\mid\mathbf X_k)
\,d\mathbf y,
\label{eq:general_diagnostic_conditional_mean}
\end{equation}
and conditional measurement covariance
\begin{align}
\boldsymbol\Sigma_{D,s}(\mathbf X_k)
\doteq{}&
\int
\left[
\mathbf y
-
\boldsymbol\mu_{D,s}(\mathbf X_k)
\right]
\left[
\mathbf y
-
\boldsymbol\mu_{D,s}(\mathbf X_k)
\right]^T
\nonumber\\
&\times
p_{D,s}
(\mathbf y\mid\mathbf X_k)
\,d\mathbf y.
\label{eq:general_diagnostic_conditional_covariance}
\end{align}
The predicted measurement mean is
\begin{equation}
\widehat{\mathbf y}_k^{\rm pred}
\doteq
\int
\boldsymbol\mu_{D,s}(\mathbf X_k)
\rho_k^{\rm pred}(\mathbf X_k)
\,d\mathbf X_k,
\label{eq:general_predicted_measurement_mean}
\end{equation}
and the law of total covariance gives
\begin{align}
\boldsymbol\Sigma_{y,k}^{\rm pred}
={}&
\int
\boldsymbol\Sigma_{D,s}(\mathbf X_k)
\rho_k^{\rm pred}(\mathbf X_k)
\,d\mathbf X_k
\nonumber\\
&+
\int
\left[
\boldsymbol\mu_{D,s}(\mathbf X_k)
-
\widehat{\mathbf y}_k^{\rm pred}
\right]
\left[
\boldsymbol\mu_{D,s}(\mathbf X_k)
-
\widehat{\mathbf y}_k^{\rm pred}
\right]^T
\nonumber\\
&\qquad\times
\rho_k^{\rm pred}(\mathbf X_k)
\,d\mathbf X_k.
\label{eq:general_predicted_measurement_covariance}
\end{align}
The first term is the conditional variability of the diagnostic at fixed physical state.
The second term is the uncertainty of the physical state propagated through the conditional diagnostic mean.
For the linear-Gaussian observation model,
\begin{equation}
\boldsymbol\mu_{D}(\mathbf X_k)
=
\mathsf H_k\mathbf X_k,
\qquad
\boldsymbol\Sigma_D(\mathbf X_k)
=
\mathsf R_k,
\label{eq:general_measurement_gaussian_specialization}
\end{equation}
and Eq.~\eqref{eq:general_predicted_measurement_covariance} reduces exactly to
\begin{equation}
\boldsymbol\Sigma_{y,k}^{\rm pred}
=
\mathsf S_k
=
\mathsf H_k
\mathsf P_k^{\rm pred}
\mathsf H_k^T
+
\mathsf R_k.
\label{eq:general_measurement_covariance_gaussian_limit}
\end{equation}
Thus \(\mathsf R_k\) is the conditional measurement-noise covariance, whereas \(\mathsf S_k\) is the total predicted-measurement uncertainty.
Equation~\eqref{eq:general_predicted_measurement_covariance} is the exact non-Gaussian counterpart of the Gaussian innovation-covariance decomposition.
The realized innovation is
\begin{equation}
\mathbf r_k
\doteq
\mathbf y_k^{\mathrm{meas}}
-
\widehat{\mathbf y}_k^{\mathrm{pred}}.
\end{equation}
Before the measurement is observed, its predictive probability density is
\begin{equation}
p_{r,s}
\left(
\mathbf r
\mid
\mathcal D_{k-1}
\right)
=
p_{D,s}^{\mathrm{pred}}
\left(
\mathbf r
+
\widehat{\mathbf y}_k^{\mathrm{pred}}
\mid
\mathcal D_{k-1}
\right).
\label{eq:non_gaussian_innovation_density}
\end{equation}
The ordinary Kalman representation retains the mean and covariance of this innovation distribution through \(\mathbf r_k\) and \(\mathsf S_k\), whereas the non-Gaussian construction retains the complete predictive density in Eq.~\eqref{eq:non_gaussian_innovation_density}.
A single large observed innovation therefore does not by itself imply increased process noise or increased measurement noise, because it can be a statistically expected realization of an asymmetric or heavy-tailed physical likelihood.
Persistent excess innovation or predictive surprise under a restricted kinetic manifold has a different interpretation: when the same kinetic discrepancy remains unresolved over repeated measurements and is absorbed into the residual channel, it can enlarge the effective measurement-noise law and consequently the inferred state uncertainty.
Appendix~\ref{app:model_induced_innovation} derives this model-induced contribution and separates it from the physical uncertainty that remains after the kinetic likelihood is adequately resolved.
For a scalar diagnostic likelihood with cumulants
\(\kappa_{r,D,s}\), define the standardized skewness and excess kurtosis by
\begin{equation}
S_{D,s}
\doteq
\frac{
\kappa_{3,D,s}
}{
\kappa_{2,D,s}^{3/2}
},
\qquad
F_{D,s}
\doteq
\frac{
\kappa_{4,D,s}
}{
\kappa_{2,D,s}^{2}
}.
\label{eq:diagnostic_skewness_kurtosis_uncertainty}
\end{equation}
These quantities characterize the shape of the measurement likelihood beyond its variance.
They are not themselves measures of state uncertainty.
Two diagnostic likelihoods can have the same mean and variance while possessing different \(S_{D,s}\), \(F_{D,s}\), tail weights, or localized structure.
The effect of this likelihood shape on the inferred state is seen directly from Bayes' rule.
Define the normalization
\begin{equation}
Z_{s,k}
\doteq
\int
p_{D,s}
\left(
\mathbf y_k^{\rm meas}\mid\mathbf X
\right)
\rho_k^{\rm pred}(\mathbf X)
\,d\mathbf X.
\label{eq:posterior_normalization}
\end{equation}
The corrected state mean is
\begin{equation}
\widehat X_{A,k}^{\rm obs}
=
\frac{1}{Z_{s,k}}
\int
X_A
p_{D,s}
\left(
\mathbf y_k^{\rm meas}\mid\mathbf X
\right)
\rho_k^{\rm pred}(\mathbf X)
\,d\mathbf X,
\label{eq:nmdf_uncertainty_updated_mean}
\end{equation}
and its covariance is
\begin{align}
P_{AB,k}^{\rm obs}
={}&
\frac{1}{Z_{s,k}}
\int
\left(
X_A-\widehat X_{A,k}^{\rm obs}
\right)
\left(
X_B-\widehat X_{B,k}^{\rm obs}
\right)
\nonumber\\
&\times
p_{D,s}
\left(
\mathbf y_k^{\rm meas}\mid\mathbf X
\right)
\rho_k^{\rm pred}(\mathbf X)
\,d\mathbf X.
\label{eq:posterior_covariance_exact}
\end{align}
Equation~\eqref{eq:posterior_covariance_exact} gives the direct connection between the physical non-Maxwellian measurement likelihood and the posterior state uncertainty.
Changing the kinetic manifold changes \(p_{D,s}\), which changes the Bayesian weighting of the predicted state density and can therefore change both the corrected covariance and the higher-order structure of the posterior.
The corresponding reduction of state uncertainty can also be written exactly before any particular measurement is realized.
For an arbitrary possible measurement \(\mathbf y\), define
\begin{equation}
Z_{s,k}(\mathbf y)
\doteq
\int
p_{D,s}
\left(
\mathbf y\mid\mathbf X
\right)
\rho_k^{\rm pred}(\mathbf X)
\,d\mathbf X
=
p_{D,s}^{\rm pred}
\left(
\mathbf y\mid\mathcal D_{k-1}
\right),
\label{eq:nmdf_uncertainty_generic_normalization}
\end{equation}
with conditional posterior
\begin{equation}
\rho_{s,k}^{\rm obs}
(\mathbf X\mid\mathbf y)
=
\frac{
p_{D,s}
\left(
\mathbf y\mid\mathbf X
\right)
\rho_k^{\rm pred}(\mathbf X)
}{
Z_{s,k}(\mathbf y)
}.
\label{eq:nmdf_uncertainty_generic_posterior}
\end{equation}
Its conditional mean is
\begin{equation}
\widehat X_{A,k}^{\rm obs}(\mathbf y)
\doteq
\int
X_A
\rho_{s,k}^{\rm obs}
(\mathbf X\mid\mathbf y)
\,d\mathbf X,
\label{eq:nmdf_uncertainty_generic_mean}
\end{equation}
and its conditional covariance is
\begin{align}
P_{AB,k}^{\rm obs}(\mathbf y)
\doteq{}&
\int
\left[
X_A
-
\widehat X_{A,k}^{\rm obs}(\mathbf y)
\right]
\left[
X_B
-
\widehat X_{B,k}^{\rm obs}(\mathbf y)
\right]
\nonumber\\
&\times
\rho_{s,k}^{\rm obs}
(\mathbf X\mid\mathbf y)
\,d\mathbf X.
\label{eq:nmdf_uncertainty_generic_covariance}
\end{align}
The law of total covariance then gives
\begin{align}
P_{AB,k}^{\rm pred}
={}&
\int
P_{AB,k}^{\rm obs}(\mathbf y)
p_{D,s}^{\rm pred}
\left(
\mathbf y\mid\mathcal D_{k-1}
\right)
\,d\mathbf y
\nonumber\\
&+
\int
\left[
\widehat X_{A,k}^{\rm obs}(\mathbf y)
-
\widehat X_{A,k}^{\rm pred}
\right]
\left[
\widehat X_{B,k}^{\rm obs}(\mathbf y)
-
\widehat X_{B,k}^{\rm pred}
\right]
\nonumber\\
&\qquad\times
p_{D,s}^{\rm pred}
\left(
\mathbf y\mid\mathcal D_{k-1}
\right)
\,d\mathbf y.
\label{eq:total_posterior_covariance}
\end{align}
The first term is the posterior state uncertainty remaining after the measurement, averaged over all possible diagnostic outcomes.
The second term is the state variance resolved by the information carried by the measurement.
Equation~\eqref{eq:total_posterior_covariance} therefore gives an exact non-Gaussian uncertainty-reduction identity.
The amount of state uncertainty removed by the diagnostic depends on the complete likelihood \(p_{D,s}\), not only on its variance or on an effective \(\mathsf R\).
The complete likelihood contains more information than its first four moments.
A local measure of the state information carried by diagnostic \(D\) is the Fisher-information matrix
\begin{equation}
\mathcal I_{AB}^{D,s}(\mathbf X)
\doteq
\int
p_{D,s}(\mathbf y\mid\mathbf X)
\frac{\partial\ln p_{D,s}}{\partial X_A}
\frac{\partial\ln p_{D,s}}{\partial X_B}
\,d\mathbf y.
\label{eq:fisher_information}
\end{equation}
For the linear-Gaussian observation law, \(\mathcal I^{D,G} = \mathsf H^T \mathsf R^{-1} \mathsf H\).
Within a local Gaussian or Laplace approximation, the corresponding precision update has the familiar form
\begin{equation}
\left(
\mathsf P_k^{\rm obs}
\right)^{-1}
\simeq
\left(
\mathsf P_k^{\rm pred}
\right)^{-1}
+
\mathcal I^{D,s}.
\label{eq:local_precision_information_update}
\end{equation}
In the intended NMDF construction, the non-Maxwellian correction does not enter the uncertainty by replacing \(R\) with an empirical effective covariance; it enters through the complete physical likelihood \(p_{D,s}\).
Appendix~\ref{app:model_induced_innovation} examines the distinct failure mode in which a restricted MDF is nevertheless forced and the resulting kinetic-model discrepancy is absorbed into an effective residual covariance.
Observed changes of skewness, excess kurtosis, shoulders, or tails are signatures of the physical likelihood structure and can consequently modify the inferred posterior uncertainty even when the measurement variance is unchanged.
There is no universal monotonic relation between \(S_{D,s}\), \(F_{D,s}\), and \(P_{AB}^{\rm obs}\); the full likelihood determines the Bayesian information.
When an additive instrumental-noise layer is represented separately from the physical diagnostic response, its role can be written explicitly as
\begin{equation}
p_{D,s}^{\mathrm{meas}}
\left(
y
\mid
\mathbf X
\right)
=
\int_{-\infty}^{\infty}
p_\nu
\left(
y-y'
\right)
p_{D,s}^{\mathrm{phys}}
\left(
y'
\mid
\mathbf X
\right)
\,dy'.
\label{eq:measurement_noise_convolution}
\end{equation}
In the Gaussian specialization, \(\mathsf R=\operatorname{Cov}(\nu)\).
Thus \(\mathsf R\) characterizes instrumental measurement noise, whereas the non-Gaussian structure generated by the kinetic state and diagnostic response belongs to the physical likelihood and should not be absorbed into \(\mathsf R\).
Likewise, \(\mathsf Q\) describes unresolved state dynamics and should not be re-fitted to reproduce the observed diagnostic tails.

\section{Kinetic entropy and Bayesian information}
\label{sec:entropy}

With the state and measurement uncertainties now separated, three quantities with similar statistical language must remain distinct.
The kinetic relative entropy belongs to the physical distribution \(f\).
The posterior entropy measures uncertainty in the inferred state.
The KL divergence produced by a measurement quantifies information gained by Bayesian conditioning.
Only the first is a thermodynamic property of the plasma.

\subsection{Physical kinetic entropy balance}
For the plasma distribution, we use the relative entropy with respect to its Maxwellian backbone,
\begin{equation}
\mathcal L_f(t)
\doteq
\int_{-\infty}^{\infty}
f_p(\mathbf v,t)
\ln\left[
\frac{
f_p(\mathbf v,t)
}{
f_M(\mathbf v,t)
}
\right]
\,d^3\vpartb{v}.
\label{eq:kinetic_entropy}
\end{equation}
Using Eq.~\eqref{eq:open_lfp}, its evolution is
\begin{equation}
\frac{d\mathcal L_f}{dt}
=
-
\mathcal D_{pp}[f_p]
+
\mathcal P_{p{\rm src}}[f_p,f_{\rm src}]
+
\mathcal B_f,
\label{eq:kinetic_entropy_balance}
\end{equation}
where the self-collisional dissipation is
\begin{equation}
\mathcal D_{pp}[f_p]
\doteq
-
\int_{-\infty}^{\infty}
C_{pp}[f_p,f_p]
\ln\left(
\frac{f_p}{f_M}
\right)
\,d^3\vpartb{v}.
\label{eq:self_collision_entropy}
\end{equation}
Because \(\ln f_M = a_0+a_iv_i+a_2|\mathbf v|^2\) for coefficients fixed by $(n,\mathbf u,T)$, conservation of particles, momentum, and energy by $C_{pp}$ gives
\begin{equation}
\int_{-\infty}^{\infty}
C_{pp}[f_p,f_p]
\ln f_M
\,d^3\vpartb{v}
=
0.
\label{eq:maxwellian_collision_invariant_projection}
\end{equation}
Consequently,
\begin{equation}
\mathcal D_{pp}[f_p]
=
-
\int_{-\infty}^{\infty}
C_{pp}[f_p,f_p]
\ln f_p
\,d^3\vpartb{v}.
\label{eq:self_collision_entropy_reduced}
\end{equation}
Symmetrizing the bilinear Landau operator under
$\mathbf v\leftrightarrow\mathbf v'$ yields
\begin{align}
\mathcal D_{pp}[f_p]
={}&
\frac{\gamma_{pp}}{2}
\int_{-\infty}^{\infty}
\int_{-\infty}^{\infty}
f_p(\mathbf v)
f_p(\mathbf v')
U_{ij}(\mathbf v-\mathbf v')
\nonumber\\
&\times
\left[
\frac{\partial\ln f_p(\mathbf v)}{\partial v_i}
-
\frac{\partial\ln f_p(\mathbf v')}{\partial v'_i}
\right]
\nonumber\\
&\times
\left[
\frac{\partial\ln f_p(\mathbf v)}{\partial v_j}
-
\frac{\partial\ln f_p(\mathbf v')}{\partial v'_j}
\right]
\,d^3\vpartb{v}'\,d^3\vpartb{v}
\geq
0,
\label{eq:self_collision_entropy_positive_form}
\end{align}
because $U_{ij}$ is positive semidefinite on the subspace perpendicular to $\mathbf v-\mathbf v'$, consistently with the collisional Landau $H$-theorem~\cite{Landau_1937,Rosenbluth_1957}.
The source-collision contribution is
\begin{equation}
\mathcal P_{p{\rm src}}[f_p,f_{\rm src}]
\doteq
\int_{-\infty}^{\infty}
C_{p{\rm src}}[f_p,f_{\rm src}]
\ln\left(
\frac{f_p}{f_M}
\right)
\,d^3\vpartb{v},
\label{eq:source_collision_entropy}
\end{equation}
and has no universal sign because $f_{\rm src}$ is an externally maintained population rather than the self-collisional equilibrium of $f_p$.
The transport and moving-reference contribution is
\begin{align}
\mathcal B_f
\doteq{}&
-
\int_{-\infty}^{\infty}
\left(
\mathbf v\cdot\nabla_{\mathbf x}f_p
+
\dot{\mathbf v}\cdot\nabla_{\mathbf v}f_p
\right)
\left[
1+
\ln\left(
\frac{f_p}{f_M}
\right)
\right]
\,d^3\vpartb{v}
\nonumber\\
&-
\int_{-\infty}^{\infty}
f_p
\frac{\partial\ln f_M}{\partial t}
\,d^3\vpartb{v}.
\label{eq:transport_entropy_term}
\end{align}
At a source-supported steady state~\cite{Izacard_2026_INMDF},
\begin{equation}
\mathcal D_{pp}[f_p]
=
\mathcal P_{p{\rm src}}[f_p,f_{\rm src}]
+
\mathcal B_f,
\label{eq:steady_entropy_balance}
\end{equation}
so a finite non-Maxwellian component is maintained when source and transport injection compensate the nonnegative self-collisional dissipation.

\subsection{Conditional-state entropy and measurement information}
We now leave the physical entropy balance and consider the two inferential quantities.
For the fixed physical state-space coordinates used here, one scalar measure of posterior spread is the Shannon differential entropy~\cite{Shannon_1948} of the conditional state density,
\begin{align}
S_\rho(t)
\doteq{}&
-
\int_{-\infty}^{\infty}
\cdots
\int_{-\infty}^{\infty}
\rho(\mathbf X,t\mid\mathcal D_t)
\ln
\rho(\mathbf X,t\mid\mathcal D_t)
\,dX_1\cdots dX_{N_X}.
\label{eq:posterior_entropy}
\end{align}
Its numerical value depends on the continuous state-coordinate representation, so it is interpreted only within the fixed physical coordinates of the present construction.
The probability current associated with the prediction equation is
\begin{equation}
J_A(\mathbf X,t)
\doteq
F_A\rho
-
\frac{1}{2}
\frac{\partial}{\partial X_B}
\left(
D_{AB}\rho
\right).
\label{eq:posterior_probability_current}
\end{equation}
Equation~\eqref{eq:state_fokker_planck} then becomes
\begin{equation}
\partial_t\rho
=
-
\frac{\partial J_A}{\partial X_A}.
\label{eq:posterior_continuity}
\end{equation}
If the state-space probability current vanishes at infinity, the prediction contribution to the conditional-state entropy rate is
\begin{align}
\left.
\frac{dS_\rho}{dt}
\right|_{\mathrm{pred}}
={}&
-
\int_{-\infty}^{\infty}
\cdots
\int_{-\infty}^{\infty}
J_A
\frac{\partial\ln\rho}{\partial X_A}
\,dX_1\cdots dX_{N_X}.
\label{eq:posterior_entropy_current}
\end{align}
This term describes transport and diffusion of conditional uncertainty in state space.
It is not the physical collisional entropy production of Eq.~\eqref{eq:self_collision_entropy}.
The second inferential quantity enters when a measurement changes that conditional state.
At a discrete measurement time $t_k$, let $\rho_k^{\mathrm{pred}}$ denote the posterior immediately before the new measurement is included and let $\rho_k^{\mathrm{obs}}$ denote the posterior immediately afterward.
The information supplied by that observation is measured through the Kullback--Leibler divergence~\cite{KullbackLeibler_1951}
\begin{equation}
I_k
\doteq
D_{\mathrm{KL}}
\left[
\rho_k^{\mathrm{obs}}
\Vert
\rho_k^{\mathrm{pred}}
\right],
\end{equation}
where
\begin{align}
D_{\mathrm{KL}}[\rho_1\Vert\rho_2]
\doteq{}&
\int_{-\infty}^{\infty}
\cdots
\int_{-\infty}^{\infty}
\rho_1(\mathbf X)
\ln\left[
\frac{
\rho_1(\mathbf X)
}{
\rho_2(\mathbf X)
}
\right]
\,dX_1\cdots dX_{N_X}.
\label{eq:KL_definition}
\end{align}
The measurement update changes posterior uncertainty through Bayesian conditioning; it must not be identified with physical entropy production.
\\
When the exact corrected posterior does not belong to the retained finite family, its finite representation follows the same two-branch closure derived in Section~\ref{sec:continuous_discrete}.
If the corrected retained moments belong to the admissible moment image, Eq.~\eqref{eq:discrete_moment_projection_update} preserves the continuous moment geometry.
When no admissible moment-matching representation exists on the retained branch, Eq.~\eqref{eq:discrete_projection_update} instead selects the positive finite-family representation by constrained relative-entropy minimization.
The latter information projection is therefore a finite-closure operation on the Bayesian posterior and must not be confused with the information gain \(I_k\) defined above.

\subsection{Weak kinetic-to-diagnostic entropy relation}
Although physical kinetic entropy and Bayesian information remain distinct, they can still be related at the level of their common physical coordinates.
For the first-INMDF branch, both the kinetic relative entropy and the diagnostic cumulants depend on the same amplitude $\Gamma$, which provides the required connection.
Specifically,
\begin{align}
f_I
=
f_M
+
\Gamma\eta G_I,
\\
\int_{-\infty}^{\infty}
\eta G_I
\,d^3\vpartb{v}
=
0,
\label{eq:weak_inmdf_entropy_branch}
\end{align}
the weak-amplitude expansion gives
\begin{align}
\mathcal L_f[f_I]
=
\frac{\Gamma^2}{2}
\mathcal A_f
+
O(\Gamma^3),
\\
\mathcal A_f
\doteq
\int_{-\infty}^{\infty}
\frac{\eta^2G_I^2}{f_M}
\,d^3\vpartb{v}
>
0.
\label{eq:weak_inmdf_relative_entropy}
\end{align}
To connect this velocity-space entropy expansion to a measurable diagnostic quantity, consider a Langmuir-probe acquisition cumulant of order $r\geq2$ and define its linear response to the physical INMDF amplitude by
\begin{equation}
\mathcal K_r(U)
\doteq
\left.
\frac{\partial
\kappa_{r,I}^{\mathrm{LP}}(U)}
{\partial\Gamma}
\right|_{\Gamma=0}.
\label{eq:probe_cumulant_linear_response}
\end{equation}
Then
\begin{equation}
\Delta\kappa_r^{\mathrm{LP}}(U)
\doteq
\kappa_{r,I}^{\mathrm{LP}}(U)
-
\kappa_{r,M}^{\mathrm{LP}}(U)
=
\Gamma\mathcal \mathcal K_r(U)
+
O(\Gamma^2).
\label{eq:probe_cumulant_weak_inmdf}
\end{equation}
Whenever $\mathcal K_r(U)\neq0$, elimination of $\Gamma$ gives the local observable relation
\begin{equation}
\mathcal L_f[f_I]
=
\frac{
A_f
}{
2K_r^2(U)
}
\left[
\Delta\kappa_{r}^{\mathrm{LP}}(U)
\right]^2
+
O
\left(
\left|
\Delta\kappa_{r}^{\mathrm{LP}}
\right|^3
\right).
\label{eq:entropy_cumulant_relation}
\end{equation}
The coefficient $\mathcal K_r(U)$ is not an independently fitted parameter: it must be evaluated from the same fixed Langmuir-probe response and the same kinetic-coordinate dynamics used for the fluctuation prediction.
This is not an equality between kinetic entropy production and Bayesian information.
It is a coordinate-level relation between a physical entropy excess and a measurable non-Maxwellian fluctuation signature in the weak first-INMDF branch.
The kinetic relative entropy, the diagnostic likelihood, and the posterior entropy remain three distinct objects.
Section~\ref{sec:inmdf_measurement} now writes explicitly the Langmuir-probe response from which the coefficient \(\mathcal K_r(U)\) is evaluated.

\section{Langmuir-probe measurement law}
\label{sec:inmdf_measurement}

We now write this Langmuir-probe diagnostic map explicitly and separate the established mean response from the new fluctuation test.
The classical collection problem originates with Mott-Smith and Langmuir~\cite{MottSmithLangmuir_1926}, and velocity-distribution reconstruction was developed further by Druyvesteyn~\cite{Druyvesteyn_1930}.
The previous INMDF kinetic-correction analysis~\cite{Izacard_2016} already reduced the diffusionless mean current to the analytic family
\begin{equation}
J_k(a,b,c)
\doteq
\int_c^{\infty}
v^k
\exp\left(
-av^2+bv
\right)
\,dv,
\label{eq:published_probe_Jk_family}
\end{equation}
and used finite combinations of \(J_k\) to obtain the Maxwellian, Kappa, bi-Maxwellian, and INMDF characteristic curves~\cite{Izacard_2016}.
That result supplies the established first-moment baseline used here.
The new question is whether the same physical INMDF coordinates are also visible in the non-Gaussian current fluctuations.
For a standalone Langmuir-probe construction, we now write the map of Section~\ref{sec:kinetic_to_measurement} explicitly for the probe geometry.
The symbol $\mathcal H_{\mathrm{LP}}$ denotes the fixed Langmuir-probe diagnostic map.
The normalized kinetic probability remains the same object defined in Eq.~\eqref{eq:conditional_instrument_response}
\begin{equation}
K_s
\left(
\mathbf v\mid
\mathbf X_s(\mathbf x_p,t)
\right)
\doteq
\frac{
f_s
\left(
\mathbf X_s(\mathbf x_p,t),
\mathbf v
\right)
}{
n_s(\mathbf x_p,t)
}.
\label{eq:probe_kinetic_probability}
\end{equation}
It satisfies
\begin{align}
&
\int_{-\infty}^{\infty}
\int_{-\infty}^{\infty}
\int_{-\infty}^{\infty}
K_s
\left(
\mathbf v\mid\mathbf X_s
\right)
\,dv_\parallel
dv_{\perp1}
dv_{\perp2}
=
1.
\label{eq:probe_kinetic_probability_normalization}
\end{align}
The diagnostic-specific physics is contained entirely in $\mathcal H_{\mathrm{LP}}$, not in a second probability kernel.
\\
In the collisionless/diffusionless probe-collection limit developed in classical sheath-probe theory~\cite{Laframboise_1966} and used in modern electron-distribution probe analysis~\cite{GodyakDemidov_2011}
\begin{align}
\mathcal H_{\mathrm{LP}}^{(0)}
\left[
K_s
\right](U)
\doteq{}&
-eS_p n_s
\int_{-\infty}^{\infty}
\int_{-\infty}^{\infty}
\int_{-\infty}^{\infty}
v_n
\mathcal T_0(U,\mathbf v)
K_s
(\mathbf v\mid\mathbf X_s)
\nonumber\\
&\hspace{8em}
\times
dv_\parallel
dv_{\perp1}
dv_{\perp2},
\label{eq:diffusionless_probe_operator}
\end{align}
where
\begin{align}
v_n
&\doteq
-\mathbf v\cdot\widehat{\mathbf n}_p,
\\
\mathcal T_0(U,\mathbf v)
&\doteq
\Theta(v_n)
\Theta\left[
\frac{mv_n^2}{2}
-
E_b(U)
\right].
\label{eq:diffusionless_probe_transmission}
\end{align}
Here $S_p$ is the collecting area, $\widehat{\mathbf n}_p$ is the outward probe normal, $U$ is the probe bias relative to the plasma potential, and $E_b(U)$ is the corresponding barrier energy.
The density $n_s$ appears because the electric current is an extensive particle-flux measurement, whereas $K_s=f_s/n_s$ remains normalized to unity.
\\
In the published isotropic one-dimensional reduction, the same response becomes
\begin{equation}
\mathcal H_{\mathrm{LP}}^{(0)}
\left[
K_s
\right](U)
=
-\Delta_{LP}
\int_{u(U)}^{\infty}
\left(
\frac{mv^2}{2}
-
eU
\right)
v
K_s(v\mid\mathbf X_s)
\,dv,
\label{eq:published_diffusionless_probe_integral}
\end{equation}
with
\begin{equation}
\Delta_{LP} \doteq \frac{2\pi eS_p n_s}{m},
\quad
u(U)
\doteq
\sqrt{\frac{2eU}{m}},
\end{equation}
for the corresponding positive barrier convention.
This is exactly the same diffusionless physical response previously evaluated for Maxwellian, Kappa, bi-Maxwellian, and INMDF velocity distributions, now written consistently in terms of the normalized kinetic probability $K_s=f_s/n_s$.
\\
The established first-INMDF mean current is
\begin{equation}
I_{e,I}^{(0)}(U\mid\mathbf X_I)
=
I_{e,M}^{(0)}(U\mid\mathbf X_M)
+
\Gamma
I_{e,\Gamma}^{(0)}
(U\mid c,W_\parallel,W_\perp),
\label{eq:inmdf_probe_current_decomposition}
\end{equation}
where every term on the right-hand side is already available from the published $J_k$ reduction.
The historical Maxwellian current with empirical diffusion is denoted by $I_{e,M}^{(\psi)}$ and is retained only as the earlier phenomenological comparator.
The deformation of the mean characteristic relative to the Maxwellian response is therefore the established physical baseline, not the new held-out result of this manuscript.
\\
The Langmuir probe provides the scalar-response counterpart to the resolved diagnostic laws of Section~\ref{sec:kinetic_to_measurement}.
For any retained physical kinetic family \(s\), including the MDF, Kappa, and INMDF examples above, the instantaneous current at fixed probe voltage $U$ is obtained by applying the same physical Langmuir-probe functional to the corresponding normalized kinetic probability
\begin{equation}
I_{e,s}^{(0)}(t;U)
\doteq
\mathcal H_{\mathrm{LP}}^{(0)}
\left[
K_s
\left(
\mathbf v\mid
\mathbf X_s(\mathbf x_p,t)
\right)
\right](U).
\label{eq:instantaneous_probe_current}
\end{equation}
Here $\mathbf x_p$ is the probe position.
The probe voltage $U$ is the controlled diagnostic setting, whereas the measured scalar quantity at fixed $U$ is the current \(y \equiv I_e\).
The kinetic coordinates and therefore $K_s$ may fluctuate in time, but the physical functional $\mathcal H_{\mathrm{LP}}^{(0)}$, probe geometry, bias definition, and calibration are kept fixed.
\\
For a stationary fixed-voltage interval $t\in[t_0,t_1]$, the corresponding current probability density is the time pushforward of this physical response,
\begin{equation}
p_{\mathrm{LP},s}(y\mid U)
\doteq
\frac{1}{t_1-t_0}
\int_{t_0}^{t_1}
\delta
\left[
y-I_{e,s}^{(0)}(t;U)
\right]
\,dt.
\label{eq:langmuir_probe_likelihood}
\end{equation}
Equations~\eqref{eq:instantaneous_probe_current} and~\eqref{eq:langmuir_probe_likelihood} separate the two physical-statistical steps explicitly: (i) the velocity-space distribution \(f_s\) determines the normalized kinetic probability \(K_s\), which the fixed Langmuir-probe response \(\mathcal H_{\mathrm{LP}}^{(0)}\) maps to the instantaneous current \(I_{e,s}^{(0)}(t;U)\), and (ii) the temporal or ensemble variation of that physical current determines \(p_{\mathrm{LP},s}(y\mid U)\) through Eq.~\eqref{eq:langmuir_probe_likelihood}.
The time dependence in Eq.~\eqref{eq:langmuir_probe_likelihood} must arise from evolution of the physical state and is not introduced through an independently fitted fluctuation law.
Time-resolved scrape-off-layer probe measurements exhibit strongly non-Gaussian intermittent current statistics~\cite{Kube_2016}, for which stochastic intermittent-pulse descriptions provide an established measurement-space baseline~\cite{Garcia_2012}.
These stochastic descriptions characterize the temporal statistics of the diagnostic signal after the measurement response; they do not replace the Maxwellian or non-Maxwellian velocity-space family $f_s(\mathbf v)$ entering $\mathcal H_{\mathrm{LP}}^{(0)}$.
In the present construction, the required time dependence is supplied by the trajectory $\mathbf X_s(\mathbf x_p,t)$, generated from the projected coordinate dynamics of Eq.~\eqref{eq:projected_state_sde} with $F_A$ obtained from the kinetic projection and $D_{AB}$ specified independently of the held-out current statistics, or supplied by an independent time-resolved reconstruction.
Once this trajectory is specified, Eq.~\eqref{eq:langmuir_probe_likelihood} determines the current PDF and hence its raw moments,
\begin{align}
M_{r,s}^{\mathrm{LP}}(U)
\doteq{}&
\int_{-\infty}^{\infty}
y^r
p_{\mathrm{LP},s}(y\mid U)
\,dy
\nonumber\\
={}&
\frac{1}{t_1-t_0}
\int_{t_0}^{t_1}
\left[
I_{e,s}^{(0)}(t;U)
\right]^r
\,dt,
\quad
r\in\mathbb N_0.
\label{eq:probe_acquisition_raw_moments}
\end{align}
In the static-state limit,
\begin{equation}
\mathbf X_s(\mathbf x_p,t)
=
\mathbf X_s(\mathbf x_p),
\end{equation}
Eq.~\eqref{eq:langmuir_probe_likelihood} collapses to
\begin{equation}
p_{\mathrm{LP},s}(y\mid U)
=
\delta
\left[
y-
I_{e,s}^{(0)}
\left(
U\mid\mathbf X_s
\right)
\right].
\end{equation}
A nontrivial current probability density therefore requires temporal or ensemble variation of the physical kinetic coordinates.
The first moment nevertheless reduces to the established mean response,
\begin{equation}
M_{1,s}^{\mathrm{LP}}(U)
=
I_{e,s}^{(0)}
\left(
U\mid\mathbf X_s
\right).
\label{eq:probe_first_moment_mean_current}
\end{equation}
The continuous-time definitions have a direct discrete counterpart for repeated sweeps: the time averages are replaced by sample averages at each voltage.
To separate the established mean response from the fluctuations around it, define the centered signal
\begin{equation}
\delta y_s
\doteq
y-
M_{1,s}^{\mathrm{LP}}.
\label{eq:probe_centered_signal}
\end{equation}
The second, third, and fourth fluctuation cumulants used in the experimental comparison are then
\begin{align}
\kappa_{2,s}^{\mathrm{LP}}(U)
\doteq{}&
\int_{-\infty}^{\infty}
\delta y_s^2
p_{\mathrm{LP},s}(y\mid U)
\,dy,
\nonumber\\
\kappa_{3,s}^{\mathrm{LP}}(U)
\doteq{}&
\int_{-\infty}^{\infty}
\delta y_s^3
p_{\mathrm{LP},s}(y\mid U)
\,dy,
\nonumber\\
\kappa_{4,s}^{\mathrm{LP}}(U)
\doteq{}&
\int_{-\infty}^{\infty}
\delta y_s^4
p_{\mathrm{LP},s}(y\mid U)
\,dy
-
3
\left[
\kappa_{2,s}^{\mathrm{LP}}(U)
\right]^2.
\label{eq:probe_acquisition_cumulants}
\end{align}
The cumulants above characterize the fluctuation statistics at each voltage separately.
A repeated sweep contains additional information because currents measured at different voltages within the same sweep remain correlated.
For the voltage sequence
\begin{equation}
\mathbf U
\doteq
(U_1,\ldots,U_{N_U}),
\label{eq:probe_voltage_vector}
\end{equation}
let $I_{s,k}(U_i)$ denote the current measured or predicted at voltage $U_i$ during sweep $k$, with $k\in\{1,\ldots,N_{\mathrm{sw}}\}$.
The corresponding cross-voltage covariance is
\begin{align}
C_{ij,s}^{\mathrm{LP}}
\doteq{}&
\frac{1}{N_{\mathrm{sw}}}
\sum_{k=1}^{N_{\mathrm{sw}}}
\left[
I_{s,k}(U_i)-M_{1,s}^{\mathrm{LP}}(U_i)
\right]
\nonumber\\
&\times
\left[
I_{s,k}(U_j)-M_{1,s}^{\mathrm{LP}}(U_j)
\right].
\label{eq:probe_cross_voltage_covariance}
\end{align}
To ensure that differences in the predicted fluctuations originate from the kinetic manifold rather than from model-specific diagnostic tuning, the same probe geometry, voltage sequence, electronic transfer function, and calibration are used for $f_M$, $f_\kappa$, and $f_I$.
Two additional constraints are imposed.
First, the mean physical coordinates $(n,\mathbf u,T,\Gamma,c,W_\parallel,W_\perp)$ are imported from the established mean-characteristic inversion or inferred from an independent reconstruction.
Second, neither kinetic nor diagnostic parameters are re-fitted to the higher current statistics.
The time-dependent coordinates entering Eq.~\eqref{eq:langmuir_probe_likelihood} are then propagated through the independently specified $F_A$ and $D_{AB}$, or supplied by an independent time-resolved reconstruction, while the same probe functional is retained.
Under these constraints, the model predicts
\begin{align}
p_{\mathrm{LP},s}(y\mid U),
\ 
\kappa_{2,s}^{\mathrm{LP}}(U),
\ 
\kappa_{3,s}^{\mathrm{LP}}(U),
\ 
\kappa_{4,s}^{\mathrm{LP}}(U),
\ 
C_{ij,s}^{\mathrm{LP}}.
\label{eq:probe_new_prediction_targets}
\end{align}
A model-specific residual-shape parameter may not be introduced after the mean response and coordinate dynamics have been fixed.
The strongest validation target is therefore not another reproduction of $I_e(U)$, but the prediction of independently measured non-Gaussian current statistics from a source-supported kinetic state with independently specified dynamics.
The Kube benchmark of Sections~\ref{sec:validation} and \ref{sec:kube_results} reaches an intermediate held-out level: the kinetic-response parameters are frozen before each target PDF is scored, while the target source and noise controls are fixed to their published values rather than independently reconstructed.
If only an averaged characteristic is available, Eq.~\eqref{eq:probe_first_moment_mean_current} can be tested but Eqs.~\eqref{eq:probe_acquisition_cumulants} and~\eqref{eq:probe_cross_voltage_covariance} cannot be validated experimentally.

\section{Validation}
\label{sec:validation}

Two validation questions are distinct.
The first is whether a kinetic manifold, propagated through the fixed diagnostic map, predicts an experimental measurement PDF that was not used to determine its response parameters.
The second is whether the resulting posterior closure tracks a time-dependent state recursively.
The present Alcator C-Mod data test the first question; the second requires chronological measurements that are not available in the published histograms.
\\
The first validation question is implemented by the experimental benchmark of Section~\ref{sec:kube_results}, using the previously published diffusionless INMDF inversion of the Langmuir-probe mean characteristic as its established physical baseline.
The analytic \(J_k\) reduction and the replacement of the empirical diffusion contribution are therefore not repeated as new results.
Instead, the benchmark compares six velocity-space manifolds under one diagnostic construction: the MDF, first-INMDF~\cite{Izacard_2016,Izacard_2017}, double-INMDF, Kappa~\cite{Vasyliunas_1968,SummersThorne_1991,PierrardLazar_2010,LazarFichtnerYoon_2016}, Tsallis, and a two-Maxwellian population.
The Maxwellian response with empirical diffusion is retained separately only as the historical mean-current comparator.
At the available level, each published histogram is treated as one complete experimental condition.
For each held-out condition, the source-to-kinetic response parameters are fitted from the remaining conditions and frozen before the target current PDF is evaluated.
The held-out source density and background-noise controls remain fixed to the published Kube values and are not re-optimized for any kinetic manifold.
We perform both a universal leave-one-condition-out test using all other conditions and a within-region test using only the other midplane or divertor conditions.
The present benchmark is therefore a held-out kinetic-response prediction conditional on published source controls, not yet a prediction from independently measured source statistics.
The present validation has two complementary outputs: predictive agreement in diagnostic space and physical admissibility in velocity space.
The first is quantified from the held-out measurement PDFs and their higher moments without re-fitting the target kinetic response.
The second requires the inferred kinetic state to remain inside its positivity-admissible domain and to be consistent with the assumed physical source and response structure.
A stronger future experiment would additionally reconstruct the source independently and use raw time traces or repeated sweeps with disjoint calibration and prediction sets.
The diagnostic calibration, mean kinetic coordinates, projected drift \(F_A\), and unresolved diffusion \(D_{AB}\) would then be fixed before the prediction set is evaluated.
Such a test would require the same kinetic state to predict the held-out measurement PDF, cumulants, cross-voltage covariance, and predictive likelihood without introducing a new residual-shape parameter.
\\
Accordingly, the present validation addresses the measurement-likelihood, held-out cross-condition, and posterior-uncertainty questions available from the published data.
A chronologically resolved data set would add the separate temporal-recursion test without changing the kinetic-manifold, diagnostic-likelihood, or Bayesian uncertainty constructions derived here.
Section~\ref{sec:kube_results} now implements the available held-out prediction test.

\section{Experimental representation and held-out prediction}
\label{sec:kube_results}

\subsection{Experimental benchmark and common forward model}
\label{sec:kube_benchmark_setup}

The experimental benchmark uses the ion-saturation-current statistics reported for the Alcator C-Mod scrape-off layer by Kube et al.~\cite{Kube_2016}.
The three horizontal-probe histograms of their Figs.~5--7 and the four divertor-probe histograms of their Figs.~12--13 provide seven distinct combinations of probe position, discharge, and intermittency level.
The analysis has two levels.
First, a joint same-condition calculation measures the representation capacity of each kinetic manifold when its permitted condition-dependent coordinates are calibrated using the seven PDFs.
Second, leave-one-condition-out calculations test whether a kinetic-response map calibrated from other plasma conditions predicts a held-out current PDF without re-fitting its kinetic coordinates to that PDF.
At both levels, the source statistics, background-noise model, normalization, and ion-saturation-current response are identical across candidate families.
The published source-shape and background-noise parameters are treated as fixed condition controls rather than as independently measured source quantities.
Because the published histograms do not retain their original time ordering, the benchmark tests kinetic-manifold discrimination and cross-condition prediction of the measurement likelihood rather than temporal recursive tracking.
\\
For each condition, let \(z\) denote the retained scalar source coordinate with fixed density \(P_z(z)\), and let
\begin{equation}
I_s(z)
\doteq
\mathcal H_{I_{\rm sat}}
\left[
f_s(\mathbf v\mid z)
\right]
\label{eq:kube_results_current_pushforward}
\end{equation}
be the current generated by kinetic family \(s\) through the same ion-saturation-current functional.
The measurement density before the fixed background-noise operation is the push-forward
\begin{equation}
P_s(I)
=
\int_{-\infty}^{\infty}
P_z(z)
\,
\delta
\left[
I-I_s(z)
\right]
\,dz.
\label{eq:kube_results_pdf_pushforward}
\end{equation}
The same normalization and the same Kube background-noise operation are then applied to every candidate.
The six analytic velocity-space families compared here are the MDF, the first-INMDF, the double-INMDF, Kappa, Tsallis, and a two-Maxwellian population.
Their normalized parallel distributions, source-response maps, and analytic half-space currents used in the numerical benchmark are given explicitly in Appendix~\ref{app:benchmark_families}.
The double-INMDF is
\begin{equation}
f_{I_2}
\doteq
f_M
+
\delta f_{I,1}
+
\delta f_{I,2},
\label{eq:double_inmdf_results_definition}
\end{equation}
where the two localized corrections have independent amplitudes, centers, widths, and source responses subject to positivity.
In the joint seven-condition fit, each channel width and source-response coefficient is shared globally, whereas its amplitude coordinate and support center are allowed to vary with experimental condition.
No sign restriction is imposed on either channel.
The first-INMDF branch is recovered continuously when the second correction vanishes.
The double-INMDF therefore tests a specific physical question: whether one localized kinetic correction is sufficient, or whether the diagnostic statistics require a second independently located channel.
\\
The digitized histograms span several decades in probability, so a vertical error at fixed \(\widetilde I\) can strongly overweight a point on a steep branch even when that point lies geometrically close to the predicted curve.
For the quantitative comparison, define
\begin{align}
x_i
&\doteq
\widetilde I_i,
\\
y_i
&\doteq
\log_{10}
P_{\rm obs}(x_i),
\\
g_s(x)
&\doteq
\log_{10}
P_s(x).
\label{eq:kube_results_log_coordinates}
\end{align}
The local normal-distance residual used in the numerical optimization is
\begin{equation}
d_{\perp,s,i}
\doteq
\frac{
g_s(x_i)-y_i
}{
\sqrt{
1+
\left[
g_s'(x_i)
\right]^2
}
},
\label{eq:kube_results_perpendicular_residual}
\end{equation}
where $i\in\{2,\ldots,N-1\}$, which is the local-tangent approximation to the shortest distance in the \((\widetilde I,\log_{10}P)\) plane.
The first and last digitized points remain visible in all figures but are excluded from the error evaluation because they are isolated extreme-tail points and can otherwise dominate an error measure while carrying negligible probability.
The total comparison metric is
\begin{equation}
\mathcal E_{\perp,s}
\doteq
\left[
\frac{1}{N-2}
\sum_{i=2}^{N-1}
d_{\perp,s,i}^2
\right]^{1/2}.
\label{eq:kube_results_error_metric}
\end{equation}
This score is used as a geometric goodness-of-fit measure rather than as a likelihood.
\begin{figure*}[t]
\centering
\includegraphics[width=0.8\textwidth]{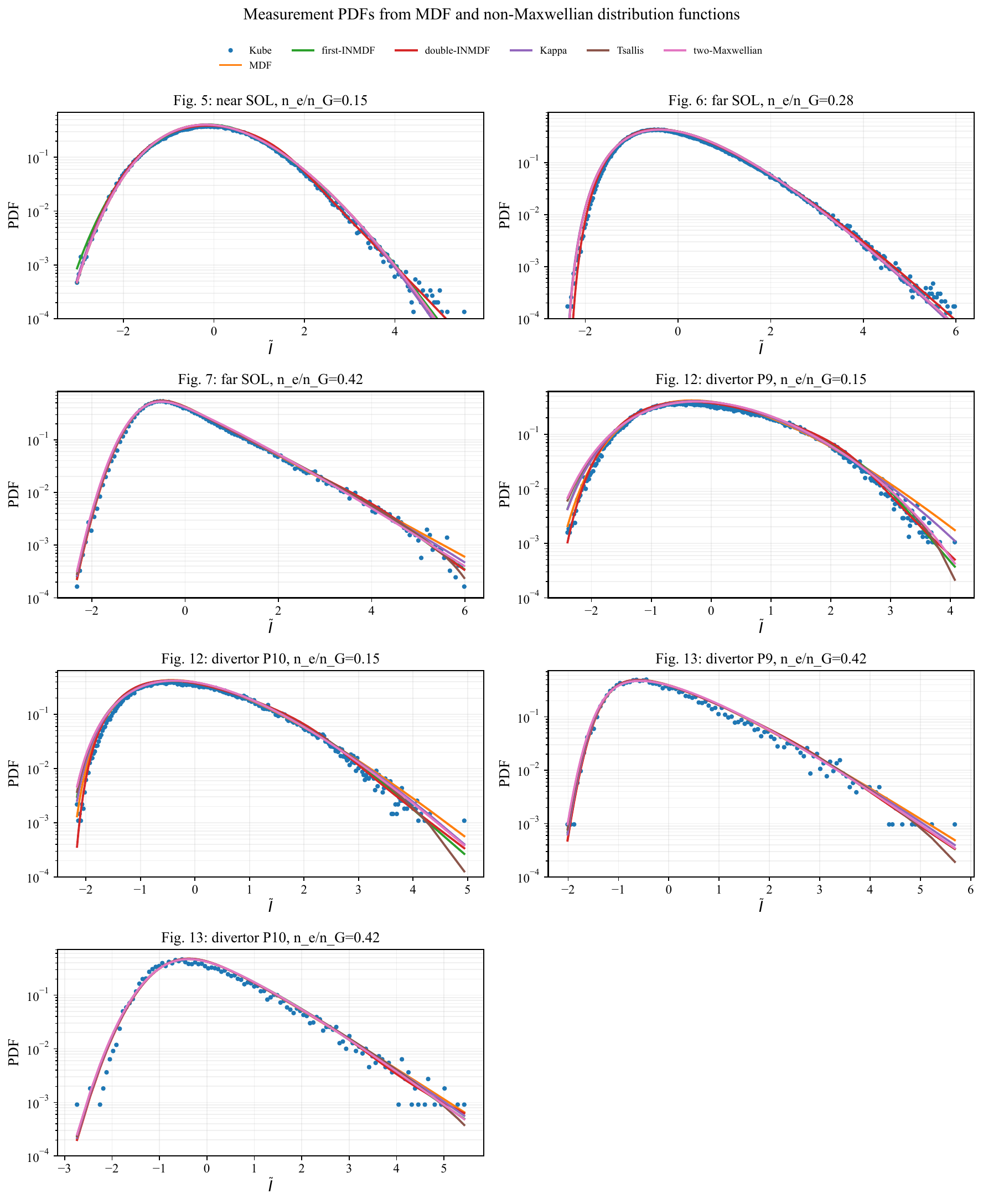}
\caption{
Measured Kube ion-saturation-current PDFs and the corresponding same-condition forward reconstructions obtained from six velocity-space families under the same fixed source statistics, background-noise model, normalization, and diagnostic response.
The seven panels cover the near-SOL and far-SOL horizontal-probe cases and the two divertor probes at low and high line-averaged density.
The curves are not independent empirical residual fits: changing the family changes the velocity-space distribution entering the same diagnostic map.
The kinetic coordinates shown here are calibrated using these experimental PDFs; held-out predictions are tested separately below.
}
\label{fig:kube_nmdf_pdf_effects}
\end{figure*}
Figure~\ref{fig:kube_nmdf_pdf_effects} shows the same-condition representation stage.
The families are similar near the PDF maximum but separate in the rising branch, shoulder, and positive tail.
The second INMDF channel produces its largest changes when one localized deformation cannot follow both sides of the measured PDF, while other conditions gain little from the additional channel.
This stage therefore measures representation capacity, not predictive preference.
\begin{table*}[t]
\caption{
Edge-trimmed local-normal PDF error \(\mathcal E_{\perp,s}\) defined by Eq.~\eqref{eq:kube_results_error_metric}.
The smallest value in each row is shown in bold.
For Fig.~13 P9, the first- and double-INMDF values differ by less than \(10^{-4}\) and are numerically indistinguishable at the precision relevant here.
The final row is the unweighted mean over the seven experimental conditions.
}
\label{tab:kube_nmdf_pdf_errors}
\centering
\begin{ruledtabular}
\begin{tabular}{lcccccc}
Condition & MDF & first-INMDF & double-INMDF & Kappa & Tsallis & two-Maxwellian \\
\hline
Fig.~5 near SOL & 0.0788 & 0.0752 & \(\mathbf{0.0601}\) & 0.0843 & 0.0793 & 0.0789 \\
Fig.~6 far SOL & 0.0605 & 0.0601 & \(\mathbf{0.0483}\) & 0.0666 & 0.0608 & 0.0605 \\
Fig.~7 far SOL & 0.0855 & 0.0644 & 0.0642 & 0.0727 & \(\mathbf{0.0637}\) & 0.0710 \\
Fig.~12 P9 & 0.1187 & 0.0656 & \(\mathbf{0.0514}\) & 0.0980 & 0.0779 & 0.0837 \\
Fig.~12 P10 & 0.0884 & 0.0704 & \(\mathbf{0.0617}\) & 0.0810 & 0.0763 & 0.0826 \\
Fig.~13 P9 & 0.1049 & \(\mathbf{0.0995}\) & \(\mathbf{0.0995}\) & 0.1021 & 0.1016 & 0.1008 \\
Fig.~13 P10 & 0.1237 & 0.1196 & \(\mathbf{0.1157}\) & 0.1223 & 0.1233 & 0.1209 \\
\hline
Mean & 0.0943 & 0.0792 & \(\mathbf{0.0715}\) & 0.0896 & 0.0833 & 0.0855
\end{tabular}
\end{ruledtabular}
\end{table*}
Table~\ref{tab:kube_nmdf_pdf_errors} quantifies this representation stage.
The double-INMDF has the smallest \(\mathcal E_\perp\) in five conditions and is tied with the first-INMDF in Fig.~13 P9; in Fig.~7, Tsallis is smaller by less than one percent.
The mean error decreases from \(0.0943\) for the MDF to \(0.0792\) for the first-INMDF and \(0.0715\) for the double-INMDF.
The double-INMDF value is therefore approximately \(24\%\) below the MDF and \(10\%\) below the first-INMDF.
The stronger question is whether these gains survive when the target PDF is withheld, which is tested in Section~\ref{sec:kube_loo_prediction}.
\\
These numbers should not be interpreted as a complexity-penalized proof that the double-INMDF is universally preferred.
The parameter count relevant to the present comparison is the dimension of the joint optimization over the \(N_c=7\) experimental conditions, rather than the number of coordinates required to describe one local realization of a kinetic manifold.
For the first-INMDF, the width and source-response coefficient are shared globally, while the amplitude coordinate and support center are condition dependent.
Its fitted dimension is therefore
\begin{equation}
k_{I_1}
=
2+2N_c
=
16.
\label{eq:first_inmdf_fit_dimension}
\end{equation}
For the double-INMDF, each localized channel contains one globally shared width and source-response coefficient together with one condition-dependent amplitude coordinate and support center, giving
\begin{equation}
k_{I_2}
=
4+4N_c
=
32.
\label{eq:double_inmdf_fit_dimension}
\end{equation}
Similarly,
\begin{align}
k_{\kappa}
&=
k_{\mathrm{Tsallis}}
=
1+N_c
=
8,
\\
k_{2M}
&=
1+2N_c
=
15,
\label{eq:other_fit_dimensions}
\end{align}
while the standardized MDF prediction contains no fitted kinetic parameter in the present comparison.
Thus the double-INMDF introduces additional flexibility, but its \(32\) fitted parameters describe a joint fit to seven distinct experimental conditions rather than a \(32\)-parameter fit to any single measured PDF.
The digitized PDF points are samples of published probability-density curves rather than independent raw counts, so a conventional information criterion is not assigned to this same-condition stage.
The parameter counts above therefore quantify representation flexibility only.
Predictive complexity is evaluated separately by the leave-one-condition-out errors, fitted response dimensions, and Pareto ranking in Section~\ref{sec:kube_loo_prediction}.
The skewness and excess kurtosis remain complementary shape diagnostics derived from the same experimental PDFs rather than independent observations.
This establishes the practical relevance of the hierarchy
\begin{equation}
f_M,
\quad
f_M+\delta f_{I,1},
\quad
f_M+\delta f_{I,1}+\delta f_{I,2},
\label{eq:kube_results_inmdf_hierarchy}
\end{equation}
without implying that the hierarchy should be extended indefinitely.
\begin{figure*}[t]
\centering
\includegraphics[width=0.8\textwidth]{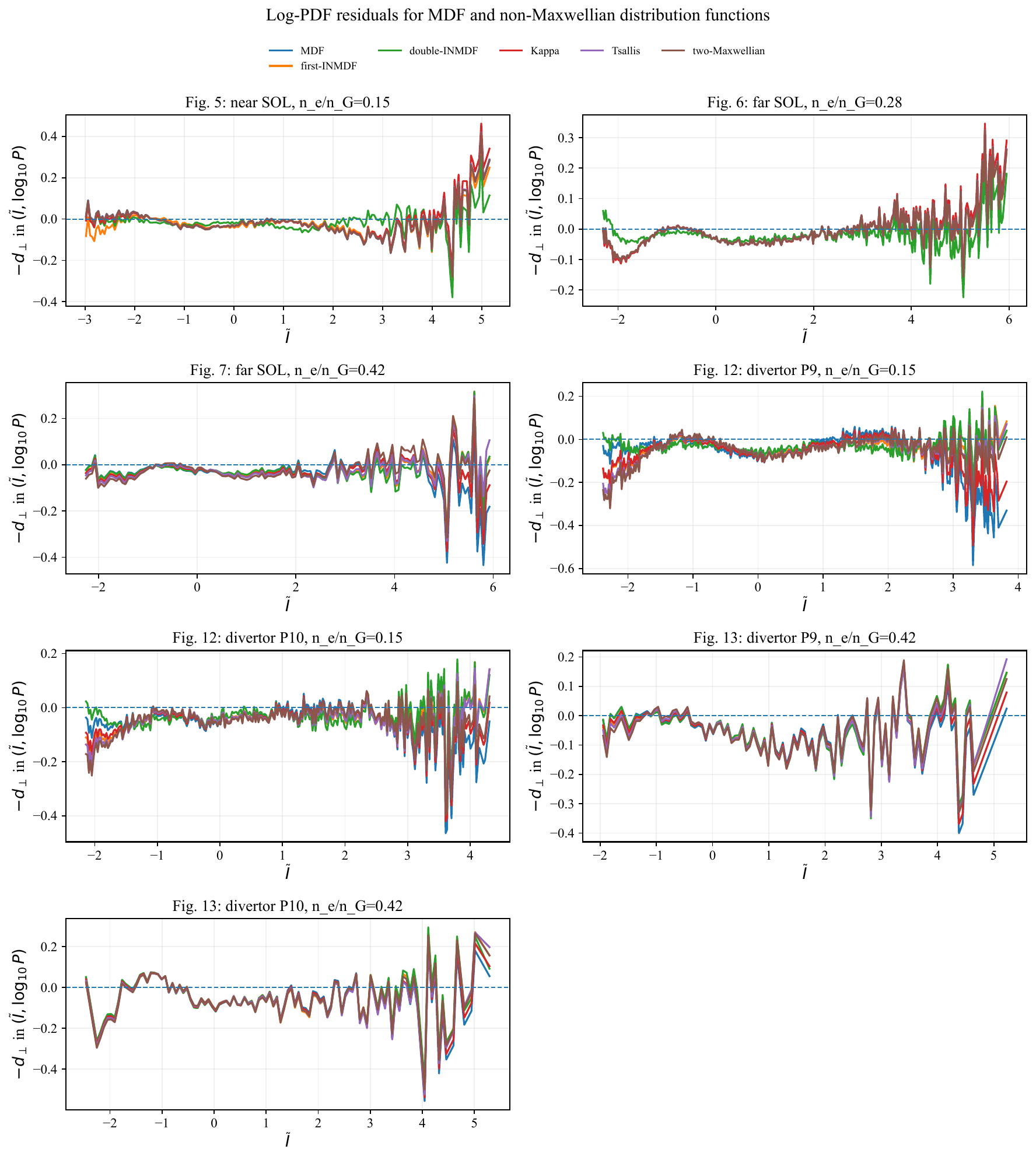}
\caption{
Local normal-distance residuals in the \((\widetilde I,\log_{10}P)\) plane for the six analytic kinetic families.
The first and last digitized points are excluded from the quantitative score but remain visible in the corresponding PDF figure.
The residual structure shows where an analytic family fails geometrically rather than only vertically at fixed \(\widetilde I\).
}
\label{fig:kube_nmdf_pdf_residuals}
\end{figure*}
Figure~\ref{fig:kube_nmdf_pdf_residuals} shows where the scalar errors originate.
The second INMDF channel removes broad residual structure in Fig.~12 P9 and across extended intervals in Figs.~5 and 7, rather than improving only an isolated tail point.
By contrast, the Fig.~13 cases leave comparable residual structure among several families.
The same-condition result therefore supports condition-dependent kinetic structure rather than one universal analytic distribution.
\\
The analytic-family comparison does not by itself show how much velocity-space flexibility a single diagnostic can support.
To provide that reference without assigning another analytic kinetic family, we also construct a numerical minimum-deformation distribution.
The parallel distribution is represented as
\begin{equation}
f_\star(v_\parallel)
=
\frac{
f_M(v_\parallel)
\exp
\left[
g_\star(v_\parallel)
\right]
}{
\displaystyle
\int_{-\infty}^{\infty}
f_M(v_\parallel')
\exp
\left[
g_\star(v_\parallel')
\right]
\,dv_\parallel'
},
\label{eq:kube_results_numerical_distribution}
\end{equation}
with \(g_\star\) represented by a smooth cubic deformation so that \(f_\star>0\) and \(f_\star\in C^2\).
The reconstruction minimizes departure from the MDF together with a curvature penalty while satisfying the diagnostic half-space flux constraint.
Because one ion-saturation-current functional constrains only a limited projection of velocity space, \(f_\star\) is not interpreted as the unique experimental distribution function.
It is a minimum-deformation parallel reference consistent with the selected diagnostic constraints and is used only to illustrate the velocity-space nonuniqueness of the diagnostic inversion; it does not enter the analytic-family parameter ranking or the leave-one-condition-out model comparison.
\\
With this numerical reference defined, the higher moments provide a complementary geometric diagnostic of the analytic-family comparison.
Here the scalar symbol \(S_s\) denotes skewness and is distinct from the Kalman innovation-covariance matrix \(S_{\mu\nu,k}\) defined in Eq.~\eqref{eq:discrete_innovation_covariance}.
For each model, the full predicted standardized current density gives
\begin{align}
S_s
\doteq{}&
\int_{-\infty}^{\infty}
\widetilde I^3
P_s(\widetilde I)
\,d\widetilde I,
\\
F_s
\doteq{}&
\int_{-\infty}^{\infty}
\widetilde I^4
P_s(\widetilde I)
\,d\widetilde I
-
3,
\label{eq:kube_results_sk_definition}
\end{align}
after the distribution has been normalized to zero mean and unit variance.
Thus every distribution entering the \(S\)-\(F\) comparison satisfies
\begin{equation}
\int_{-\infty}^{\infty}
\widetilde I
P_s(\widetilde I)
\,d\widetilde I
=
0,
\qquad
\int_{-\infty}^{\infty}
\widetilde I^2
P_s(\widetilde I)
\,d\widetilde I
=
1.
\label{eq:kube_results_standardized_first_two_moments}
\end{equation}
The separation among the measured and predicted distributions in the \(S\)-\(F\) plane therefore occurs entirely through statistical structure that remains after their first two moments have been made identical.
This discrimination is consequently invisible to a Gaussian measurement representation characterized only by its mean and covariance.
These moments were not used to define the analytic PDF-distance metric in Eq.~\eqref{eq:kube_results_error_metric}.
They therefore provide a complementary test of the tails and of the global asymmetry.
\begin{figure*}[t]
\centering
\includegraphics[width=0.98\textwidth]{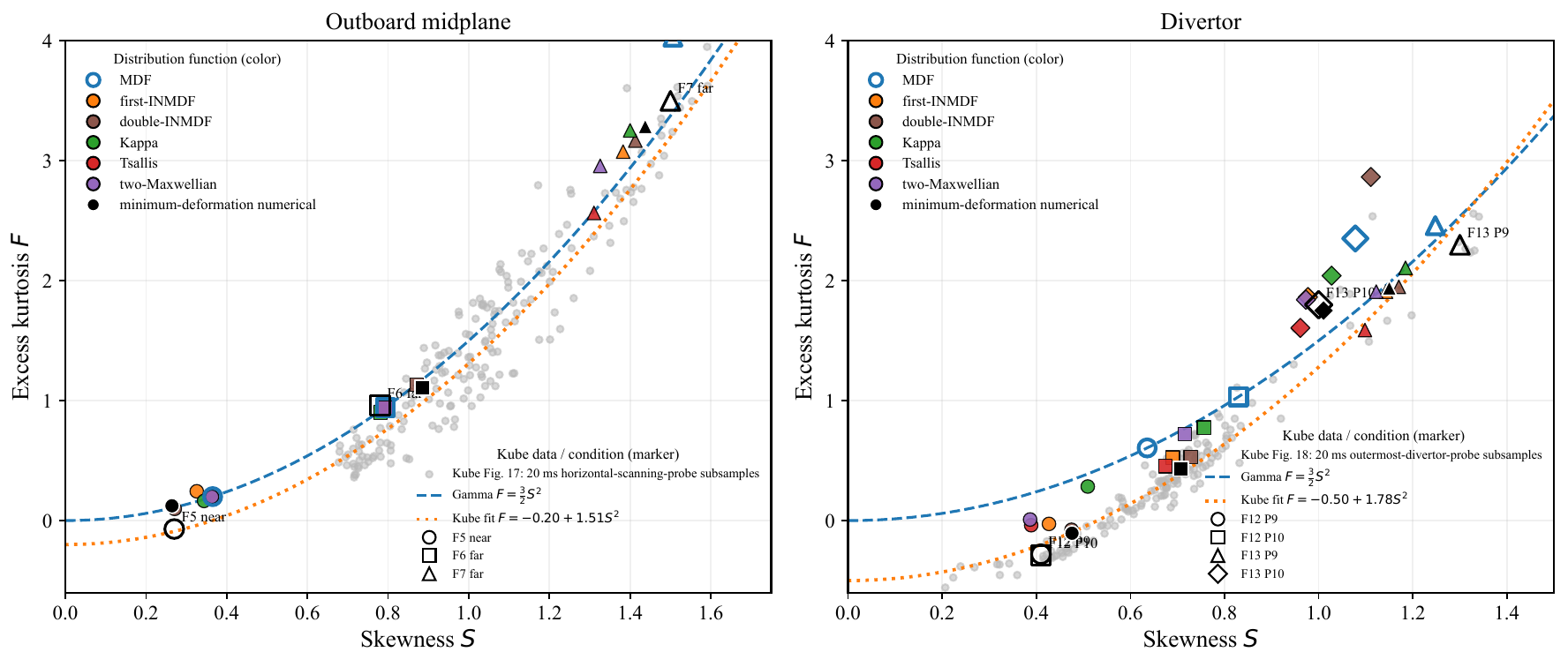}
\caption{
Skewness \(S\) and excess kurtosis \(F\) for the measured Kube data and for the kinetic-family predictions.
Gray points are the 20-ms subsample clouds reported by Kube et al.; open black condition markers denote the published whole-series values.
The dashed curve is the pure Gamma relation \(F=3S^2/2\), and the dotted curves are the empirical Kube fits for the midplane and divertor data.
The same kinetic family does not remain closest to the experimental point in every probe and regime.
}
\label{fig:kube_sk_map}
\end{figure*}
\begin{figure*}[t]
\centering
\includegraphics[width=0.98\textwidth]{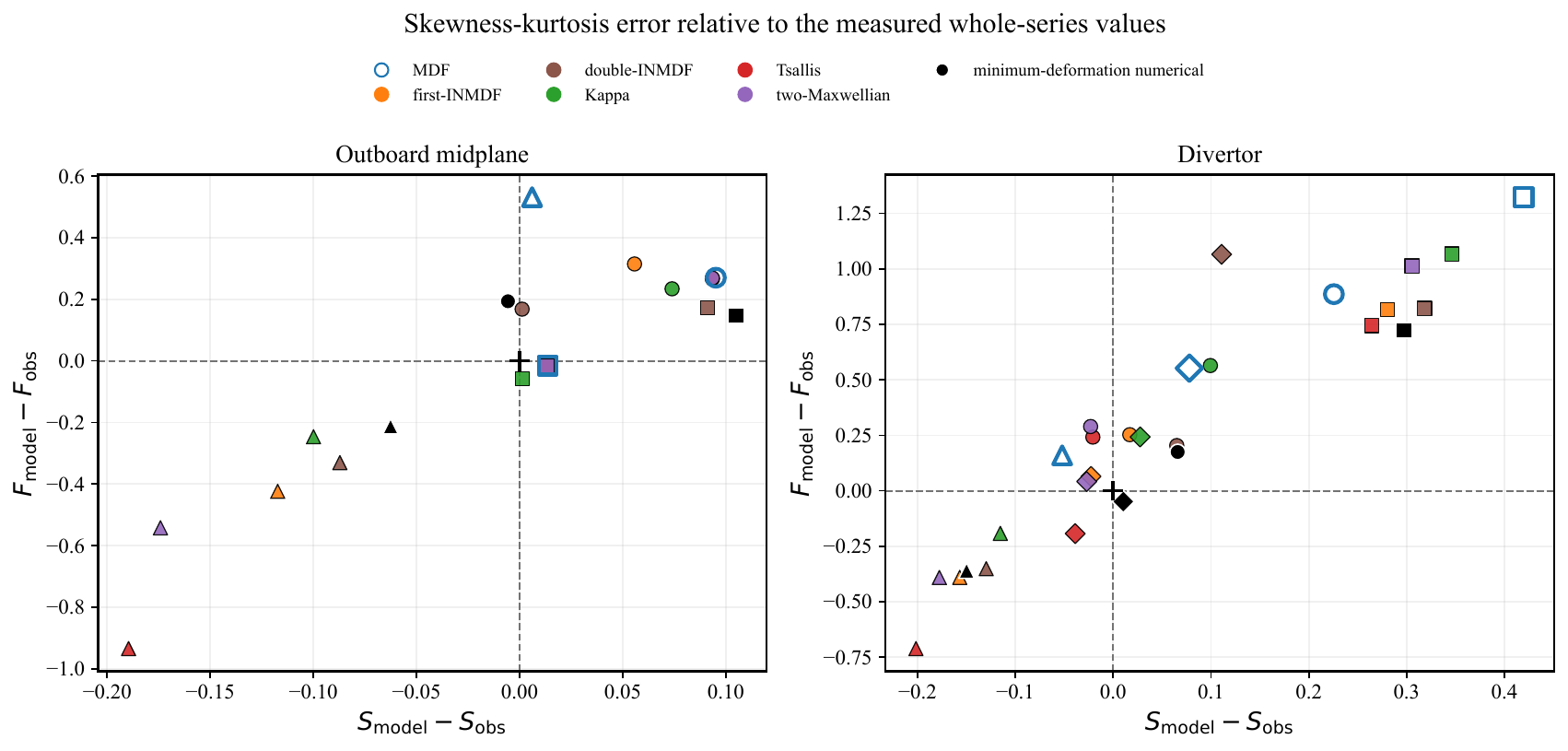}
\caption{
Residual form of Fig.~\ref{fig:kube_sk_map}.
Each point is displaced by the corresponding published whole-series value, so the origin represents exact agreement in both skewness and excess kurtosis.
The family ordering changes with diagnostic location and discharge condition.
}
\label{fig:kube_sk_residual_map}
\end{figure*}
Figures~\ref{fig:kube_sk_map} and \ref{fig:kube_sk_residual_map} show that the \(S\)-\(F\) ranking is not identical to the PDF-distance ranking.
The agreement is strongest for Fig.~12 P9, where the double-INMDF improves both measures, whereas several other conditions favor different families in PDF geometry and higher moments.
This difference is expected because \(S\) and \(F\) weight the low-probability tails more strongly than the local PDF metric.
The combined comparison therefore provides complementary shape information but does not define one universal family ranking.
\\
This condition-dependent discrimination has a direct consequence for the observation likelihood used by the filter.
The diagnostic response is unchanged among the candidate families; therefore the condition-dependent changes of the measured probability density originate from the kinetic distribution entering that response.
A single universal non-Gaussian residual law cannot represent this behavior.
Instead, the physical likelihood remains the state-dependent diagnostic map defined in Eq.~\eqref{eq:event_resolved_pushforward}: \(p_{D,s}(y\mid\mathbf X_s)\) changes with the kinetic state while \(\mathcal H_D\) remains fixed.
The experimental PDF therefore tests the physical measurement likelihood generated by the kinetic manifold, while \(S\) and \(F\) provide complementary sensitivity to asymmetry and tails.
As shown in Section~\ref{sec:nmdf_uncertainty}, these higher-order statistics are not themselves state uncertainties.
Rather, they are observable signatures that the likelihood \(p_{D,s}\) contains structure beyond its variance.
Because the same likelihood enters the Bayesian correction, kinetic families that generate different \(S\)-\(F\) structure can also generate different corrected posterior uncertainties even when their measurement means and variances are similar.

\subsection{Held-out prediction with frozen kinetic-response maps}
\label{sec:kube_loo_prediction}

The same-condition comparison above establishes representation capacity, but does not by itself determine whether the inferred kinetic response generalizes to an experimental condition that was not used to determine its response parameters.
We therefore perform leave-one-condition-out prediction.
For kinetic family \(s\) and held-out condition \(j\), let \(\boldsymbol{\theta}_s\) denote the globally shared kinetic-response parameters.
They are determined from the remaining conditions through
\begin{equation}
\widehat{\boldsymbol{\theta}}_s^{(-j)}
\doteq
\underset{\boldsymbol{\theta}_s}{\operatorname{argmin}}
\sum_{\substack{i=1\\i\neq j}}^{N_c}
\sum_{\ell=2}^{N_i-1}
d_{\perp,s,i,\ell}^2,
\label{eq:kube_loo_training}
\end{equation}
subject to the physical admissibility conditions of the retained family.
For the INMDF branches this includes positivity over the retained source and velocity domain, and for every analytic family the single-branch push-forward additionally requires a positive source-to-current Jacobian over every source domain on which the frozen response is evaluated.
Equation~\eqref{eq:kube_loo_training} weights the retained digitized curve points directly, so its training weight follows the sampling of the published PDF curves.
It is therefore used as a common geometric response-calibration objective applied identically to all candidate families, not as a likelihood, Bayesian evidence, or equal-condition statistical weight.
No parameter in \(\widehat{\boldsymbol{\theta}}_s^{(-j)}\) is determined from the held-out current PDF.
The held-out measurement density is then generated as
\begin{equation}
P_{s,j}^{\mathrm{pred}}
(\widetilde I)
\doteq
P_s
\left(
\widetilde I
\mid
P_{z,j},
\epsilon_j,
\widehat{\boldsymbol{\theta}}_s^{(-j)}
\right),
\label{eq:kube_loo_prediction}
\end{equation}
where \(P_{z,j}\) is the fixed source density for condition \(j\) and \(\epsilon_j\) is its fixed published background-noise variance ratio.
The measured PDF of condition \(j\) enters only after this forward density has been generated, when Eq.~\eqref{eq:kube_results_error_metric} and the predicted higher moments are evaluated.
Because \(P_{z,j}\) and \(\epsilon_j\) are fixed from the published Kube stochastic model rather than from an independent source diagnostic, Eq.~\eqref{eq:kube_loo_prediction} is a held-out kinetic-response prediction conditional on published source controls.
Unlike the joint same-condition fit of Eqs.~\eqref{eq:first_inmdf_fit_dimension}--\eqref{eq:other_fit_dimensions}, the predictive response contains no condition-specific fitted kinetic coordinate.
The fitted dimensions in each leave-one-out training fold are
\begin{equation}
\left(
k_M,
k_{I_1},
k_{I_2},
k_{\kappa},
k_{\mathrm{Tsallis}},
k_{2M}
\right)
=
\left(
0,4,8,2,2,3
\right).
\label{eq:kube_loo_parameter_counts}
\end{equation}
Thus the double-INMDF remains the most flexible analytic candidate, but its predictive comparison uses eight shared parameters rather than the \(32\) parameters of the seven-condition representation fit.
\begin{figure*}[t]
\centering
\includegraphics[width=0.97\textwidth]{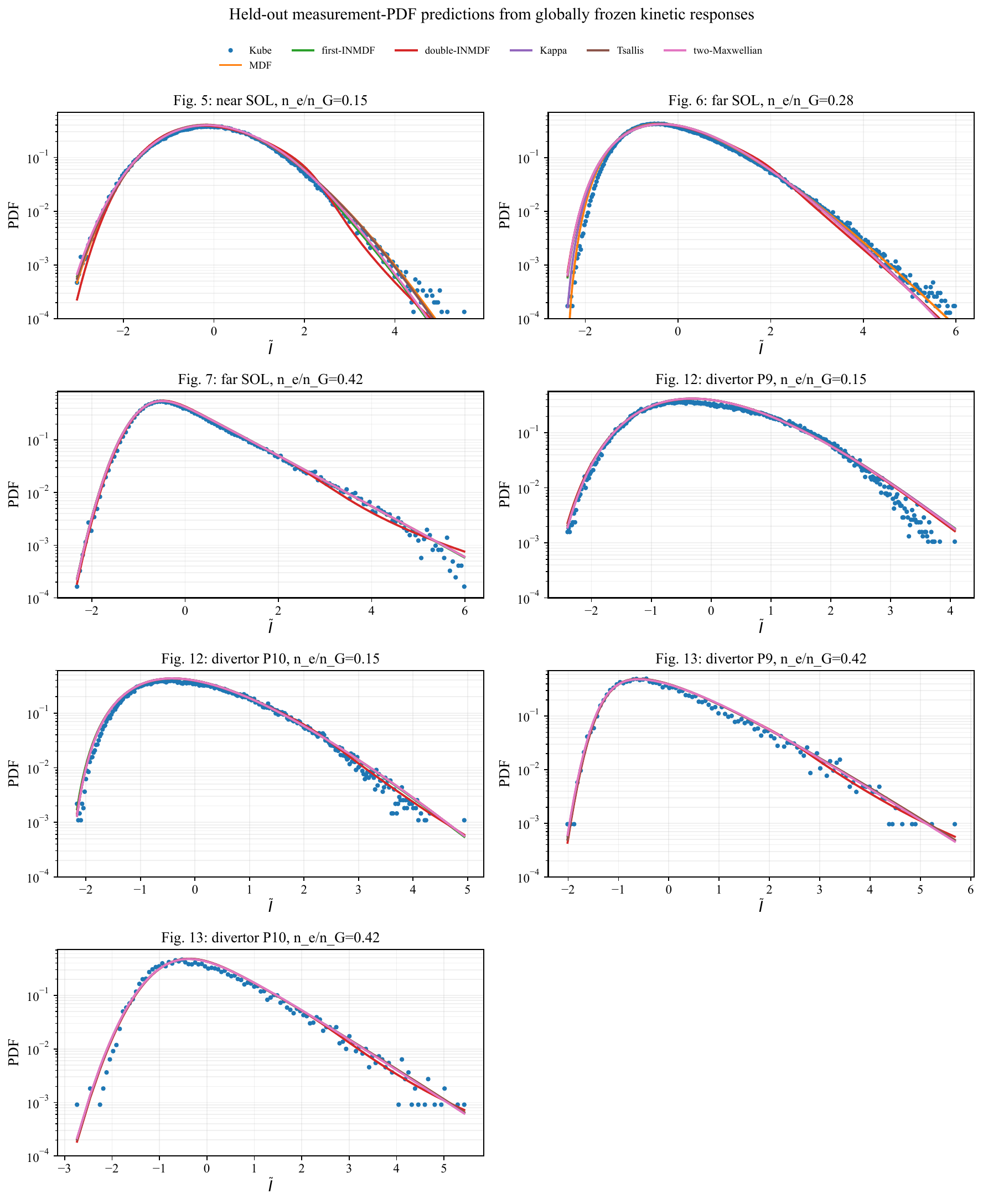}
\caption{
Universal leave-one-condition-out measurement-PDF prediction.
For each displayed condition, the kinetic-response parameters are fitted using the other six experimental conditions and then frozen before the displayed PDF is evaluated.
The source-density and background-noise controls of the held-out condition are fixed to the published values and are identical for every candidate family.
}
\label{fig:kube_loo_pdf_predictions}
\end{figure*}
\begin{figure*}[t]
\centering
\includegraphics[width=0.98\textwidth]{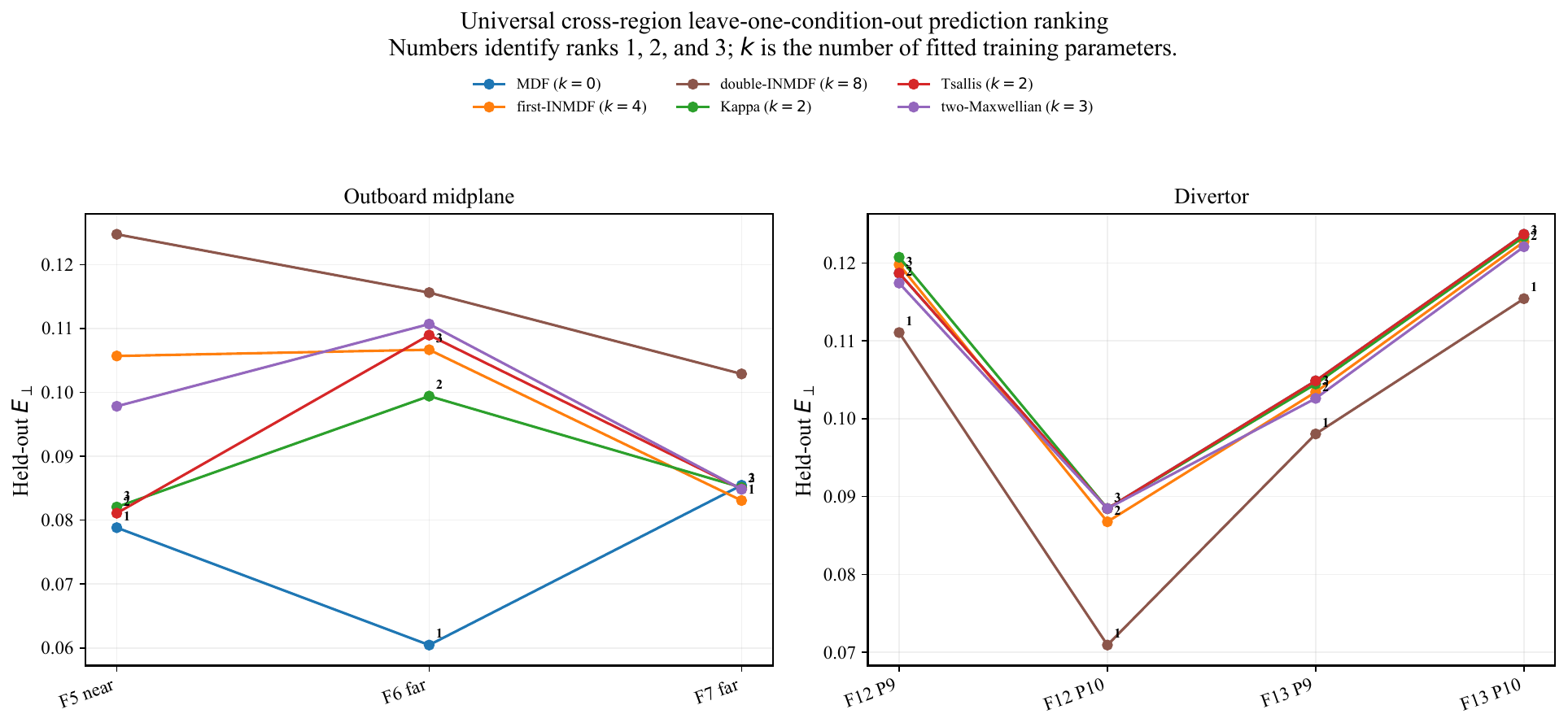}
\caption{
Ranking of the universal leave-one-condition-out predictions.
All six analytic kinetic families are shown.
The numbers attached to the curves identify the first-, second-, and third-ranked held-out PDF errors for each experimental condition, while \(k\) denotes the number of fitted response parameters in each training fold.
}
\label{fig:kube_loo_ranking}
\end{figure*}
The universal leave-one-out test deliberately asks whether one response law can span both the outboard-midplane and divertor measurements.
For the four divertor conditions, the double-INMDF ranks first in every held-out fold.
Its mean divertor error is
\begin{equation}
\left\langle
\mathcal E_{\perp,I_2}^{\mathrm{pred}}
\right\rangle_{\mathrm{div}}
=
0.0989,
\quad
\left\langle
\mathcal E_{\perp,M}^{\mathrm{pred}}
\right\rangle_{\mathrm{div}}
=
0.1089,
\label{eq:kube_loo_universal_divertor_mean}
\end{equation}
corresponding to a reduction of approximately \(9.2\%\) in the ratio of the two mean errors.
The double-INMDF prediction is also Pareto efficient with respect to held-out error and fitted parameter count in all four universal divertor folds.
The midplane result is different: the MDF has the lowest universal mean error, \(0.0749\), whereas the double-INMDF gives \(0.1144\).
A single universal response law therefore does not describe the two probe regions with equal efficiency.
\\
To test whether this difference originates from the physical diagnostic region rather than from a failure of the kinetic families themselves, a second leave-one-out calculation restricts the training set to the same probe class as the held-out condition.
For \(g\in\{\mathrm{mid},\mathrm{div}\}\),
\begin{equation}
\widehat{\boldsymbol{\theta}}_{s,g}^{(-j)}
\doteq
\underset{\boldsymbol{\theta}_s}{\operatorname{argmin}}
\sum_{\substack{i\in g\\i\neq j}}
\sum_{\ell=2}^{N_i-1}
d_{\perp,s,i,\ell}^2.
\label{eq:kube_group_loo_training}
\end{equation}
Each midplane prediction is therefore trained on the other two midplane conditions, while each divertor prediction is trained on the other three divertor conditions.
The held-out PDF remains excluded from the response fit.

\begin{figure*}[t]
\centering
\includegraphics[width=0.95\textwidth]{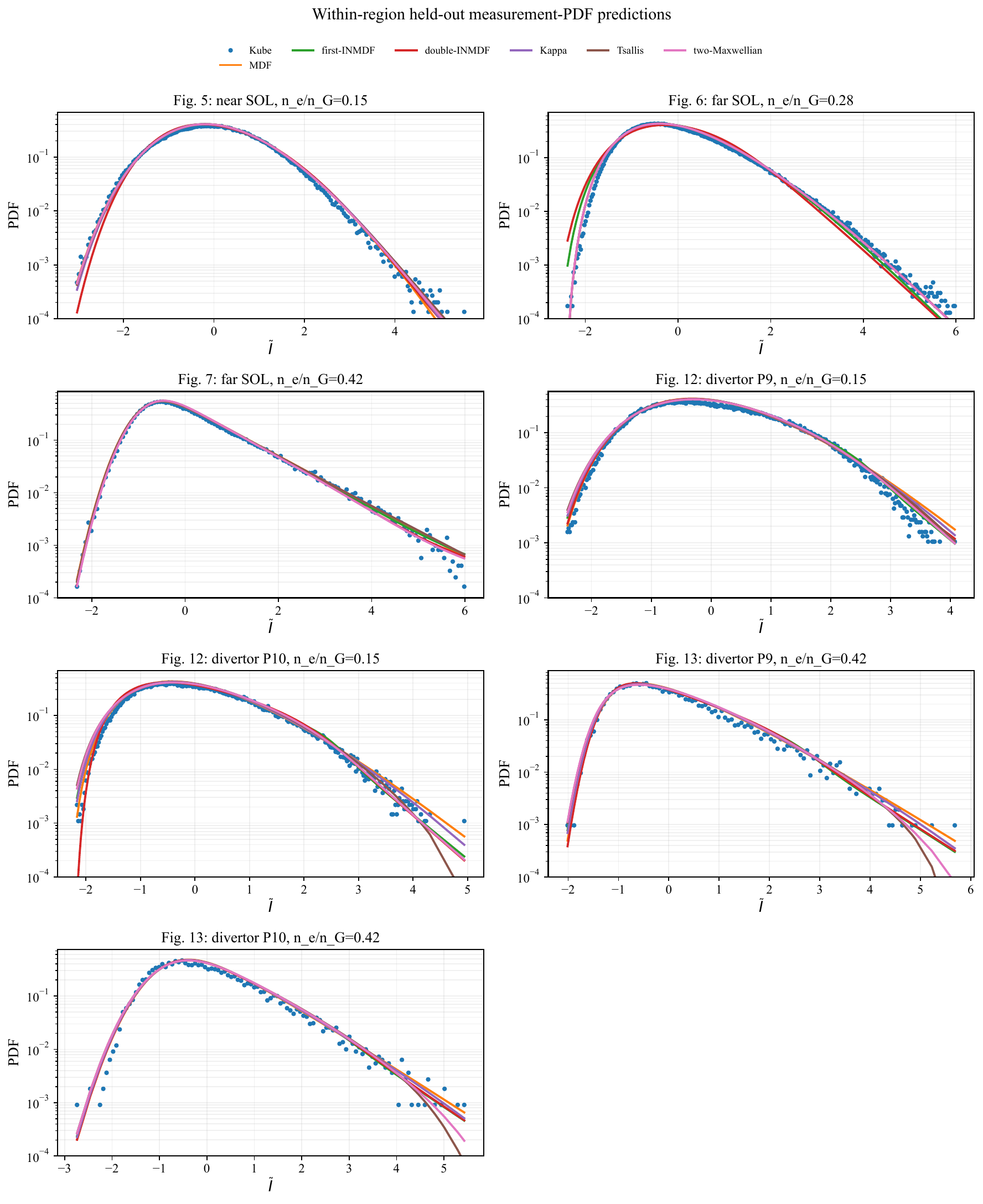}
\caption{
Within-region leave-one-condition-out PDF predictions.
A held-out midplane condition is predicted from the other two midplane conditions, and a held-out divertor condition is predicted from the other three divertor conditions.
No kinetic-response parameter is fitted to the displayed target PDF.
}
\label{fig:kube_group_loo_pdf_predictions}
\end{figure*}

\begin{figure*}[t]
\centering
\includegraphics[width=0.98\textwidth]{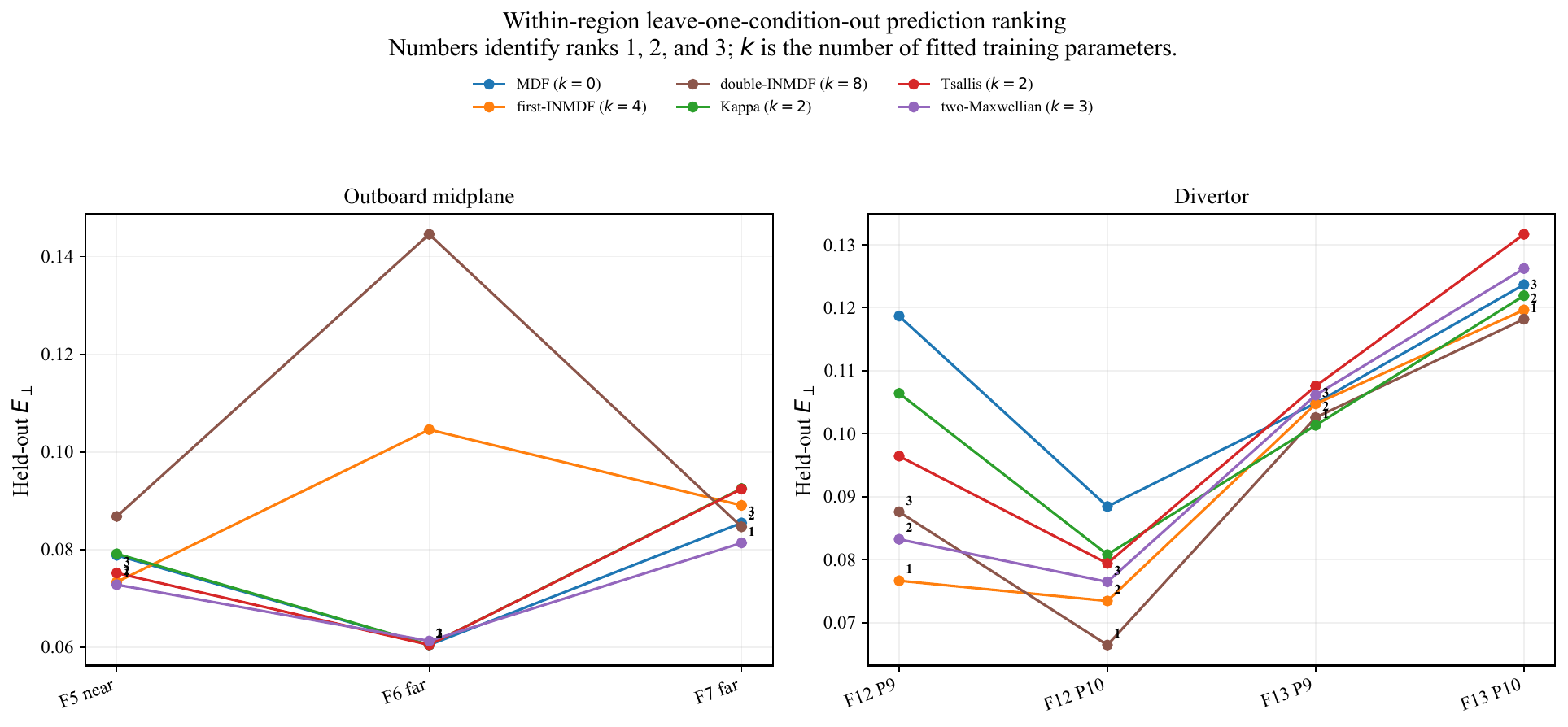}
\caption{
Within-region leave-one-condition-out ranking.
All analytic families are shown; numerical labels identify ranks one, two, and three for each held-out condition.
The same fitted parameter counts of Eq.~\eqref{eq:kube_loo_parameter_counts} apply.
}
\label{fig:kube_group_loo_ranking}
\end{figure*}

\begin{table*}[t]
\caption{
First three kinetic families ranked by held-out edge-trimmed local-normal PDF error.
The universal calculation trains on all other conditions, whereas the within-region calculation trains only on the other conditions belonging to the same midplane or divertor class.
Here \(I_1\) and \(I_2\) denote the first- and double-INMDF, \(q\) denotes Tsallis, and \(2M\) denotes the two-Maxwellian family.
}
\label{tab:kube_loo_top3}
\centering
\scriptsize
\begin{ruledtabular}
\begin{tabular}{lccc|ccc}
& \multicolumn{3}{c|}{Universal leave-one-out} &
\multicolumn{3}{c}{Within-region leave-one-out} \\
Condition & First & Second & Third & First & Second & Third \\
\hline
Fig.~5 near SOL
& MDF (0.0788) & \(q\) (0.0811) & \(\kappa\) (0.0821)
& \(2M\) (0.0728) & \(I_1\) (0.0734) & \(q\) (0.0752) \\
Fig.~6 far SOL
& MDF (0.0605) & \(\kappa\) (0.0994) & \(I_1\) (0.1067)
& MDF (0.0605) & \(q\) (0.0605) & \(\kappa\) (0.0605) \\
Fig.~7 far SOL
& \(I_1\) (0.0831) & \(2M\) (0.0848) & \(q\) (0.0849)
& \(2M\) (0.0814) & \(I_2\) (0.0847) & MDF (0.0855) \\
Fig.~12 P9
& \(I_2\) (0.1111) & \(2M\) (0.1174) & MDF (0.1187)
& \(I_1\) (0.0767) & \(2M\) (0.0833) & \(I_2\) (0.0876) \\
Fig.~12 P10
& \(I_2\) (0.0709) & \(I_1\) (0.0868) & MDF (0.0884)
& \(I_2\) (0.0665) & \(I_1\) (0.0735) & \(2M\) (0.0765) \\
Fig.~13 P9
& \(I_2\) (0.0980) & \(2M\) (0.1026) & \(I_1\) (0.1034)
& \(\kappa\) (0.1014) & \(I_2\) (0.1026) & \(I_1\) (0.1047) \\
Fig.~13 P10
& \(I_2\) (0.1154) & \(2M\) (0.1221) & \(I_1\) (0.1227)
& \(I_2\) (0.1182) & \(I_1\) (0.1197) & \(\kappa\) (0.1219)
\end{tabular}
\end{ruledtabular}
\end{table*}

The within-region result changes the interpretation of the analytic hierarchy.
For the divertor,
\begin{align}
\left\langle
\mathcal E_{\perp,I_1}^{\mathrm{pred}}
\right\rangle_{\mathrm{div}}
&=
0.0936,
\\
\left\langle
\mathcal E_{\perp,I_2}^{\mathrm{pred}}
\right\rangle_{\mathrm{div}}
&=
0.0937,
\label{eq:kube_group_loo_divertor_means}
\end{align}
so the first- and double-INMDF have nearly identical mean predictive error once the midplane and divertor response maps are separated.
The double-INMDF nevertheless has the better mean rank, \(1.75\) compared with \(2.00\) for the first-INMDF, and wins two of the four individual divertor folds.
The first-INMDF is top three in all four folds with only four fitted parameters and is Pareto efficient in three of four folds, compared with two of four for the eight-parameter double-INMDF.
The present divertor data therefore support localized INMDF structure strongly, while they do not require the second localized channel universally once region dependence is resolved.
The midplane ordering is different.
The two-Maxwellian gives the smallest mean within-region error,
\begin{equation}
\left\langle
\mathcal E_{\perp,2M}^{\mathrm{pred}}
\right\rangle_{\mathrm{mid}}
=
0.0718,
\quad
\left\langle
\mathcal E_{\perp,M}^{\mathrm{pred}}
\right\rangle_{\mathrm{mid}}
=
0.0749,
\label{eq:kube_group_loo_midplane_means}
\end{equation}
and wins two of the three midplane folds, while the MDF wins the third.
The first- and double-INMDF mean errors are \(0.0890\) and \(0.1053\), respectively.
Because each midplane prediction is calibrated from only two other conditions, this ranking should be interpreted as evidence for regional response structure rather than as a definitive selection among all possible midplane kinetic models.
\\
The comparison of Figs.~\ref{fig:kube_loo_ranking} and \ref{fig:kube_group_loo_ranking} separates two effects that cannot be distinguished by the same-condition fit alone.
The second INMDF channel provides the strongest generalization when one response law is forced to span heterogeneous midplane and divertor conditions, but much of this advantage disappears when the response is calibrated within the same physical region.
The predictive evidence therefore supports a region-dependent source-to-kinetic mapping and a hierarchy in which additional localized channels are retained when they improve held-out prediction, rather than a universal requirement for the highest-dimensional analytic manifold.
This region dependence motivates more general composite kinetic manifolds, but no such composite is fitted in the present benchmark.
We return to this implication in Section~\ref{sec:conclusion}.
\begin{figure*}[t]
\centering
\includegraphics[width=0.8\textwidth]{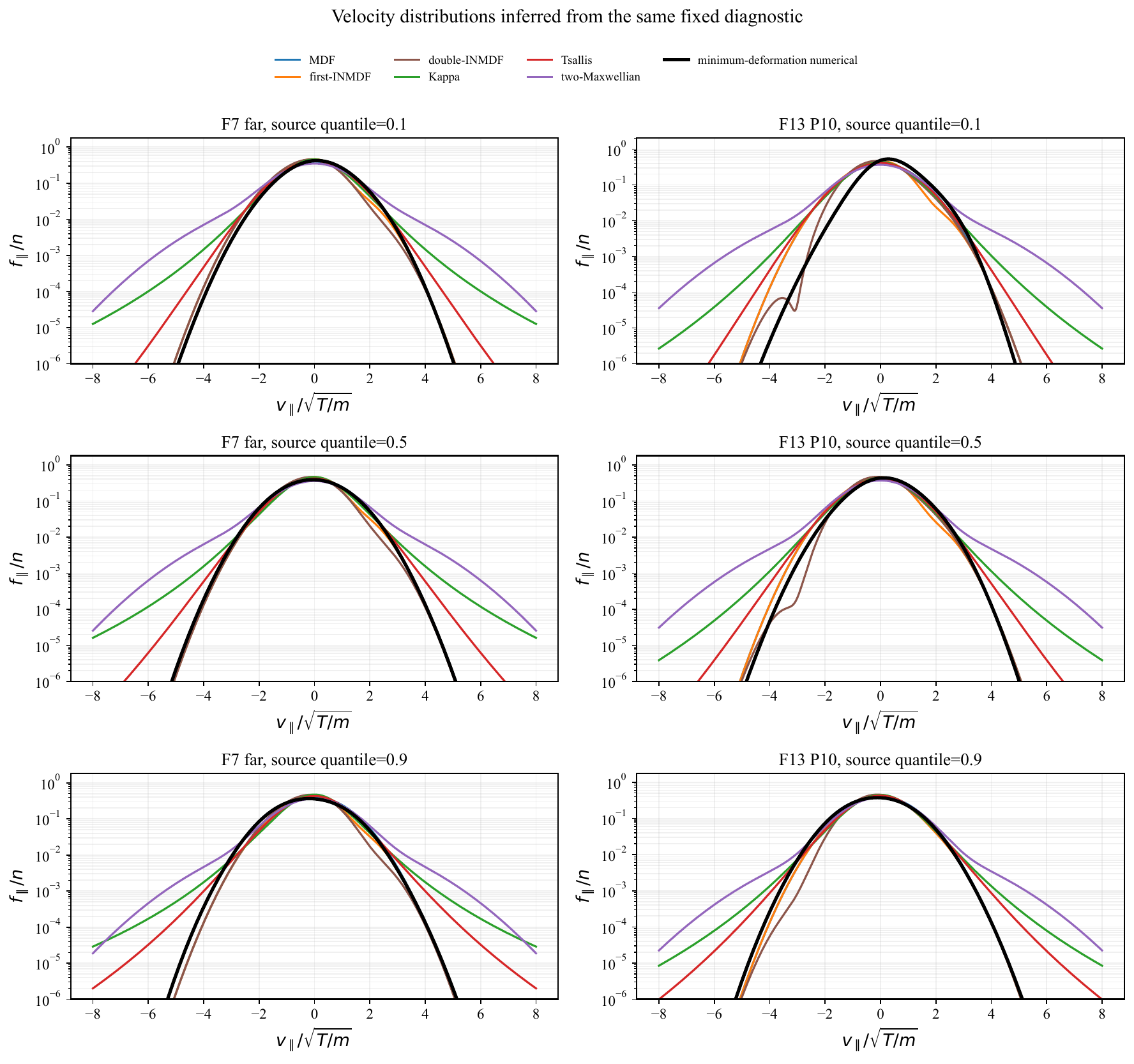}
\caption{
Representative parallel velocity distributions for the far-SOL high-density condition of Fig.~7 and divertor P10 of Fig.~13 at source quantiles \(0.1\), \(0.5\), and \(0.9\).
All analytic families and the numerical minimum-deformation reference are propagated through the same diagnostic construction.
The figure illustrates that similar measurement PDFs can correspond to substantially different localized or tail-dominated velocity-space structures.
}
\label{fig:kube_velocity_distribution_comparison}
\end{figure*}
The representative velocity distributions in Fig.~\ref{fig:kube_velocity_distribution_comparison} explain why the measurement-space family ordering changes with condition.
The diagnostic integrates selected sectors of the physical distribution rather than measuring an abstract skewness parameter, so source-dependent changes of those sectors alter the current through the same fixed \(\mathcal H_{I_{\rm sat}}\).
A localized INMDF channel can therefore be prominent in one regime and weak in another, while a broad Kappa or two-population deformation can dominate elsewhere.
The held-out results sharpen this interpretation: the additional localized INMDF coordinates improve cross-region prediction, but the lower-dimensional first-INMDF reaches essentially the same mean divertor accuracy once regional response structure is resolved.
This condition dependence also exposes a separate inverse problem: one diagnostic need not identify one velocity-space distribution.
More generally, a single diagnostic may admit several kinetically distinct distributions
\begin{equation}
\left\{
f^{(a)}(\mathbf v)
\right\}_{a=1}^{N_{\mathrm{sol}}}
\label{eq:multiple_admissible_kinetic_solutions}
\end{equation}
whose forward responses are indistinguishable within the uncertainty of that diagnostic.
Such solutions should not be collapsed into a unique reconstruction solely from the first diagnostic.
They can instead be compared through additional independent forward responses
\begin{equation}
p_{D_j}^{(a)}(y\mid\mathbf X^{(a)})
=
\mathcal H_{D_j}
\left[
K^{(a)}
(\mathbf v\mid\mathbf X^{(a)})
\right](y),
\label{eq:cross_diagnostic_likelihoods}
\end{equation}
with \(j\in\{1,\ldots,N_D\}\) where the same kinetic solution \(\mathbf X^{(a)}\) must be propagated through every diagnostic without diagnostic-specific retuning of its kinetic coordinates.
Cross-diagnostic consistency therefore provides an identifiability test: two distributions that are nearly degenerate under one probe can be distinguished when their predictions separate under a second diagnostic functional.
\\
When simultaneous diagnostic measurements are conditionally independent for a specified physical state, their joint likelihood entering the recursive correction is
\begin{equation}
p(\mathbf y\mid\mathbf X^{(a)})
=
\prod_{j=1}^{N_D}
p_{D_j}^{(a)}
(y_j\mid\mathbf X^{(a)}).
\label{eq:cross_diagnostic_joint_likelihood}
\end{equation}
For correlated diagnostic uncertainties, Eq.~\eqref{eq:cross_diagnostic_joint_likelihood} is replaced by the corresponding explicitly specified joint diagnostic probability law.
A physical kinetic reconstruction is therefore strengthened not by forcing uniqueness from one projection, but by requiring one admissible distribution to remain consistent across independent diagnostic maps.
For the filtering theory, the experimental result has two consequences.
First, the observation likelihood should not be assigned a universal non-Gaussian family independently of the kinetic state.
The same diagnostic location can move among MDF-like, localized-INMDF-like, broad-tail, and multi-population behavior as the plasma regime changes.
Second, the retained kinetic manifold itself can be treated as part of the state representation.
A practical recursive filter may therefore use a fixed multi-channel manifold with inactive coordinates when they are not needed, or a bank of admissible kinetic manifolds whose posterior weights are updated with the measurements.
The present results do not yet choose between those implementations.
They establish the prerequisite for either approach: the measured current statistics contain enough information to discriminate physically distinct velocity-space manifolds under a common diagnostic response and to test whether the associated response map generalizes to an unseen experimental condition.
The held-out comparison shows that a second localized INMDF channel is particularly effective when one response law is required to span heterogeneous probe regions, whereas within the divertor the first- and double-INMDF have nearly identical mean predictive errors and the midplane favors simpler MDF or two-population structure.
The retained kinetic hierarchy should therefore be selected from predictive physical evidence rather than increased automatically with model dimension.

\section{Discussion and conclusion}
\label{sec:conclusion}

The construction keeps three levels distinct: the kinetic state \(f_s\) and \(K_s\), the measurement likelihood \(p_{D,s}=\mathcal H_D[K_s]\), and the posterior \(\rho\) used to infer that state.
The projected kinetic dynamics evolve the physical coordinates, while the diagnostic map converts those coordinates into measurements.
When the posterior is Gaussian, the dynamics are affine, and the observation model is linear Gaussian, the construction reduces to the ordinary Kalman system.
\\
The continuous and discrete formulations share the same physical construction.
Continuous projection supplies the kinetic drift \(F_A\), unresolved diffusion \(D_{AB}\), and moment-projected posterior evolution.
Discrete acquisition uses the same physical likelihood \(p_{D,s}\), preserves the same retained moment map when the corrected moments remain admissible, and uses positivity-constrained KL projection when a finite Bayesian correction cannot be represented on that branch.
Both reduce to the ordinary Kalman equations in the Gaussian posterior and Gaussian-diffusion observation limit.
\\
The Alcator C-Mod benchmark then separates representation from prediction.
In the same-condition stage, the double-INMDF reduces the mean PDF error by approximately \(24\%\) relative to the MDF and \(10\%\) relative to the first-INMDF, but the different fitted dimensions make this a representation result rather than predictive model selection.
In the universal leave-one-condition-out test, the double-INMDF ranks first for all four divertor conditions and reduces the mean divertor error from \(0.1089\) to \(0.0989\).
After conditioning the response on probe region, however, the first- and double-INMDF give nearly identical mean divertor errors, \(0.0936\) and \(0.0937\), with four and eight fitted parameters, while the two-Maxwellian gives the smallest mean midplane error, \(0.0718\).
The second localized channel therefore earns its additional flexibility most clearly when one response must span heterogeneous plasma regions; it is not universally required once regional response structure is resolved.
Because the source and noise controls remain fixed to the published Kube stochastic-model values rather than independently reconstructed source measurements, this result is a held-out kinetic-response prediction conditional on published source controls rather than the final independently measured-source test.
\\
The region-dependent ranking also motivates composite kinetic manifolds.
A Kappa distribution supplies a broad algebraic-tail deformation, whereas an INMDF channel supplies a localized source-centered deformation; there is no requirement that one structure dominate every velocity sector.
One smooth positive example is
\begin{align}
f_{\rm comp}(v_\parallel)
\doteq{}&
\frac{1}{Z}
\exp
\left\{
w_-(v_\parallel)
\ln
f_\kappa(v_\parallel)
\right.
\nonumber\\
&\left.
+
\left[
1-w_-(v_\parallel)
\right]
\ln
f_{I_2}(v_\parallel)
\right\},
\label{eq:kube_results_composite_distribution}
\end{align}
with
\begin{equation}
w_-(v_\parallel)
\doteq
\frac{1}{2}
\left[
1-
\tanh
\left(
\frac{
v_\parallel-v_c
}{
\Delta v
}
\right)
\right].
\label{eq:kube_results_composite_weight}
\end{equation}
Here \(Z\) normalizes the distribution.
For \(v_\parallel\ll v_c\) the model approaches the Kappa sector, whereas for \(v_\parallel\gg v_c\) it approaches the double-INMDF sector.
The logarithmic interpolation preserves positivity and a smooth \(w_-\) preserves derivative continuity.
This composite is not fitted here; it is a concrete direction suggested by the regional ranking rather than an additional result.
\\
The same construction clarifies how the experimentally observed non-Gaussian measurement statistics enter uncertainty estimation.
The variance, skewness, and excess kurtosis of \(p_{D,s}\) characterize different aspects of the diagnostic likelihood and must not be identified with the posterior covariance of the inferred physical state.
Instead, the complete likelihood reweights \(\rho^{\mathrm{pred}}\) through Bayes' rule and therefore determines the corrected posterior \(\rho^{\mathrm{obs}}\), its covariance, and its higher-order structure.
In the linear-Gaussian limit this dependence reduces to \(\mathsf S=\mathsf H\mathsf P^{\mathrm{pred}}\mathsf H^T+\mathsf R\) and the ordinary Kalman covariance update, whereas outside that limit the likelihood shape contains information that cannot be represented by \(\mathsf R\) or \(\mathsf S\) alone.
The exact total-covariance identities further separate state-transition noise from propagated state uncertainty and conditional diagnostic variability from state uncertainty mapped into measurement space.
The complete prediction--measurement loop adds a second distinction.
The relevant sequence is: (i) a kinetic manifold predicts the complete diagnostic likelihood, (ii) the acquired measurement determines its realized innovation and predictive probability, (iii) a forced MDF that leaves diagnostically visible non-Maxwellian structure unresolved produces persistent excess innovation or surprise, and (iv) if an adaptive estimator assigns this unresolved structure to its residual channel, the resulting effective measurement noise and posterior uncertainty are artificially increased.
Appendix~\ref{app:model_induced_innovation} derives this mechanism and its Gaussian covariance limit.
The present Alcator C-Mod benchmark deliberately prevents this compensation by keeping the published source statistics and background-noise controls fixed across all kinetic families, so the unresolved MDF contribution remains visible as predictive error rather than being hidden inside a fitted noise level.
Once this model-induced contribution is controlled, the remaining uncertainty becomes physically interpretable.
It contains the robustness of the inferred state to the realized measurement together with the genuine diagnostic noise, propagated state uncertainty, unresolved dynamics, diagnostic observability, and possible transient or instability-driven evolution.
Its variation across plasma conditions is therefore not only a confidence measure: it can identify regimes in which the same diagnostic more or less constrains the retained kinetic state.
This interpretation connects directly to the \(S\)-\(F\) result.
Every PDF entering the Alcator C-Mod \(S\)-\(F\) analysis is standardized to zero mean and unit variance, so the observed family separation lies entirely in likelihood structure absent from a Gaussian covariance-only measurement model.
The NMDF therefore provides a physical generator of the higher-order likelihood structure that determines the Bayesian uncertainty update, while the condition dependence of the resulting uncertainty provides an additional measure of inference robustness, state observability, and plasma-regime dependence.
\\
The five-coordinate state-space INMDF provides one explicit non-Gaussian realization through its Gaussian core, localized channel, exact moment map, invertibility criterion, and admissible positivity domain.
The Alcator C-Mod benchmark shows that the corresponding kinetic response can be tested on held-out measurement PDFs while separating predictive benefit from model dimension and regional heterogeneity.
The present prediction remains conditional on the published source statistics, so an independently reconstructed source would provide a stronger future test of the complete source-to-kinetic-to-diagnostic chain.
Temporal recursive tracking is a separate extension requiring chronologically resolved measurements and is not required for the measurement-likelihood and uncertainty results established here.
Beyond plasma velocity space, the microscopic generator, retained physical manifold, diagnostic map, and posterior closure can be replaced by their system-specific counterparts while preserving the same sequence from physical dynamics to measurement likelihood, Bayesian correction, and state-uncertainty inference.

\appendix
\pdfbookmark[1]{Appendices}{appendices}
\section*{Appendices}
\addcontentsline{toc}{section}{Appendices}

\section{Finite-moment inversion and projected dynamics of the first state-space INMDF}
\label{app:projection}

For the coordinate ordering \(\mathbf U_I = (Y_G,P_G,\Gamma_X,c_X,W_X)\), direct differentiation of Eq.~\eqref{eq:state_inmdf_raw_moments} gives
\begin{widetext}
\begin{equation}
\mathsf J^{(I)}
=
\left(
\begin{array}{ccccc}
1
&
0
&
1
&
0
&
0
\\
2Y_G
&
1
&
2c_X
&
2\Gamma_X
&
0
\\
3(Y_G^2+P_G)
&
3Y_G
&
3(c_X^2+W_X)
&
6\Gamma_Xc_X
&
3\Gamma_X
\\
4Y_G^3+12Y_GP_G
&
6(Y_G^2+P_G)
&
4c_X^3+12c_XW_X
&
12\Gamma_X(c_X^2+W_X)
&
12\Gamma_Xc_X
\\
5Y_G^4+30Y_G^2P_G+15P_G^2
&
10Y_G^3+30Y_GP_G
&
5c_X^4+30c_X^2W_X+15W_X^2
&
20\Gamma_X(c_X^3+3c_XW_X)
&
30\Gamma_X(c_X^2+W_X)
\end{array}
\right).
\label{eq:explicit_state_inmdf_jacobian}
\end{equation}
To write the inverse compactly, introduce the unit-determinant row-operation matrix
\begin{equation}
\mathsf L
=
\left(
\begin{array}{ccccc}
1&0&0&0&0\\
-2Y_G&1&0&0&0\\
3Y_G^2-3P_G&-3Y_G&1&0&0\\
12P_GY_G-4Y_G^3&6Y_G^2-6P_G&-4Y_G&1&0\\
15P_G^2-30P_GY_G^2+5Y_G^4
&
30P_GY_G-10Y_G^3
&
10Y_G^2-10P_G
&
-5Y_G
&
1
\end{array}
\right),
\quad
\det\mathsf L
=
1.
\label{eq:moment_jacobian_row_transform}
\end{equation}
With $\Delta_c$ and $\Delta_P$ defined by Eq.~\eqref{eq:moment_map_differences},
\begin{equation}
\widetilde{\mathsf J}
\doteq
\mathsf L\mathsf J^{(I)}
=
\left(
\begin{array}{cc}
\mathsf I_2 & \mathsf B\\
\mathsf 0 & \mathsf S
\end{array}
\right),
\label{eq:block_moment_jacobian}
\quad\quad\quad
\mathsf B
=
\left(
\begin{array}{ccc}
1&0&0\\
2\Delta_c&2\Gamma_X&0
\end{array}
\right),
\end{equation}
and
\begin{equation}
\mathsf S
=
\left(
\begin{array}{ccc}
-3(\Delta_P-\Delta_c^2)
&
6\Delta_c\Gamma_X
&
3\Gamma_X
\\
-4\Delta_c(3\Delta_P-\Delta_c^2)
&
-12\Gamma_X(\Delta_P-\Delta_c^2)
&
12\Delta_c\Gamma_X
\\
5(3\Delta_P^2-6\Delta_P\Delta_c^2+\Delta_c^4)
&
-20\Delta_c\Gamma_X(3\Delta_P-\Delta_c^2)
&
-30\Gamma_X(\Delta_P-\Delta_c^2)
\end{array}
\right).
\label{eq:block_moment_jacobian_S}
\end{equation}
Its determinant is \(\det\mathsf S = -60\Gamma_X^2\mathcal D_J = \det\mathsf J^{(I)}\).
The inverse of the reduced block is
\begin{equation}
\mathsf S^{-1}
=
\frac{1}{\mathcal D_J}
\left(
\begin{array}{ccc}
-2(3\Delta_P^2+\Delta_c^4)
&
2\Delta_c^3
&
-\dfrac{3}{5}(\Delta_P+\Delta_c^2)
\\[0.8ex]
\dfrac{\Delta_c(3\Delta_P^2-2\Delta_P\Delta_c^2+\Delta_c^4)}{\Gamma_X}
&
-\dfrac{3\Delta_P^2-6\Delta_P\Delta_c^2+5\Delta_c^4}{4\Gamma_X}
&
\dfrac{2\Delta_c^3}{5\Gamma_X}
\\[1.0ex]
-\dfrac{9\Delta_P^3+9\Delta_P^2\Delta_c^2-3\Delta_P\Delta_c^4+\Delta_c^6}{3\Gamma_X}
&
\dfrac{\Delta_c(3\Delta_P^2-2\Delta_P\Delta_c^2+\Delta_c^4)}{2\Gamma_X}
&
-\dfrac{3\Delta_P^2+\Delta_c^4}{5\Gamma_X}
\end{array}
\right).
\label{eq:explicit_reduced_moment_inverse}
\end{equation}
\end{widetext}
Therefore the exact inverse of the original raw-moment Jacobian is
\begin{equation}
\mathsf J^{(I)-1}
=
\left(
\begin{array}{cc}
\mathsf I_2
&
-\mathsf B\mathsf S^{-1}
\\
\mathsf 0
&
\mathsf S^{-1}
\end{array}
\right)
\mathsf L.
\label{eq:explicit_factorized_moment_inverse}
\end{equation}
Equations~\eqref{eq:explicit_reduced_moment_inverse} and~\eqref{eq:explicit_factorized_moment_inverse} expose the two independent sources of ill conditioning: $1/\Gamma_X$ in the support-coordinate rows and $1/\mathcal D_J$ on the additional singular surface.
The numerical implementation evaluates the scaled condition number of Eq.~\eqref{eq:moment_jacobian_condition_number} before applying this inverse.
The Hessian $\mathsf H_{rAB}^{(I)}$ is obtained by differentiating Eq.~\eqref{eq:explicit_state_inmdf_jacobian}; it is retained in the It\^{o} correction of Eq.~\eqref{eq:inmdf_coordinate_drift} rather than omitted in the finite-coordinate dynamics.

\section{Analytic kinetic families used in the Kube benchmark}
\label{app:benchmark_families}

For reproducibility of the numerical comparison in Section~\ref{sec:kube_results}, we state here the exact one-dimensional parallel distributions and source-response maps used for the six analytic families.
Let
\begin{equation}
\xi
\doteq
\frac{v_\parallel-u_\parallel}{\sqrt{T/m}},
\end{equation}
and let all support centers and widths below be expressed in this dimensionless parallel coordinate.
The normalized Maxwellian reference is
\begin{equation}
\varphi_M(\xi)
\doteq
\frac{1}{\sqrt{2\pi}}
\exp\left(
-\frac{\xi^2}{2}
\right).
\label{eq:benchmark_maxwellian_parallel}
\end{equation}
For one or two localized INMDF channels,
\begin{equation}
\varphi_{I_L}(\xi\mid z)
=
\varphi_M(\xi)
+
\sum_{\ell=1}^{L}
g_\ell(z)
\frac{\xi-c_\ell}
{\sqrt{2\pi}W_\ell^{3/2}}
\exp\left[
-\frac{(\xi-c_\ell)^2}{2W_\ell}
\right],
\label{eq:benchmark_inmdf_parallel}
\end{equation}
where \(L\in\{1,2\}\) with the bounded source response
\(g_\ell(z) \doteq g_{\max} \tanh\left(a_{\Gamma,\ell}+b_{\Gamma,\ell}z\right)\).
Here \(g_{\max}\) is the fixed positivity-control amplitude used throughout the numerical analysis rather than an additional fitted coordinate.

For the Kappa family,
\begin{align}
\kappa(z)
&\doteq
\frac{3}{2}
+
\exp\left(
a_\kappa+b_\kappa z
\right),
\\
w_\kappa(z)
&\doteq
2\kappa(z)-3,
\end{align}
and
\begin{equation}
\varphi_\kappa(\xi\mid z)
=
\frac{
\Gamma\!\left[\kappa(z)\right]
}{
\Gamma\!\left[\kappa(z)-\frac{1}{2}\right]
\sqrt{\pi w_\kappa(z)}
}
\left[
1+
\frac{\xi^2}{w_\kappa(z)}
\right]^{-\kappa(z)}.
\label{eq:benchmark_kappa_parallel}
\end{equation}

For the Tsallis family,
\begin{align}
q(z)
&\doteq
1+
\frac{
q_\Delta
}{
1+
\exp\left[
-\left(
a_q+b_qz
\right)
\right]
},
\\
p(z)
&\doteq
\frac{1}{q(z)-1},
\\
w_q(z)
&\doteq
2p(z)-5,
\end{align}
where \(q_\Delta\) is a fixed admissible-span constant.
The normalized parallel distribution is
\begin{equation}
\varphi_q(\xi\mid z)
=
\frac{
\Gamma\!\left[p(z)-1\right]
}{
\Gamma\!\left[p(z)-\frac{3}{2}\right]
\sqrt{\pi w_q(z)}
}
\left[
1+
\frac{\xi^2}{w_q(z)}
\right]^{-\left[p(z)-1\right]}.
\label{eq:benchmark_tsallis_parallel}
\end{equation}

For the two-Maxwellian family, define
\begin{align}
r(z)
&\doteq
\frac{1}{
1+
\exp\left[
-\left(
a_r+b_rz
\right)
\right]
},
\\
\tau
&\doteq
1+\exp(a_T),
\end{align}
where \(r(z)\) is the tail-population fraction and \(\tau=T_{\rm tail}/T_{\rm core}>1\).
Then
\begin{equation}
\varphi_{2M}(\xi\mid z)
=
\left[
1-r(z)
\right]
\varphi_M(\xi)
+
\frac{
r(z)
}{
\sqrt{2\pi\tau}
}
\exp\left(
-\frac{\xi^2}{2\tau}
\right).
\label{eq:benchmark_two_maxwellian_parallel}
\end{equation}
The benchmark therefore compares three distinct non-Maxwellian mechanisms: localized odd deformations through the INMDF channels, symmetric broad tails through the Kappa and Tsallis families, and a second thermal population through the two-Maxwellian family.
All are then propagated through the same dimensionless half-space collection functional,
\begin{equation}
J_s(z)
\doteq
n(z)
\int_{-\infty}^{0}
(-\xi)
\varphi_s(\xi\mid z)
\,d\xi.
\label{eq:benchmark_halfspace_current}
\end{equation}
Writing \(\Phi\) for the standard-normal cumulative distribution function, the analytic currents evaluated in the numerical implementation are
\begin{align}
J_M(z)
={}&
\frac{n(z)}{\sqrt{2\pi}},
\\
J_{I_L}(z)
={}&
n(z)
\left[
\frac{1}{\sqrt{2\pi}}
+
\sum_{\ell=1}^{L}
g_\ell(z)
\left(
\Phi\left[
\frac{c_\ell}{\sqrt{W_\ell}}
\right]
-
1
\right)
\right],
\\
J_\kappa(z)
={}&
n(z)
\frac{
\sqrt{2\kappa(z)-3}
\,
\Gamma[\kappa(z)]
}{
2\sqrt{\pi}
[\kappa(z)-1]
\Gamma[\kappa(z)-\frac{1}{2}]
},
\\
J_q(z)
={}&
n(z)
\frac{
\sqrt{2p(z)-5}
\,
\Gamma[p(z)]
}{
2\sqrt{\pi}
[p(z)-1]
[p(z)-2]
\Gamma[p(z)-\frac{3}{2}]
},
\\
J_{2M}(z)
={}&
\frac{n(z)}{\sqrt{2\pi}}
\left[
1-r(z)
+
r(z)\sqrt{\tau}
\right].
\label{eq:benchmark_analytic_currents}
\end{align}
These distributions are propagated through the same source density, background-noise operation, standardization, and diagnostic comparison.
In the held-out calculations, all fitted kinetic-response coordinates appearing above are shared over the training conditions and frozen before the target PDF is generated.
The same analytic maps also supply the uncertainty calculation: combined with the common \(P_z(z)\) and background-noise operation, they generate the family-dependent predicted measurement density entering Eq.~\eqref{eq:predicted_measurement_density}.
Because the benchmark standardizes the resulting PDFs to the same mean and variance, differences in \(S\) and \(F\) isolate higher-order measurement structure that cannot be represented by a redefined \(\mathsf R\) or \(\mathsf S\).
Thus the formulas above specify not only candidate kinetic responses but distinct physically generated measurement-information channels for the Bayesian state update in Eq.~\eqref{eq:total_posterior_covariance}.

\section{Model-induced innovation inflation and physical regime-specific uncertainty}
\label{app:model_induced_innovation}

Section~\ref{sec:nmdf_uncertainty} separated the physical diagnostic likelihood from the additive instrumental-noise layer in Eq.~\eqref{eq:measurement_noise_convolution}, defined the complete predictive measurement density in Eq.~\eqref{eq:predicted_measurement_density}, and showed through Eqs.~\eqref{eq:posterior_covariance_exact} and~\eqref{eq:total_posterior_covariance} how that likelihood determines posterior state uncertainty.
The remaining question is what part of the inferred uncertainty can be created artificially when the retained kinetic manifold itself is too restrictive.
Consider the same experimental realization \(y_k^{\mathrm{meas}}\) evaluated with two kinetic descriptions.
Let \(M\) denote the MDF restriction and \(N\) an NMDF description that retains the non-Maxwellian structure required by the diagnostic in the regime being considered.
Their predictive probabilities are already generated by Eq.~\eqref{eq:predicted_measurement_density}.
When
\begin{equation}
p_{D,M}^{\mathrm{pred}}
\left(
y_k^{\mathrm{meas}}
\mid
\mathcal D_{k-1}
\right)
<
p_{D,N}^{\mathrm{pred}}
\left(
y_k^{\mathrm{meas}}
\mid
\mathcal D_{k-1}
\right),
\label{eq:app_mdf_lower_predictive_probability}
\end{equation}
the same observed measurement is less compatible with the MDF prediction.
Define its predictive surprisal under kinetic family \(s\) by
\begin{equation}
\mathcal J_{s,k}
\doteq
-
\ln
p_{D,s}^{\mathrm{pred}}
\left(
y_k^{\mathrm{meas}}
\mid
\mathcal D_{k-1}
\right).
\label{eq:app_predictive_surprisal}
\end{equation}
Equation~\eqref{eq:app_mdf_lower_predictive_probability} then implies \(\mathcal J_{M,k} > \mathcal J_{N,k}\).
This comparison does not yet modify the diagnostic noise; it states only that the restricted kinetic likelihood assigns less probability to the observed measurement.
A single large innovation remains compatible with a correctly specified asymmetric or heavy-tailed likelihood, so kinetic-model discrepancy is identified only when excess surprisal persists over repeated acquisitions or independent held-out conditions.
In that case the discrepancy contains systematic measurement structure that the forced MDF does not explain.
The effect need not arise from a displaced mean.
The MDF and NMDF likelihoods may have the same mean and variance while assigning different probabilities to the same measurement through their skewness, excess kurtosis, localized structure, or tail probability.
The standardized Alcator C-Mod PDFs provide exactly this situation because their first two moments are fixed while their higher-order likelihood structures remain different among kinetic families.
The predictive probability in Eq.~\eqref{eq:app_mdf_lower_predictive_probability}, rather than only the distance from the predicted mean, therefore provides the appropriate non-Gaussian measure of the kinetic discrepancy.
For the scalar-response construction used in the present benchmark, the kinetic discrepancy can be written directly at the diagnostic level.
Let \(y_{D,s}(\mathbf X_k)\) denote the noise-free scalar response already defined in Eq.~\eqref{eq:scalar_diagnostic_signal}, and let \(N\) denote the adequately resolved NMDF response for the regime under consideration.
The discrepancy produced by forcing the MDF is
\begin{equation}
\delta_{M,k}^{\mathrm{kin}}
\doteq
y_{D,N}(\mathbf X_k)
-
y_{D,M}(\mathbf X_k).
\label{eq:app_kinetic_discrepancy}
\end{equation}
If this discrepancy is not represented explicitly and the residual channel is allowed to compensate, the effective MDF residual becomes
\begin{equation}
\nu_{M,k}^{\mathrm{eff}}
\doteq
\nu_k
+
\delta_{M,k}^{\mathrm{kin}},
\label{eq:app_effective_measurement_residual}
\end{equation}
where \(\nu_k\) is the genuine instrumental-noise variable of Eq.~\eqref{eq:measurement_noise_convolution}.
Its covariance is
\begin{align}
\mathsf R_{M,k}^{\mathrm{eff}}
={}&
\mathsf R_k
+
\operatorname{Cov}
\left(
\delta_{M,k}^{\mathrm{kin}}
\right)
+
\operatorname{Cov}
\left(
\nu_k,
\delta_{M,k}^{\mathrm{kin}}
\right)
+
\operatorname{Cov}
\left(
\delta_{M,k}^{\mathrm{kin}},
\nu_k
\right).
\label{eq:app_effective_measurement_covariance}
\end{align}
When the instrumental noise and kinetic discrepancy are independent,
\begin{equation}
\mathsf R_{M,k}^{\mathrm{eff}}
=
\mathsf R_k
+
\operatorname{Cov}
\left(
\delta_{M,k}^{\mathrm{kin}}
\right).
\label{eq:app_effective_measurement_covariance_independent}
\end{equation}
Because \(\delta_{M,k}^{\mathrm{kin}}\) is generally state dependent, \(\mathsf R_{M,k}^{\mathrm{eff}}\) is not a new physical instrumental covariance.
It is the covariance representation obtained when that unresolved state-dependent kinetic discrepancy is absorbed into an additive residual channel.
A systematic nonzero mean of \(\delta_{M,k}^{\mathrm{kin}}\) additionally appears as an innovation bias and therefore remains distinguishable from the centered covariance contribution in Eq.~\eqref{eq:app_effective_measurement_covariance_independent}.
This additive decomposition is the Gaussian or mean-response specialization of the more general likelihood mismatch measured by Eq.~\eqref{eq:app_predictive_surprisal}.
It does not require the MDF and NMDF likelihoods to differ in their mean.
Two likelihoods can have the same mean and variance while assigning very different probabilities to the same measurement through their skewness, tails, localized structure, or multimodality.
In that case the difference is visible in \(\mathcal J_{s,k}\) even when a mean-level residual discrepancy is small.
A Gaussian covariance \(\mathsf R\) can broaden a likelihood, but it cannot reproduce arbitrary higher-order kinetic structure.
For the adequately resolved NMDF, define the analogous discrepancy \(\delta_{N,k}^{\mathrm{kin}}\) and effective residual by the same construction.
When the retained NMDF resolves the relevant kinetic structure, \(\delta_{N,k}^{\mathrm{kin}}\simeq0\), so that its effective covariance satisfies \(\mathsf R_{N,k}^{\mathrm{eff}}\simeq\mathsf R_k\).
For fixed \(\mathsf P_k^{\mathrm{pred}}\) and local diagnostic sensitivity \(\mathsf H_k\), suppose that
\begin{equation}
\mathsf R_{M,k}^{\mathrm{eff}}
\succeq
\mathsf R_{N,k}^{\mathrm{eff}}.
\label{eq:app_effective_noise_ordering}
\end{equation}
The Gaussian correction of Eqs.~\eqref{eq:discrete_innovation_covariance}--\eqref{eq:discrete_corrected_covariance} then gives
\begin{equation}
\mathsf P_{M,k}^{\mathrm{obs}}
\succeq
\mathsf P_{N,k}^{\mathrm{obs}}.
\label{eq:app_model_induced_covariance_ordering}
\end{equation}
The physical sequence is therefore: (i) the forced MDF leaves part of the measured kinetic structure unexplained, (ii) repeated unexplained structure enters the innovation statistics, (iii) an estimator that assigns this discrepancy to its residual channel infers an enlarged or more structured effective noise law, and (iv) the reduced information assigned to the measurement increases the posterior state uncertainty.
The increase is model induced: the instrumental noise itself has not changed.
The complete non-Gaussian update remains the Bayesian correction already derived in Eqs.~\eqref{eq:posterior_normalization}--\eqref{eq:posterior_covariance_exact}, and its local information content remains described by Eq.~\eqref{eq:fisher_information}.
No additional posterior or Fisher-information definition is required here.
The important distinction is that \(\mathsf R_{M,k}^{\mathrm{eff}}\) is only the Gaussian covariance representation of a more general residual compensation; outside that limit, the complete predictive likelihood and its surprisal retain the information that a single effective covariance cannot represent.
\\
The present Alcator C-Mod benchmark deliberately suppresses this compensation mechanism by holding the source statistics, background-noise operation, normalization, and diagnostic response fixed across all candidate kinetic families.
The MDF therefore cannot improve a deficient physical prediction by fitting a larger residual variance.
Differences among the MDF, INMDF, Kappa, Tsallis, and two-Maxwellian measurement PDFs are generated by their kinetic responses under the same prescribed noise controls, so unresolved MDF structure remains visible directly in the held-out PDF error and in the predictive probability assigned to the measurement.
In a recursive estimator that instead adapts its residual statistics from the innovations, this persistent discrepancy would be available to inflate the effective residual law.
The fixed-noise benchmark therefore exposes the physical origin of a contribution that a conventional adaptive filter could otherwise hide inside its inferred measurement noise.
The role of the NMDF likelihood is not to reduce the actual instrumental noise, but to prevent physically generated non-Gaussian structure from being transferred into that residual channel.
This distinction is also important for interpreting the MDF limit.
An NMDF description is not rendered inferior when the plasma happens to be close to Maxwellian.
For the INMDF hierarchy, the Maxwellian response is recovered when the additional localized kinetic coordinates become inactive.
In that regime the kinetic discrepancy associated with forcing the MDF vanishes, and the corresponding model-induced contribution to the effective residual vanishes with it.
The broader kinetic theory therefore recovers the MDF result when the additional kinetic structure is physically unnecessary, while remaining able to resolve departures from that limit when the diagnostic becomes sensitive to them.
\\
Once model-induced residual inflation has been separated from the physical likelihood, the remaining uncertainty has a direct plasma interpretation.
The predicted covariance already separates uncertainty inherited from the previous state estimate from uncertainty introduced by unresolved state dynamics through Eq.~\eqref{eq:total_state_covariance}.
A transient or instability can therefore increase \(\mathsf P_k^{\mathrm{pred}}\) when the physical state evolves more rapidly or through more unresolved channels, even though the diagnostic noise is unchanged.
The measurement correction contains complementary information: a change of kinetic regime modifies \(p_{D,s}(y\mid\mathbf X)\) and can make the same diagnostic more or less sensitive to the retained physical coordinates.
Posterior uncertainty can consequently change because of true instrumental noise, unresolved dynamics, state degeneracy, diagnostic sensitivity, transient evolution, or instability-driven kinetic restructuring.
In this regime the uncertainty includes a robustness score for the inference while its condition dependence provides additional information about the plasma physics.
For the standardized source coordinate \(z\) used in the present uncertainty analysis, let
\begin{equation}
V_z^{\mathrm{pred}}
\doteq
\operatorname{Var}_{P_z}(z)
\label{eq:app_prior_source_variance}
\end{equation}
denote the prior source variance, where \(P_z(z)\) remains the source probability density defined in Section~\ref{sec:kube_results}.
For a possible measurement \(y\), let
\begin{equation}
V_{z\mid y,s}^{\mathrm{obs}}
\doteq
\operatorname{Var}
\left(
z\mid y,s
\right)
\label{eq:app_conditional_source_variance}
\end{equation}
denote the corresponding posterior variance.
The local remaining-variance ratio is
\begin{equation}
U_s(y)
\doteq
\frac{
V_{z\mid y,s}^{\mathrm{obs}}
}{
V_z^{\mathrm{pred}}
}.
\label{eq:app_local_uncertainty_ratio}
\end{equation}
Its probability-weighted mean is
\begin{equation}
\overline U_s
\doteq
\int_{-\infty}^{\infty}
U_s(y)
p_{D,s}^{\mathrm{pred}}
\left(
y\mid\mathcal D_{k-1}
\right)
\,dy.
\label{eq:app_mean_uncertainty_ratio}
\end{equation}
Equation~\eqref{eq:total_posterior_covariance} immediately gives
\begin{equation}
0
\le
\overline U_s
\le
1.
\label{eq:app_mean_uncertainty_bounds}
\end{equation}
Thus \(\overline U_s\) measures the state variance remaining on average after the diagnostic update, while \(U_s(y)\) measures the robustness of that contraction with respect to the particular measurement realization.
A local value \(U_s(y)>1\) is compatible with Eq.~\eqref{eq:app_mean_uncertainty_bounds}: a particular measurement can broaden the conditional variance even though the diagnostic reduces variance on average over all possible measurements.
After model-induced residual inflation has been controlled, the dependence of \(U_s(y)\), \(\overline U_s\), and the complete posterior on plasma condition becomes physically informative.
A simultaneous displacement in the measured \(S\)-\(F\) plane and in the inferred uncertainty identifies a change in the non-Gaussian measurement-information channel; an increase of \(\mathsf P_k^{\mathrm{pred}}\) with unchanged instrumental statistics instead points toward unresolved state evolution; and a large \(U_s(y)\) localized to particular measurements identifies weak state observability or multiple physical states that remain compatible with the same diagnostic realization.
The uncertainty therefore retains its role as a robustness measure while also providing a regime-dependent probe of diagnostic observability, transient dynamics, instabilities, and the underlying kinetic state.

\section*{AUTHOR DECLARATIONS}

\subsection*{Conflict of Interest}

The author has no conflicts to disclose.

\subsection*{Author Contributions}

Olivier Izacard: Conceptualization; Data curation; Formal analysis; Investigation; Methodology; Software; Validation; Visualization; Writing -- original draft; Writing -- review \& editing.

\section*{Use of Artificial Intelligence}

OpenAI ChatGPT was used interactively for manuscript editing, mathematical and notational consistency checks, and literature organization.
AI-generated suggestions were treated as provisional and independently checked by the author.
The author is solely responsible for the scientific content, theoretical derivations, numerical methodology, interpretation, and conclusions.
No generative AI was used to generate, modify, or fabricate research data or experimental results.

\section*{DATA AVAILABILITY}

The experimental data analyzed in this study are available in Ref.~[35].
The digitized data and numerical data generated in this study are
available from the corresponding author upon reasonable request.

\phantomsection
\section*{References}
\addcontentsline{toc}{section}{References}
\pdfbookmark[1]{References}{references}
\begingroup
\sloppy

\endgroup

\end{document}